\documentclass[journal]{new-aiaa}

\usepackage{textcomp}
\usepackage{graphicx}
\usepackage{tabularx,array}
\usepackage{makecell}
\usepackage{subcaption}
\usepackage[percent]{overpic}
\usepackage{amsmath}
\usepackage[version=4]{mhchem}
\usepackage{siunitx}
\usepackage{longtable}
\usepackage{xcolor}
\usepackage{tikz}
\usetikzlibrary{shapes.geometric}
\usepackage{float}
\usepackage{placeins}
\newcommand{\legendsquare}[3]{%
  \begingroup
  \definecolor{tmpdraw}{HTML}{#1}%
  \definecolor{tmpfill}{HTML}{#2}%
  \tikz[baseline=-2.6]{
    \coordinate (A) at (-#3,#3);
    \coordinate (B) at (#3,#3);
    \coordinate (C) at (#3,-#3);
    \coordinate (D) at (-#3,-#3);
    \fill[tmpfill] (A)--(B)--(C)--(D)--cycle;
    \draw[tmpdraw, line width=0.5 ] (A)--(B)--(C)--(D)--cycle;
  }
  \endgroup
}
\newcommand{\legendsquarethick}[3]{%
  \begingroup
  \definecolor{tmpdraw}{HTML}{#1}%
  \definecolor{tmpfill}{HTML}{#2}%
  \tikz[baseline=-2.6]{
    \coordinate (A) at (-#3,#3);
    \coordinate (B) at (#3,#3);
    \coordinate (C) at (#3,-#3);
    \coordinate (D) at (-#3,-#3);
    \fill[tmpfill] (A)--(B)--(C)--(D)--cycle;
    \draw[tmpdraw, line width=1 ] (A)--(B)--(C)--(D)--cycle;
  }
  \endgroup
}
\newcommand{\legendcircle}[3]{%
  \begingroup
  \definecolor{tmpdraw}{HTML}{#1}%
  \definecolor{tmpfill}{HTML}{#2}%
  \tikz[baseline=-2.4]{
    \fill[tmpfill] (0,0) circle (#3);
    \draw[tmpdraw, line width=0.5 ] (0,0) circle (#3);
  }
  \endgroup
}
\newcommand{\legendcirclethick}[3]{%
  \begingroup
  \definecolor{tmpdraw}{HTML}{#1}%
  \definecolor{tmpfill}{HTML}{#2}%
  \tikz[baseline=-2.4]{
    \fill[tmpfill] (0,0) circle (#3);
    \draw[tmpdraw, line width=1 ] (0,0) circle (#3);
  }
  \endgroup
}
\newcommand{\legendhexagon}[3]{%
  \begingroup
  \definecolor{tmpdraw}{HTML}{#1}%
  \definecolor{tmpfill}{HTML}{#2}%
  \tikz[baseline=-2.4]
    \node[draw=tmpdraw, fill=tmpfill, regular polygon, regular polygon sides=6,
          rotate=30,
          minimum size=#3, inner sep=0 ] {};
  \endgroup
}
\newcommand{\legenduptriangle}[3]{%
  \begingroup
  \definecolor{tmpdraw}{HTML}{#1}%
  \definecolor{tmpfill}{HTML}{#2}%
  \tikz[baseline=-2.4] 
    \node[draw=tmpdraw, fill=tmpfill, regular polygon, regular polygon sides=3,
          minimum size=#3, inner sep=0, line width=0.8 ] {};
  \endgroup
}
\newcommand{\legenddowntriangle}[3]{%
  \begingroup
  \definecolor{tmpdraw}{HTML}{#1}%
  \definecolor{tmpfill}{HTML}{#2}%
  \tikz[ baseline=-2.4] \node[draw=tmpdraw, fill=tmpfill, regular polygon, regular polygon sides=3,
        rotate=180, minimum size=#3, inner sep=0 ] {};
  \endgroup
}

\newcommand{\legendx}[2]{%
  \begingroup
  \definecolor{tmpdraw}{HTML}{#1}%
  \tikz[ baseline=-2.4] {
    \draw[tmpdraw, line width=1.2 ] (-#2/2,-#2/2) -- (#2/2,#2/2);
    \draw[tmpdraw, line width=1.2 ] (-#2/2,#2/2) -- (#2/2,-#2/2);
  }
  \endgroup
}

\newcommand{\legendy}[2]{%
  \begingroup
  \definecolor{tmpdraw}{HTML}{#1}%
  \tikz[ baseline=-2.4]{
    \pgfmathsetmacro{\H}{#2}       
    \pgfmathsetmacro{\L}{\H/8}     
    \draw[tmpdraw, line width=1.2 ] (0,0) -- (0,-\L);                 
    \draw[tmpdraw, line width=1.2 ] (0,0) -- ({\L*sin(55)},{\L*cos(55)}); 
    \draw[tmpdraw, line width=1.2 ] (0,0) -- ({-\L*sin(55)},{\L*cos(55)});
  }
  \endgroup
}
\newcommand{\legendcross}[2]{%
  \begingroup
  \definecolor{tmpdraw}{HTML}{#1}%
  \tikz[ baseline=-2.4] {
    \draw[tmpdraw, line width=0.8 ] (-#2/2,0) -- (#2/2,0);
    \draw[tmpdraw, line width=0.8 ] (0,-#2/2) -- (0,#2/2);
  }
  \endgroup
}
\newcommand{\legendcrossthick}[2]{%
  \begingroup
  \definecolor{tmpdraw}{HTML}{#1}%
  \tikz[ baseline=-2.4] {
    \draw[tmpdraw, line width=1.2 ] (-#2/2,0) -- (#2/2,0);
    \draw[tmpdraw, line width=1.2 ] (0,-#2/2) -- (0,#2/2);
  }
  \endgroup
}

\title{Component-wise outer similarity of turbulent boundary layers with strong adverse-pressure gradient}

\author{Yu-Xuan Shi\footnote{PhD Student, Institute of Fluid Mechanics.}, Mingze Ma\footnote{PhD, Institute of Fluid Mechanics.} and Xi Chen\footnote{Professor, Institute of Fluid Mechanics. Corresponding author. Email: \texttt{chenxi97@outlook.com}}}

\affil{Key Laboratory of Fluid Mechanics of Ministry of Education, Beihang University (Beijing University of Aeronautics and Astronautics), Beijing 100191, PR China}
\begin{document}
\maketitle

\begin{abstract}
The adverse-pressure-gradient (APG) effect challenges the classical scaling of turbulent boundary layers (TBLs) over a flat plate, and it remains unclear whether outer-layer similarity holds across a wide range of APG strengths.
We assemble a benchmark comprising the present direct numerical simulation of a suction--blowing-induced separation bubble and published profiles, spanning Clauser parameters $\beta_e=0.04$--$180$ and Reynolds numbers $Re_\theta=530$--$19000$.
Across this broad $(\beta_e,Re_\theta)$ parameter range, existing APG normalizations do not simultaneously collapse the two mean-velocity defects and four Reynolds-stress components, leaving open whether outer similarity persists.
Instead of defining new scalings, we normalize each quantity by its value at a common location $y_s$.
Results are shown using $y_s=0.5\delta$, beyond the outer stress peaks.
For each of the six quantities, the normalized profiles exhibit a common outer shape across the conditions, with collapse emerging for $y/\delta\gtrsim0.2$ for the velocity defects and $y/\delta\gtrsim0.4$ for the Reynolds stresses.
The collapse-error analysis shows that outer self-similarity is retained over a broad range of $y_s$ within the outer region and is not restricted to $y_s=0.5\delta$.
Furthermore, explicit expressions are provided for the normalized outer profiles.
These results support component-wise outer similarity under strong APGs and provide compact references for APG turbulence-model assessment and conditional profile reconstruction.
\end{abstract}

\section{\label{introduction}Introduction}
Turbulent boundary layers (TBLs) subjected to adverse pressure gradients (APGs) are central to many aerodynamic flows, including those over airfoils and wings, in diffusers and inlets, and through turbomachinery passages.
In these configurations, APG effects strongly influence skin friction, pressure recovery, aerodynamic losses, and the onset of separation.
Accurate prediction remains challenging because APGs alter not only the near-wall region but also the outer-layer velocity defect and Reynolds stresses.
Capturing these outer-layer changes requires extending the classical wake-law description established for zero-pressure-gradient (ZPG) flows to APG mean-defect and stress profiles.

For ZPG TBLs, the underlying scaling problems have been studied in depth, including the value of the von Karman constant \citep{cite-3,cite-1,she_quantifying_2017,cite-4} and the scaling of Reynolds normal stresses \citep{perry_theoretical_1986,chen_reynolds_2021,Pirozzoli_2024,Nagib_Marusic_2025}.
These ZPG-based descriptions provide a useful baseline, but they do not transfer directly to APG TBLs, for which our understanding remains relatively limited \citep{wei_integral_2017, coleman_numerical_2018,cite-13}.
APGs introduce strong non-equilibrium effects, while upstream history further influences the outer-layer response, causing the profiles to deviate from conventional self-similar behavior.
Meanwhile, there remains no consensus on the choice of APG similarity variables, especially under conditions of strong APG, which further limits the accuracy of turbulence modeling \citep{cite-44}.
Resolving these issues first requires comparisons across a broad range of experimental and numerical data; more importantly, it requires an APG description that develops naturally from the classical wake-law framework.

The classical wake-law framework is rooted in the ZPG formulation of \citet{coles_law_1956}, which represents the outer-layer velocity defect through a wake function:

\begin{equation}\label{eq:wall-wake-formula}
    \frac{U}{u_\tau} = \frac{1}{\kappa} \ln \left( \frac{y u_\tau}{\nu}\right) + B + \frac{\Pi(x)}{\kappa}\omega\left(\frac{y}{\delta}\right),
\end{equation}
where $U$ is the streamwise velocity, and $y$ is the wall normal distance.
For the wall part, $u_\tau=\sqrt{\tau_w/\rho}$ is the friction velocity defined by the wall shear stress $\tau_w$ and density $\rho$, $\nu$ is the kinematic viscosity and $\kappa$ (the Karman constant), $B$ are coefficients in the classical logarithmic law.
For the wake part, $\Pi$ is the wake strength parameter, $\delta$ is the boundary layer thickness and $\omega$ is the wake function.
Subsequent studies \citep{Coles1969PROCEEDINGSCO,granville_amodified_1976,lewkowicz_animporved_1982} have proposed alternative forms of the wake function, reflecting continued interest in accurately characterizing the outer layer behavior.
More recently, \citet{krug_revisiting_2017} reviewed the classical wake functions and proposed a two-state wake model that distinguishes between ``pure wall flow" and ``free stream" states.
The enduring focus on the wake function underscores its fundamental importance in characterizing the structural organization of turbulent boundary layers, particularly in capturing the self-similar features of the outer region.

However, in the case of APG TBLs, the validity of the ZPG similarity profile is severely challenged, and neither the classical wall unit scaling $u_\tau$ nor the conventional freestream velocity scaling $U_e$ yields collapsed profiles.
The primary manifestation of the APG effects is an increasingly pronounced wake region that is observed by numerous researchers, as summarized by \citet{vinuesa_pressure_2017}.
The evolution of the wake strength parameter $\Pi$ with increasing pressure gradient was systematically examined by \citet{perry_streamwise_2002}.
Thus, the question is not only whether the original ZPG wall-wake formula remains applicable, but also whether a wake-law-type outer defect shape can still be recovered when the APG-dependent velocity scale is chosen appropriately.
More recently, \citet{cite-15} and \citet{cite-14} proposed a composite mixing length model for turbulent wall flows, from which the mean velocity profile can be integrated.
This model is organized based on the pipe, channel and ZPG TBLs and captures the characteristics of the APG through parameter adjustment of the outer flow mixing length function.
Meanwhile, the inner velocity profile also deviates from the logarithmic law and leads to the controversy on whether the logarithmic law breaks down ``generally" \citep{cite-12} or ``progressively" \citep{galbraith_mixing_1977}.
The former indicates a complete failure of the logarithmic function and hence requires a new form for the velocity profile, while the latter means a gradual shrinkage of the classical logarithmic region as APG grows stronger \citep{cite-13}.
In addition, for Reynolds shear stresses that are thought to scale with $u_\tau^2$ in ZPG, the $-\langle uv\rangle$ profiles in APG increase significantly in magnitude and exhibit an outward shift of the peak, while Reynolds normal stress $\left<uu\right>$ not only shows a $Re$-dependent inner peak, but also develops a clear outer peak that has not been confirmed for ZPG TBLs \citep{alfredsson_new_2011,chen_reynolds_2023}.
For Reynolds-averaged and large-eddy simulation approaches, such changes complicate the assessment of modeled Reynolds stresses and the construction of outer-layer or wall-model closures in APG aerodynamic boundary layers.
These issues  have been discussed by \citet{cite-32} and \citet{Ma2025} and call for a different consideration of APG scaling.

Various attempts have been made to establish APG scaling based on local velocity and length characteristics.
Earlier work by \citet{mellor_equilibrium_1966} introduced a ``pressure velocity" derived from the local pressure gradient and displacement thickness to scale the velocity defect under strong APG.
\citet{cite-34} proposed a velocity scale through the square root of the maximum Reynolds shear stress, whose thickness away from the wall is adopted as the corresponding length scale.
To describe the similarity of APG TBLs approaching separation, \citet{perry_study_1975} introduced a constant vorticity layer that leads to a linear description of velocity defect with two common parameters.
Moreover, \citet{cite-35} introduced the mean velocity at the maximum Reynolds shear stress location to scale the velocity defect.
Similarly, \citet{gungor_analysis_2014, gungor_scaling_2016} chose the mean velocity defect at half the boundary layer thickness as the velocity scale, while \citet{cite-37}, motivated by inflection instability, chose the outermost inflection point as the reference location for outer scaling.
In addition, integral properties across the boundary layer thickness have also been postulated to scale the APG outer profiles.
\citet{zagarola_smits_1998} (hereafter abbreviated as ZS) developed an integral scaling by the ratio between displacement and boundary layer thicknesses, originally proposed for velocity defects in pipes, now extended to channel and APG TBL \citep{buschmann_recent_2006,castillo_similarity_2001, maciel_self-similarity_2006}.
The ZS proposal, although it works well for the mean velocity, is not applicable to Reynolds stresses \citep{cite-18}.
To address this limitation, \citet{han_consistent_2024} introduced the distance responsive to the non-equilibrium conditions of APG TBLs, producing an improved form of ZS scaling.
Mixed scaling schemes combining friction velocity and a pressure gradient based velocity scale have also been proposed for Reynolds stresses \citep{cite-40, cite-41}.
Moreover, \citet{castillo_similarity_2001} examined the equilibrium functions and suggested the freestream velocity as the outer scaling.
Recently, \citet{cite-42} summarized relations of coordinate transform and scaling equivalence through self-similar analysis.
Approaches of asymptotic expansion are also used for the scaling of APG TBLs, which can be found in \citet{durbin_scaling_1992}, \citet{Afzal2008}, \citet{cite-18} and \citet{cite-20}.
Nevertheless, a unified description of outer profiles remains limited, especially for strong non-equilibrium APG flows where mean velocities and Reynolds stresses require separate but physically compatible scaling treatments.

Despite these efforts, establishing general scaling for APG TBLs remains challenging, primarily due to the breakdown of self-similarity in the outer region.
This breakdown stems from two inherent difficulties: non-equilibrium and history effects.
A strictly equilibrium flow, where all normalized mean quantities become streamwise-invariant \citep{townsend_shear_flow, rotta_turbulent_1962, cite-3, cite-23}, is rarely achieved outside of idealized cases such as sink flow \citep{dixit_pressure_2008, chen_similarity_2017}.
Regarding the influence of the flow history \citep{vinuesa_revisiting_2017,cite-28,cite-30, vinuesa_experiments_2014}, even when Reynolds numbers and pressure gradient parameters are matched, the outer velocity profiles may still deviate from each other due to differences in inflow conditions \citep{bobke_history_2017}.
\citet{gungor_response_2024} showed that the outer flow responds with a considerable delay to changes in pressure forcing and that parameters based solely on local variables cannot fully describe non-equilibrium TBLs.
For weak-to-strong APG TBLs, \citet{mahajan_upstream_2026} further demonstrated the influence of upstream pressure-gradient history and local disequilibration on the scale-dependent outer-layer energy.
Consequently, suitable scaling for APG TBLs across a wide range of flow conditions and pressure gradients is still being sought.

The direct simulation in the present study aims to provide statistics of a TBL with strong APG effects to examine outer similarity and normalization strategies for both mean velocities ($U$ and $V$) and the full Reynolds stress components ($\langle uu \rangle$, $\langle vv \rangle$, $\langle ww\rangle$ and $\langle uv\rangle$).
The imposed pressure-gradient condition and the moderate Reynolds-number range make the strong APG effects readily identifiable while permitting direct comparison with the LB case of \citet{abe_reynolds_2017,abe_direct_2019}, thereby providing a benchmark for testing APG normalizations and outer similarity.
Previous simulations of APG TBLs with separation bubbles include those of \citet{cite-6}, \citet{abe_reynolds_2017}, \citet{coleman_numerical_2018}, and \citet{ambrogi_characterization_2022}, while cases of near-equilibrium APG TBLs have also been extensively studied in numerical simulations \citep{bobke_large-eddy_2016, bobke_history_2017, kitsios_direct_2017,cite-29,rkein_direct_2023}.
In addition, experimental measurements of APG TBLs have also been presented by, \textit{e.g.}, \citet{perry_study_1975}, \citet{Dengel_Fernholz_1990}, \citet{marusic_wall_wake_1995}, \citet{sanmiguel_vila_experimental_2020}, \citet{cite-13}, and \citet{cite-27}.
Using these data, we first demonstrate the inadequacy of previously proposed outer scalings, which calls into question the existence of outer-layer similarity under strong APG conditions.
Then, using component-wise reference values at a common outer location, we obtain consistent data collapses in the outer flow region, which suggest
\begin{equation}\label{eq:sim-define}
\begin{aligned}
    \frac{U_e-U}{U_s} = f_1 \left(\frac{y}{\delta}\right),&~~~~\frac{V_e-V}{V_s} = f_2 \left(\frac{y}{\delta}\right),\\
    \frac{\langle uv\rangle}{\langle uv\rangle_s} = g_0 \left(\frac{y}{\delta}\right), &~~~~\frac{\langle uu\rangle}{\langle uu\rangle_s} = g_1 \left(\frac{y}{\delta}\right),\\
    \frac{\langle vv\rangle}{\langle vv\rangle_s} = g_2 \left(\frac{y}{\delta}\right), &~~~~\frac{\langle ww\rangle}{\langle ww\rangle_s} = g_3 \left(\frac{y}{\delta}\right).
\end{aligned}
\end{equation}
Here, the subscript $s$ denotes normalization by a reference value sampled from each quantity at a common wall-normal height (e.g., $y_s = \delta/2$), and $f_i$ and $g_i$ are self-similar functions.
This normalization strategy differs from the conventional global scaling (e.g., $u_\tau$ or $U_e$) used for ZPG TBLs and reveals component-wise outer similarity under strong APG.
The outer-region collapse covers Clauser parameters $\beta_e=0.04$--$180$ and Reynolds numbers $Re_\theta=530$--$19000$.
Also, \citet{gungor_scaling_2016} previously demonstrated collapse of the mean velocity defect using a half-thickness scale that is equivalent, up to a constant factor, to the present $U_s$ at $y_s=0.5\delta$.
The present study examines outer similarity over a much wider APG range and across a broader set of quantities, including the wall-normal velocity and Reynolds stresses.
Finally, we provide explicit expressions for these self-similar functions and show their agreement with DNS \cite{coleman_numerical_2018,coleman_numerical_2021,rkein_direct_2023} and experimental data \cite{marusic_wall_wake_1995,Dengel_Fernholz_1990}.

In the following material, section \ref{sec:DNS_setup} describes the present numerical setup; section \ref{sec:DNS_results} presents simulation results and provides a detailed comparison with published results.
In section \ref{sec:self-sim}, evidence of component-wise outer similarity is presented through reference-value normalization, which is compared with existing APG normalizations, and functional expressions are given in the subsequent discussions for velocities and Reynolds stresses.
In section \ref{sec:further-verif}, further verifications with more experimental and numerical data sets are provided.
Conclusions are drawn in section \ref{conclusion}.

\section{\label{sec:DNS_setup}Numerical Setup}
\begin{figure}[!htbp]
    \centering
    \begin{subfigure}{0.95\textwidth}
        \centering
        \begin{overpic}[width=\linewidth, trim = 50 150 50 50, clip]{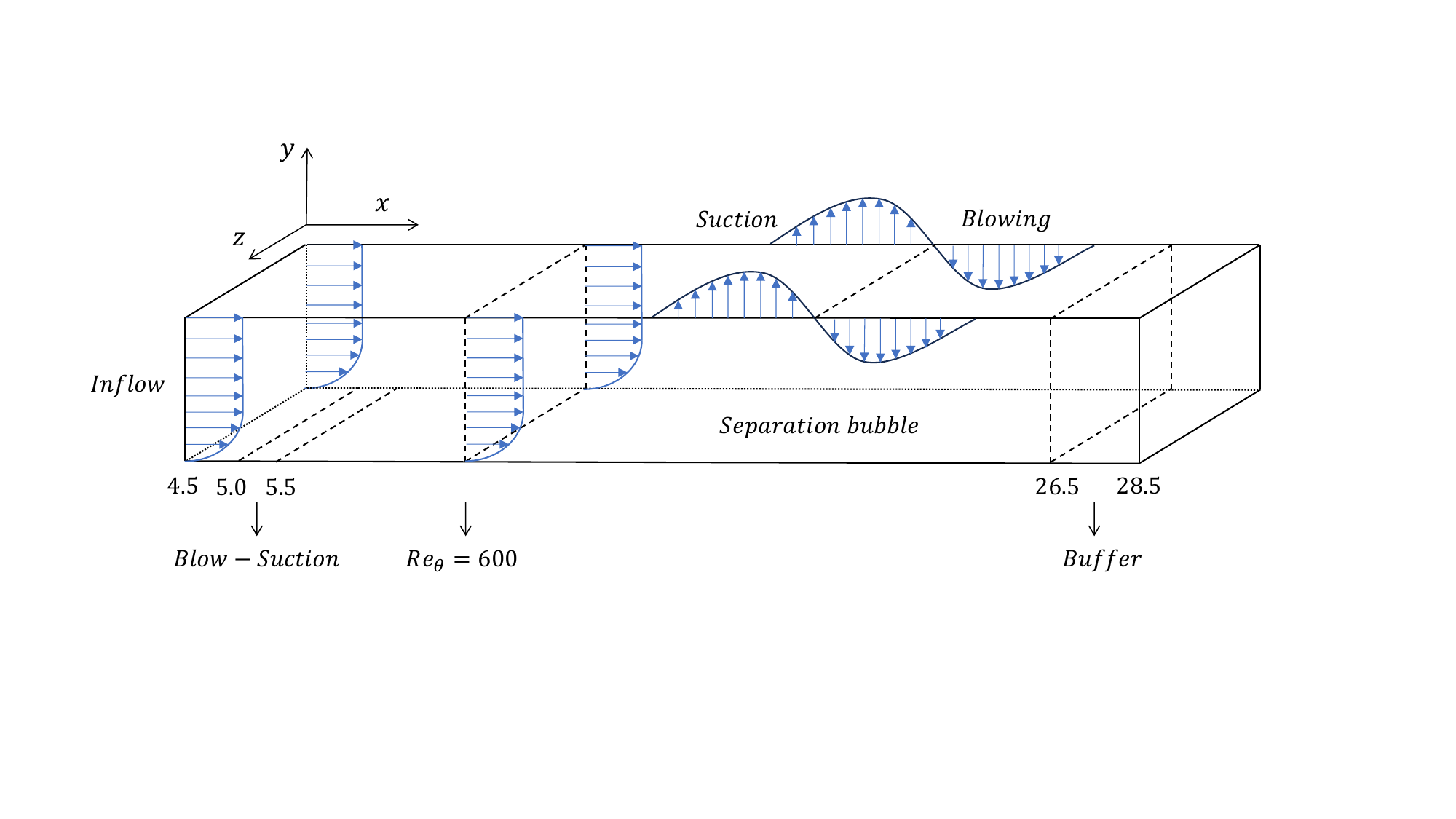}
            \put(0,38){\small(\textit{a})}
        \end{overpic}
    \end{subfigure}\\[0.1em]
    \begin{subfigure}{0.95\textwidth}
        \centering
        \begin{overpic}[width=\linewidth]{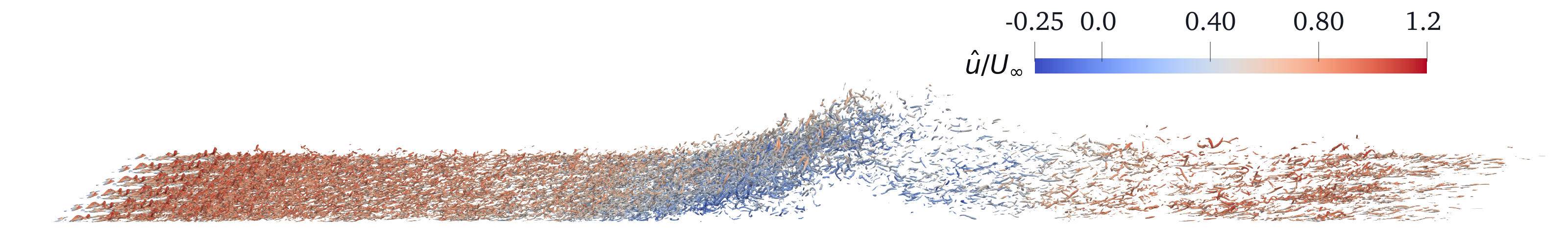}
            \put(0,14){\small(\textit{b})}
        \end{overpic}
    \end{subfigure}
    \caption{Computational setup and instantaneous vortical structures of the turbulent boundary layer.}
    \label{fig:cfd-domain}
\end{figure}

We perform a DNS of a flat-plate turbulent boundary layer subjected to an APG to examine the mean-velocity and Reynolds-stress statistics in detail.
The Reynolds number is kept moderate because Reynolds-number effects resemble those induced by a pressure gradient \cite{cite-28}, whereas at higher Reynolds numbers the APG features may become less distinct \cite{cite-29}.
Since the present study seeks common outer-layer behavior over a broad range of strong APGs, we consider a spatially developing TBL that evolves from a ZPG state to an APG sufficiently strong to cause separation.
The overall configuration is designed to reproduce conditions similar to those of the LB case of \citet{abe_reynolds_2017}.

As illustrated in Fig.~\ref{fig:cfd-domain}(\textit{a}), a laminar profile at a momentum-thickness Reynolds number of $Re_\theta=190$ is prescribed at the inlet, and wall-normal velocity disturbances are applied shortly downstream to trigger transition.
Variations in the flow velocity are negligible throughout the upstream development region, which extends over more than $200\theta_0$, where $\theta_0$ is the momentum thickness at $Re_\theta=600$.
After transition, the fully turbulent boundary layer develops under approximately ZPG conditions to $Re_\theta\approx600$ before entering the APG region.
Farther downstream, suction at the upper boundary produces an APG strong enough to cause separation.
The suction subsequently changes to blowing, causing the separated TBL to reattach and recover under a favorable pressure gradient (FPG).
The instantaneous $Q$-criterion visualization in Fig.~\ref{fig:cfd-domain}(\textit{b}) shows three-dimensional vortical structures throughout the developed boundary layer and the APG region.

The present DNS is conducted using the open-source solver OpenCFD-SC, which solves the compressible Navier--Stokes equations and has been verified for a wide variety of wall flows \citep{cite-20,cite-53,cite-55,cite-56,cite-54}.
The governing equations are presented as follows:

\begin{align}
    \frac{\partial \hat{\rho}}{\partial t} + \frac{\partial (\hat{\rho} \hat{u}_j)}{\partial x_j}&=0,\label{eq:goveq-1}\\
    \frac{\partial \hat{u}_i}{\partial t}+\hat{u}_j\frac{\partial \hat{u}_i}{\partial x_j}&=-\frac{1}{\hat{\rho}}\frac{\partial \hat{p}}{\partial x_i} + \frac{1}{\hat{\rho}}\frac{\partial \hat{\sigma}_{ij}}{\partial x_j},\\
\hat{\rho}
\frac{\partial}{\partial t}
\left(
\hat{e}
+
\frac{\hat{u}_i \hat{u}_i}{2}
\right)
+
\hat{\rho}\hat{u}_j
\frac{\partial}{\partial x_j}
\left(
\hat{e}
+
\frac{\hat{u}_i \hat{u}_i}{2}
\right)
&=
-
\frac{\partial (\hat{p}\hat{u}_j)}{\partial x_j}
+
\frac{\partial}{\partial x_i}
\left(
\hat{\sigma}_{ij}\hat{u}_j
\right)
+
\frac{\partial}{\partial x_i}
\left(
\hat{\lambda}
\frac{\partial \hat{T}}{\partial x_i}
\right).
\end{align}
Here, the hat denotes instantaneous values.
$\hat{\rho}$, $\hat{u}_i$, $\hat{p}$, $\hat{e}$, and $\hat{T}$ represent the density, velocity, pressure, specific internal energy, and temperature of the gas, respectively.
$\hat{\lambda}$ is the thermal conductivity and $\hat{\sigma}_{ij}$ is the stress tensor:

\begin{equation}
    \hat{\sigma}_{ij}=\hat{\mu}\left(\frac{\partial \hat{u}_i}{\partial x_j}+\frac{\partial \hat{u}_j}{\partial x_i}\right)-\frac{2}{3}\hat{\mu}\frac{\partial \hat{u}_k}{\partial x_k}\delta_{ij}, \label{eq:tauij}
\end{equation}
where $\hat{\mu}$ is the molecular viscosity and $\delta_{ij}$ is the Kronecker delta.
For an ideal gas, the equation of state can be expressed as $\hat{p}=(\hat{\rho} \hat{T})/(\gamma Ma^2)$, in which $Ma$ is the Mach number and $\gamma=c_p/c_v$ (taken as 1.4 in the present simulation) is the specific heat ratio.
Thermal conductivity $\hat{\lambda}$ is computed using a constant Prandtl number $Pr=\hat{\mu} c_p/\hat{\lambda}=0.7$.
Meanwhile, the Sutherland equation is adopted to compute viscosity:

\begin{equation}
    \hat{\mu}=\mu_\infty \left(\frac{\hat{T}}{T_\infty}\right)^{\frac{3}{2}}\left(\frac{T_\infty+T_s}{\hat{T}+T_s}\right),
\end{equation}
where $T_\infty$ and $\mu_\infty$ are the reference temperature and viscosity of the inlet freestream and $T_s=110.4\text{K}$ is the Sutherland constant of the air.

The Reynolds number based on the global reference velocity ($U_\infty=1$) and length ($L=1$) is $Re_\infty=2\times10^4$.
The Mach number is set to $0.3$ to approximate the incompressible conditions of \citet{abe_reynolds_2017}.
The variations in density $\hat{\rho}$, dynamic viscosity $\hat{\mu}$, and kinematic viscosity $\hat{\nu}=\hat{\mu}/\hat{\rho}$ remain within 2\% and are therefore neglected.
A seventh-order upwind difference scheme is adopted for the inviscid terms, and an eighth-order central difference scheme is used for the viscous terms.
A three-step Runge--Kutta method is used for time integration.
The computational domain size $L_x\times L_y\times L_z$ is set to $24\times2.52\times5$, and the total number of grid points is $1900\times360\times1096$.
Details of the coordinate ranges, grid numbers, and grid spacings are presented in table \ref{tab:grid_domain}.

\begin{table}
    \centering
    \caption{Coordinate ranges, grid-point counts, and mean grid spacings in the computational subdomains.}
    \label{tab:grid_domain}
    \begin{tabular}{cccccc}
      \hline
      $x/L$ & $N_x\times N_y\times N_z$ & $\Delta x$ & $\Delta y$ & $\Delta z$ \\\hline
      $4.5 \sim 5.0$ & $30 \times 360 \times 1096$ & 0.0167 & $4.6\times10^{-4}\sim0.0142$ & $4.56\times10^{-3}$ \\
      $5.0\sim5.5$ & $40 \times 360 \times 1096$ & 0.0125 & $4.6\times10^{-4}\sim0.0142$ & $4.56\times10^{-3}$ \\
      $5.5\sim26.5$ & $1780 \times 360 \times 1096$ & 0.0118 & $4.6\times10^{-4}\sim0.0142$ & $4.56\times10^{-3}$ \\
      $26.5\sim28.5$ & $50 \times 360 \times 1096$ & 0.04 & $4.6\times10^{-4}\sim0.0142$ & $4.56\times10^{-3}$ \\\hline
    \end{tabular}
\end{table}

No-slip and isothermal conditions are imposed at the wall, and a periodic boundary condition is used in the spanwise $z$-direction.
To trigger transition to turbulence, wall-normal blowing and suction disturbances are imposed over $x/L=5.0\sim5.5$ (Fig.~\ref{fig:cfd-domain}).
The pressure gradient and separation bubble farther downstream are generated using the upper-boundary transpiration velocity distribution of \citet{abe_direct_2019}:

\begin{equation}\label{eq:trans-vel}
    V_{\text{top}}\left(x\right)=\sqrt{2}V_{\text{max}}\frac{x_c-x}{\sigma}\exp\left[\psi-\left(\frac{x_c-x}{\sigma}\right)^2\right]
\end{equation}
together with conditions of vanishing spanwise vorticity and spanwise velocity gradient:

\begin{equation}
    \left.\frac{\partial U}{\partial y}\right\vert_{\text{top}}=\frac{dV_{\text{top}}}{dx}, \quad\quad \frac{\partial W}{\partial y}=0.
\end{equation}
Here, the parameters $V_{\text{max}}=0.33525$, $\sigma=2.65$, and $\psi=0.95$ are identical to those of the LB case of \citet{abe_direct_2019}.
The center location, $x_c=17.89$, is selected such that the center of the suction--blowing distribution is located $195\theta_0$ downstream of the station where $Re_\theta=600$, as in the LB case.
Compared with the LB case of Abe, the principal differences are as follows.
First, the compressible equations are solved at $Ma=0.3$ in the present simulation, whereas the incompressible equations were used by Abe.
Second, the incoming turbulent boundary layer is generated through disturbance-induced transition from a laminar inlet, in contrast to the recycling--rescaling method \citep{lund_generation_1998} used by \citet{abe_reynolds_2017}.
Finally, an extrapolation condition is imposed at the present outlet, while a convective condition was adopted by Abe.
Accordingly, a buffer region with coarser grids is placed over $x/L=26.5\sim28.5$ immediately upstream of the outlet to reduce numerical reflections and instability.

\section{\label{sec:DNS_results} DNS Results}

Statistics are obtained here by performing time and spanwise averaging of 2000 instantaneous fields that are sampled from a simulation period of $20L/U_\infty$.
In the following of this paper, the brackets denote the mean.
Unless otherwise specified, $U$, $V$ and $W$ denote the mean velocity component in the $x$, $y$ and $z$ directions, respectively, and the lowercase $u$, $v$ and $w$ are the velocity fluctuations in the corresponding directions.
$P$ is the mean pressure and $p$ represents the pressure fluctuation.
Moreover, $\tilde{x}=x-x_c$ is the distance to the center of the separation bubble and the momentum thickness $\theta_0$ is used to normalize both coordinates $x$ and $y$.

\begin{figure}[!htbp]
    \centering
    \begin{subfigure}{0.9\textwidth}
        \centering
        \begin{overpic}[width=\linewidth, trim=0 30pt 0 0, clip]{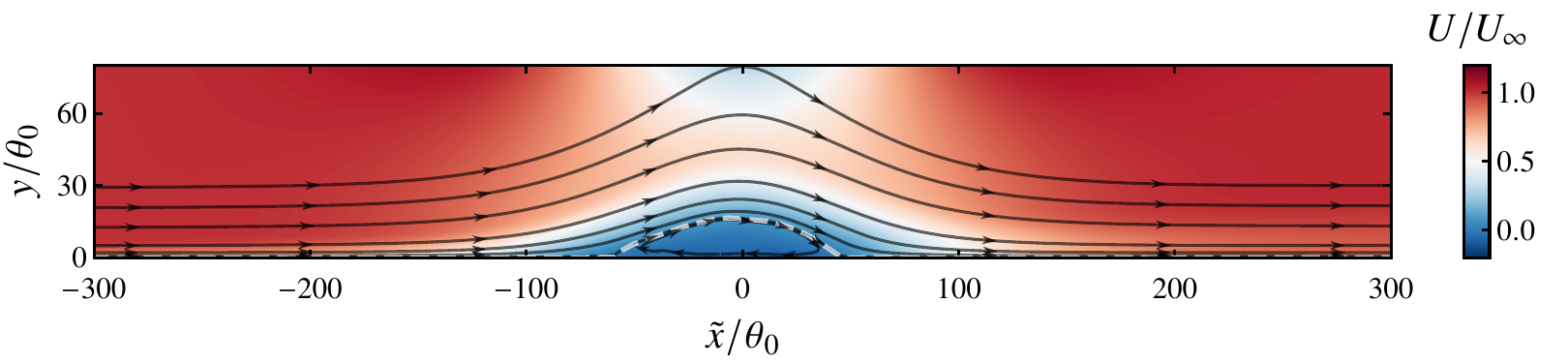}
            \put(-1,15){\small(\textit{a})}
        \end{overpic}
    \end{subfigure}
    \begin{subfigure}{0.9\textwidth}
        \centering
        \begin{overpic}[width=\linewidth, trim=0 30pt 0 0, clip]{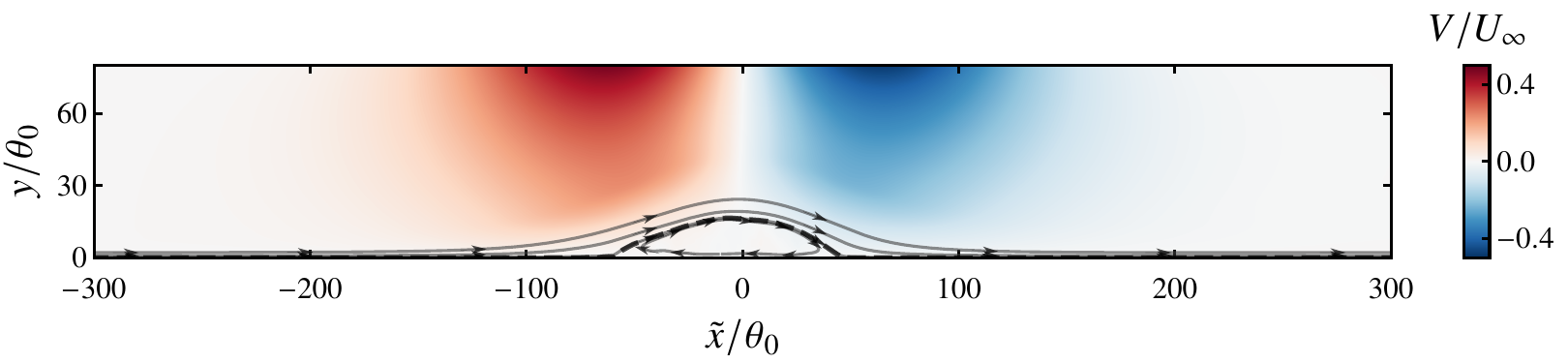}
            \put(-1,15){\small(\textit{b})}
        \end{overpic}
    \end{subfigure}\\
    \begin{subfigure}{0.9\textwidth}
        \centering
        \begin{overpic}[width=\linewidth, trim=0 30pt 0 0, clip]{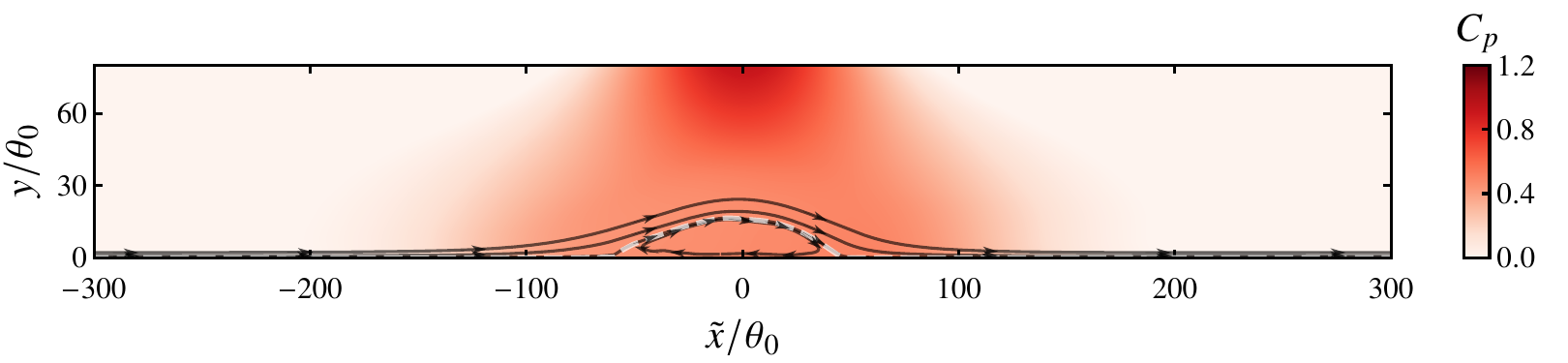}
            \put(-1,15){\small(\textit{c})}
        \end{overpic}
    \end{subfigure}\\
    \begin{subfigure}{0.9\textwidth}
        \centering
        \begin{overpic}[width=\linewidth, trim=0 0 0 0, clip]{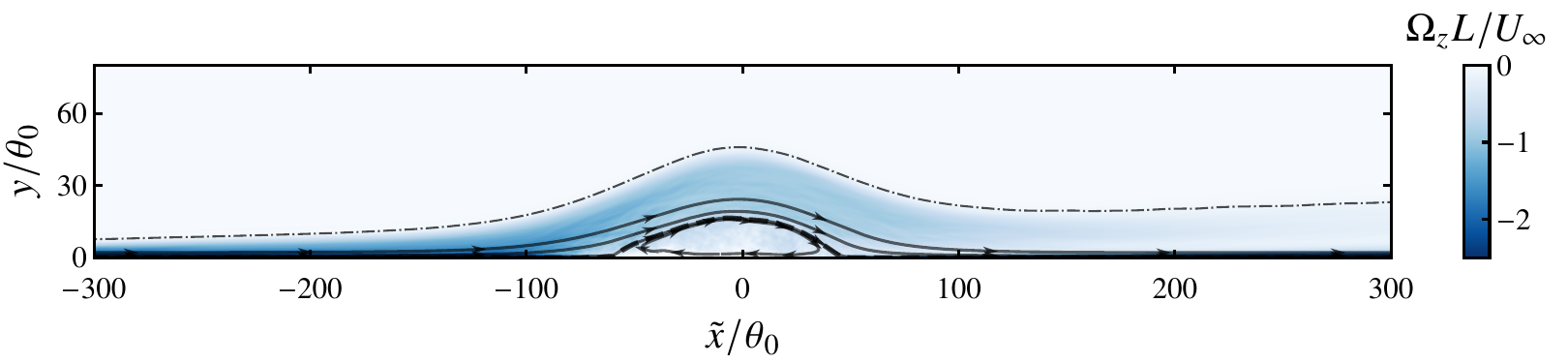}
            \put(-1,19){\small(\textit{d})}
        \end{overpic}
    \end{subfigure}
    \caption{Contour plots of mean velocities, pressure coefficient, and mean spanwise vorticity.}
    \label{fig:uvcpoz}
\end{figure}

Fig. \ref{fig:uvcpoz} shows the overall development of the mean velocities, the mean pressure coefficient and the mean spanwise vorticity.
For the streamwise mean velocity $U/U_\infty$ in Fig. \ref{fig:uvcpoz}(\textit{a}), there are blue regions ($-100\lesssim\tilde{x}/\theta_0\lesssim100$) that indicate low speed zones  where suction and blowing are applied at the upper boundary.
Also, the mean velocity $U$ decelerates before $\tilde{x}/\theta_0=0$ and accelerates after, showing nonparallel streamlines near the separation bubble.
For the wall normal velocity $V/U_\infty$ (Fig. \ref{fig:uvcpoz}(\textit{b})), it changes sign at the upper boundary when crossing $\tilde{x}/\theta_0=0$, and the streamlines near the wall clearly display the separation bubble.
To better determine the dimensions of the bubble, contour line of the zero stream function is also marked in Fig. \ref{fig:uvcpoz} with a thick dashed line, through which it is clearly seen that the bubble measures $120\theta_0$ in length and $17\theta_0$ in height.
In Fig.~\ref{fig:uvcpoz}(\textit{c}), the pressure coefficient $C_{p}=(P-P_{\infty})/\rho U_\infty^2$ identifies an APG region before separation and a favorable-pressure gradient (FPG) region after reattachment ($P_{\infty}$ is the mean pressure at the flow inlet).
In addition, the contours of large $C_p$ correspond to the low speed zones of $U/U_\infty$ in Fig. \ref{fig:uvcpoz}(\textit{a}), and $C_p$ outside the boundary layer differ significantly from those of the wall due to the strong convection effect.
Moreover, the spanwise vorticity $\Omega_z$ plotted in Fig. \ref{fig:uvcpoz}(\textit{d}) exhibits large negative values at the wall due to the shear inside the boundary layer, and approaches zero near the upper boundary as a consequence of the zero-vorticity condition imposed there.
A distinct stratification is observed between the vorticity-enriched region and zero-vorticity region, indicated by a thin dash-dotted line.
This vorticity layer gradually thickens downstream until it encounters the separation bubble and detaches from the wall, leaving the flow inside the bubble almost irrotational.
Then the vorticity layer continues to develop along the upper edge of the bubble and eventually returns to the wall after reattachment.
Observing Fig. \ref{fig:uvcpoz}(\textit{a}) and Fig. \ref{fig:uvcpoz}(\textit{d}), it is clear that compared to the streamwise velocity component, the distribution of spanwise vorticity exhibits a more pronounced ``layer" under strong pressure gradients.
Therefore, in the subsequent analysis, $\Omega_z$ is used instead of $U$ to define a series of streamwise parameters.

In Fig.~\ref{fig:uvcpoz}, the thick dashed lines represent the zero stream function and mark the separation bubble.
The thin dash-dotted line in Fig.~\ref{fig:uvcpoz}(\textit{d}) denotes the vorticity-enriched layer at $\Omega_z=-0.1U_\infty/L$.

\subsection{Determination of the boundary-layer edge}

An unambiguous boundary-layer edge is required before the local and integral quantities governing the streamwise development can be evaluated.
In a canonical ZPG TBL, the velocity at the boundary-layer edge is identified with the freestream velocity and the thickness is commonly defined by $U(\delta_{99})=0.99U_e$.
In the present flow, however, the upper-boundary transpiration produces non-parallel outer streamlines, such that $\partial U/\partial y$ need not vanish immediately outside the turbulent region.
A threshold applied directly to $U$ therefore does not uniquely identify the boundary-layer edge.

To accommodate the non-parallel outer flow, \citet{lighthill_introduction} introduced the ``generalized velocity,'' which has subsequently been applied to pressure-gradient TBLs \citep{spalart_watmuff_1993,coleman_numerical_2018}:

\begin{equation}
    \widetilde{U}\left(x,y\right)=-\int_{0}^{y}{\Omega_z\left(x,y^\prime\right)dy^\prime}.
\end{equation}

Because $\partial\widetilde{U}/\partial y=-\Omega_z$, $\widetilde{U}$ reaches a plateau once the external flow becomes approximately irrotational.
The freestream velocity $U_e=\widetilde{U}(x,y\rightarrow\infty)$ can be evaluated from this plateau, and the thickness is defined by $\widetilde{U}(x,\delta_p)=pU_e$.
For a nearly parallel ZPG TBL, $\partial V/\partial x$ is negligible, such that $\Omega_z\approx-\partial U/\partial y$ and $\widetilde{U}\approx U$; the generalized-velocity definition therefore recovers the conventional velocity-based definition.
It nevertheless retains the choice of a threshold $p$ and requires the computational domain to contain an approximately irrotational external region in which $\widetilde{U}$ reaches a plateau.

To avoid an arbitrary threshold, \citet{lozier_defining_2025} instead defined the thickness from the outermost negative-to-positive zero crossing of the streamwise-velocity skewness.
This thickness, denoted here by $\delta_L$, characterizes the intermittent passage of the turbulent/non-turbulent interface.
However, most external datasets considered below do not report the third-order moment required to evaluate $\delta_L$, preventing its consistent application across the present benchmark.

\begin{figure*}[!htbp]
    \centering
    \begin{subfigure}{0.49\textwidth}
        \centering
        \begin{overpic}[width=\linewidth]{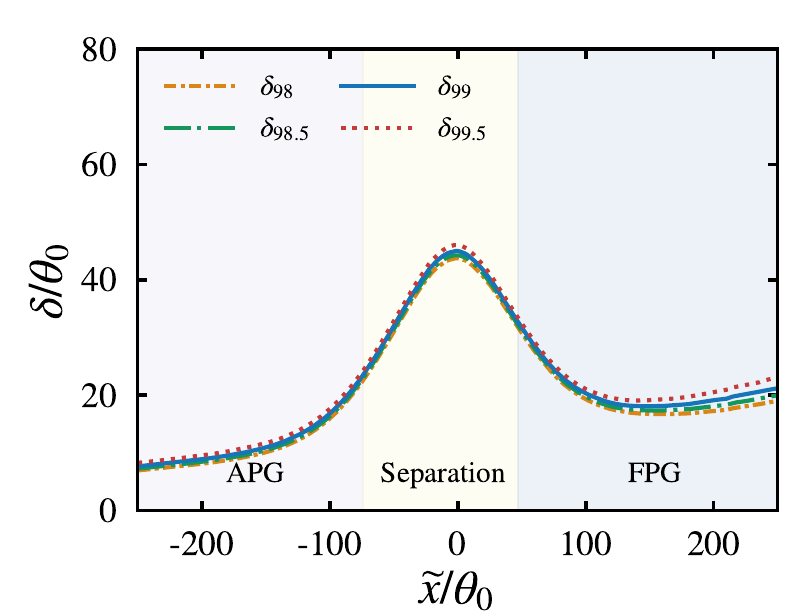}
            \put(0,72){\small(\textit{a})}
        \end{overpic}
    \end{subfigure}
    \begin{subfigure}{0.49\textwidth}
        \centering
        \begin{overpic}[width=\linewidth]{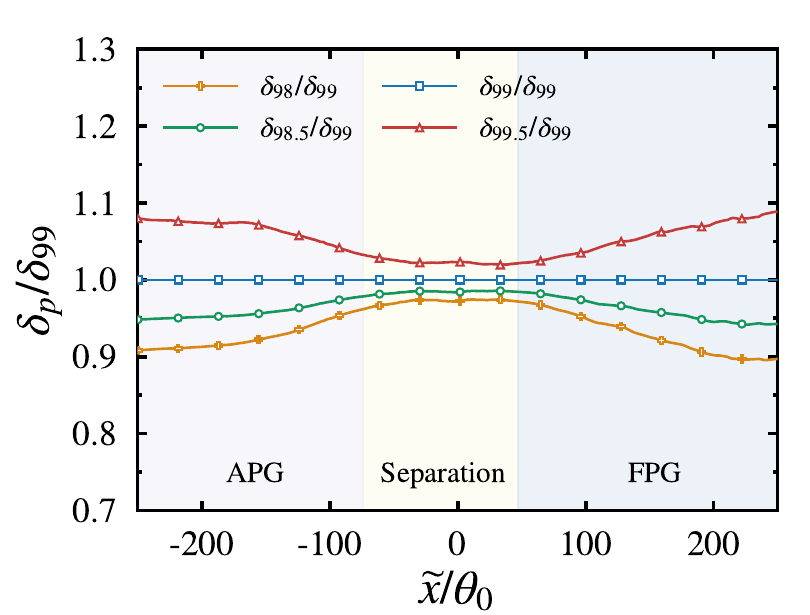}
            \put(0,72){\small(\textit{b})}
        \end{overpic}
    \end{subfigure}
    \caption{Sensitivity of the generalized-velocity boundary-layer thickness to the threshold: (\textit{a}) thicknesses and (\textit{b}) thickness ratios.}
    \label{fig:boundary-layer-edge}
\end{figure*}

The lavender, pale-yellow, and pale-blue backgrounds in Fig.~\ref{fig:boundary-layer-edge} and the subsequent streamwise plots denote the APG region, separation bubble, and post-reattachment FPG region, respectively.
Fig.~\ref{fig:boundary-layer-edge}(\textit{a}) compares boundary-layer thicknesses obtained with different generalized-velocity thresholds.
The four thicknesses defined from thresholds of generalized velocity are close in magnitude and exhibit similar streamwise trends.

Sensitivity to the generalized-velocity threshold is assessed in Fig.~\ref{fig:boundary-layer-edge}(\textit{b}) using $p=0.98$, $0.985$, $0.99$, and $0.995$.
Over the APG stations selected from the present dataset, the variations of $\delta_{98}/\delta_{99}$, $\delta_{98.5}/\delta_{99}$, and $\delta_{99.5}/\delta_{99}$ relative to their respective mean values are $5.2\%$, $2.8\%$, and $4.0\%$, respectively.
These modest variations indicate that the threshold definitions provide essentially the same boundary-layer length scale over the APG range considered.
The ratios become even closer in the strong-APG region because the larger velocity defect produces a sharper approach of $\widetilde{U}$ to its edge value.
Accordingly, $\delta$ denotes the generalized-velocity thickness $\delta_{99}$ throughout the remainder of this paper.

\subsection{Streamwise variation of integral parameters}

With $U_e$ and $\delta=\delta_{99}$ established above, the generalized displacement thickness $\widetilde{\delta}^\ast$ and momentum thickness $\widetilde{\theta}$ are defined as

\begin{equation}
    \widetilde{\delta}^\ast=\int_0^\infty{\left(1-\frac{\widetilde{U}}{U_e}\right)}dy,\quad\quad \widetilde{\theta}=\int_0^\infty{\frac{\widetilde{U}}{U_e}\left(1-\frac{\widetilde{U}}{U_e}\right)}dy,
\end{equation}
and the shape factor is $H={\widetilde{\delta}}^\ast/\widetilde{\theta}$.
The same generalized integral definitions were adopted by \citet{coleman_numerical_2018} and \citet{abe_direct_2019}.

\begin{figure}[!htbp]
    \centering
    \begin{subfigure}{0.49\textwidth}
        \centering
        \begin{overpic}[width=\linewidth, trim=-10pt 0 -10pt 0, clip]{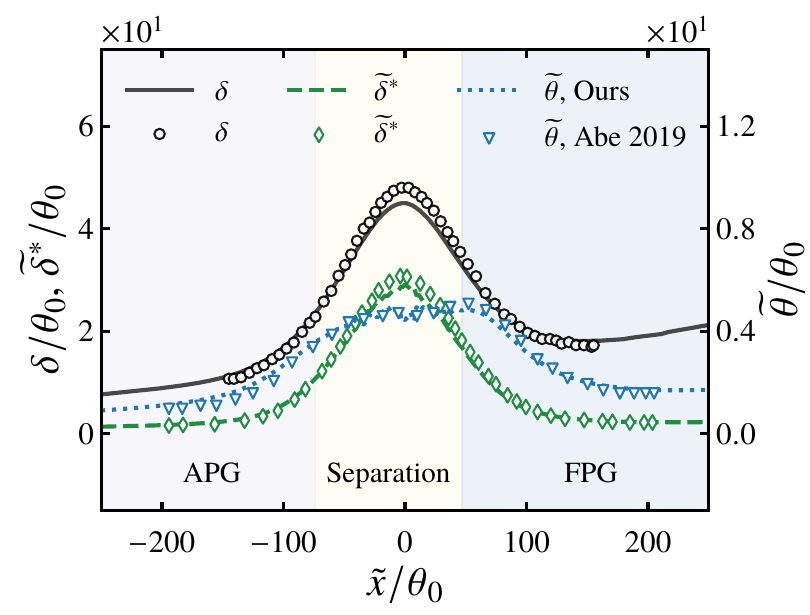}
            \put(0,68){\small(\textit{a})}
        \end{overpic}
    \end{subfigure}
    \begin{subfigure}{0.49\textwidth}
        \centering
        \begin{overpic}[width=\linewidth, trim=-10pt 0 -10pt 0, clip]{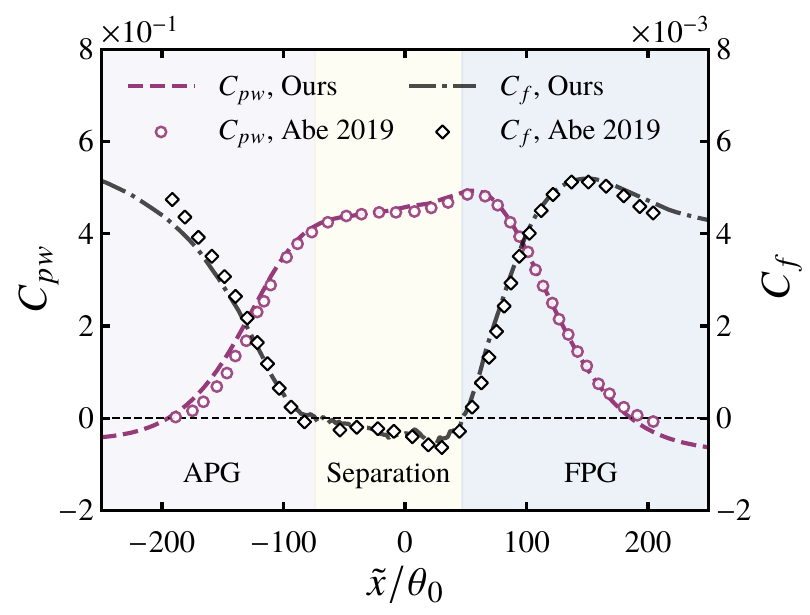}
            \put(0,68){\small(\textit{b})}
        \end{overpic}
    \end{subfigure}\\
    \begin{subfigure}{0.49\textwidth}
        \centering
        \begin{overpic}[width=\linewidth, trim=-10pt 0 -10pt 0, clip]{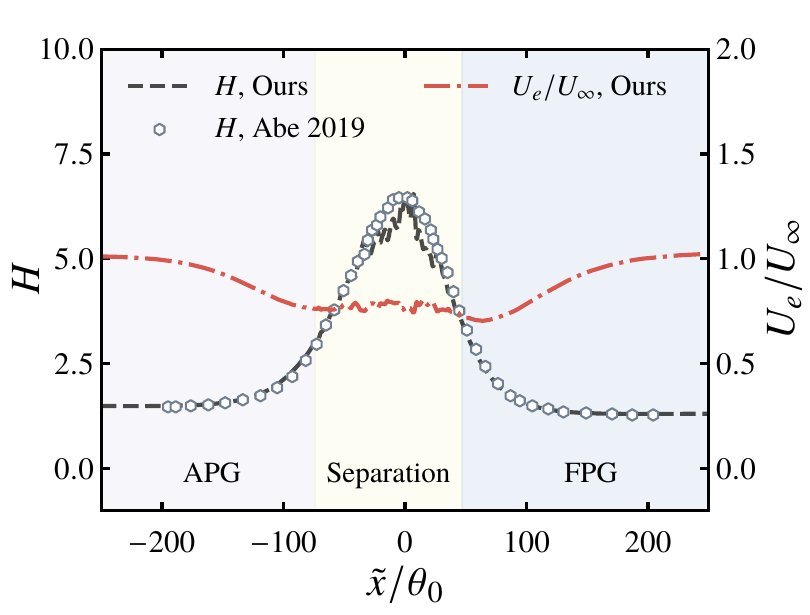}
            \put(0,68){\small(\textit{c})}
        \end{overpic}
    \end{subfigure}
    \begin{subfigure}{0.49\textwidth}
        \centering
        \begin{overpic}[width=\linewidth, trim=-10pt 0 -10pt 0, clip]{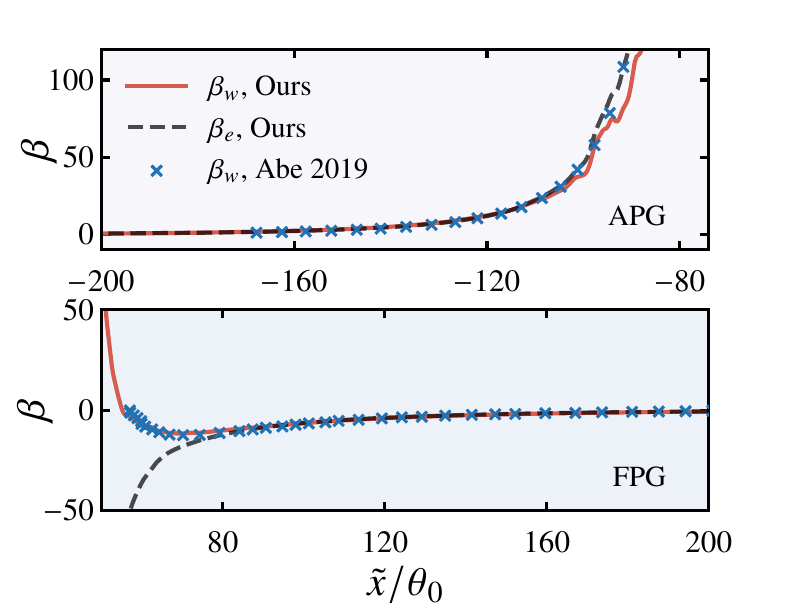}
            \put(0,68){\small(\textit{d})}
        \end{overpic}
    \end{subfigure}
    \caption{Streamwise variation of integral parameters and pressure-gradient measures.}
    \label{fig:streamwise-local-param}
\end{figure}

Fig.~\ref{fig:streamwise-local-param}(\textit{a}) shows the streamwise evolution of $\delta$, $\widetilde{\delta}^\ast$, and $\widetilde{\theta}$.
Comparisons are made with the data of \citet{abe_reynolds_2017,abe_direct_2019}.
All three thicknesses increase through the APG region, attain maxima near the switch point ($\tilde{x}/\theta_0=0$), and then decrease under the FPG.
Unlike $\delta$ and $\widetilde{\delta}^\ast$, $\widetilde{\theta}$ exhibits a plateau over $-50\lesssim\tilde{x}/\theta_0\lesssim50$, coincident with the interval of vanishing $C_f$ in Fig.~\ref{fig:streamwise-local-param}(\textit{b}).
The resulting mean separation and reattachment locations are $\tilde{x}/\theta_0\approx-74$ and $47$, respectively.
The wall-pressure coefficient $C_{pw}=(P_w-P_{\infty})/\rho U_\infty^2$ also forms a plateau across the separated region, delineating the upstream APG and downstream FPG zones.
Fig.~\ref{fig:streamwise-local-param}(\textit{c}) further shows the shape factor $H=\widetilde{\delta}^\ast/\widetilde{\theta}$ and freestream velocity $U_e$.
As the APG strengthens, $H$ rises from its ZPG value of approximately $1.4$ to a maximum of approximately $7$ at $\tilde{x}/\theta_0=0$, while $U_e$ decreases by approximately 30\%.
The close agreement between the present curves and the reference symbols supports the accuracy of the simulation.

\subsection{Pressure gradient parameters}

To quantify the pressure gradient effect, \citet{clauser_turbulent_1954} introduced the pressure gradient parameter

\begin{equation}
    \beta_e(x)=\left(\frac{\widetilde{\delta}^\ast}{\tau_w}\right)\left(\frac{dP_e}{dx}\right),
\end{equation}
of which a constant value independent of $x$ defines the so-called ``equilibrium layers".
Taking into account the interest of wall pressure, a similar definition has been introduced by \citet{abe_direct_2019}:

\begin{equation}
    \beta_w(x)=\left(\frac{\widetilde{\delta}^\ast}{\tau_w}\right)\left(\frac{dP_w}{dx}\right).
\end{equation}
As shown in Fig. \ref{fig:streamwise-local-param}(\textit{d}), both $\beta_w$ and $\beta_e$ increase dramatically before separation and are also close to each other, implying a similar streamwise variation between $P_w$ and $P_e$.
However, after reattachment, $\beta_w$ drops to zero from large positive values, whereas $\beta_e$ increases to zero from negative values.
Note that $\beta_w>0$ indicates APG on the wall, while $\beta_e<0$ indicates FPG on the edge of the boundary layer.
This difference is induced by the mean convection velocity $V$, which plays an important role in the development of APG TBL.

\subsection{Wall normal profiles}

\begin{figure}[!htbp]
    \centering
    \begin{subfigure}{0.45\textwidth}
        \begin{overpic}[width=\linewidth, trim=0 0 0 -20pt, clip]{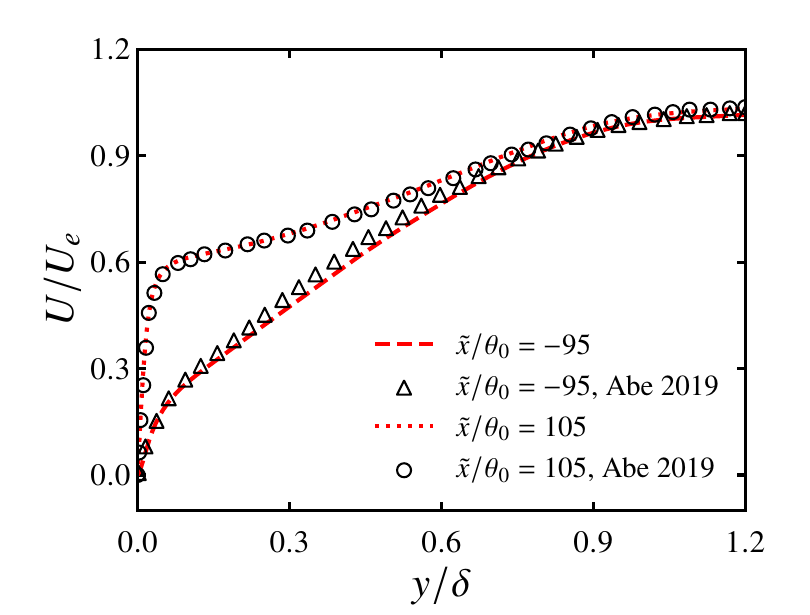}
                \put(0,70){\small(\textit{a})}
        \end{overpic}
    \end{subfigure}
    \begin{subfigure}{0.45\textwidth}
        \begin{overpic}[width=\linewidth, trim=0 0 0 -20pt, clip]{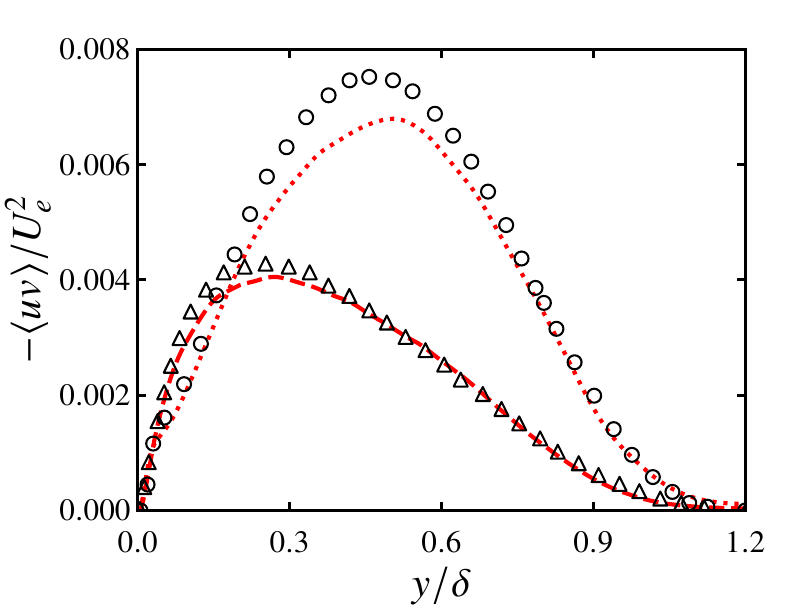}
                \put(0,70){\small(\textit{b})}
        \end{overpic}
    \end{subfigure}
    \caption{Wall-normal profiles compared with the data of \citet{abe_direct_2019}.}
    \label{fig:U-uv-sections}
\end{figure}

Fig.~\ref{fig:U-uv-sections} validates the profiles of mean $U/U_e$ and Reynolds shear stress $-\langle uv\rangle/U_e^2$ with those under comparable conditions in \citet{abe_direct_2019}.
Lines denote the present simulation and symbols denote the reference data; the APG station $\tilde{x}/\theta_0=-95$ is shown by dashed lines and triangles, and the reattached station $\tilde{x}/\theta_0=105$ by dotted lines and circles.
Before separation, the mean velocity at $\tilde{x}/\theta_0=-95$ grows gently throughout the entire boundary layer, while in the reattached region ($\tilde{x}/\theta_0=105$), it exhibits a sharp increase to $0.6 U_e$ within $0.1 {\delta}$ near the wall (Fig.~\ref{fig:U-uv-sections}a).
Also, the APG velocity profile shows a larger defect from the local freestream than the FPG one.
In contrast, the Reynolds shear stress is much lower before separation, with its peak closer to the wall ($0.3 \delta$) compared to the reattached one showing the peak at a further position of $0.5 \delta$.
Again, data from \citet{abe_direct_2019} at the same two streamwise locations are included in Fig. \ref{fig:U-uv-sections}, showing close agreement with the present simulation for both the mean velocity and Reynolds shear stress.
We note a slightly lower peak of Reynolds shear stress at $\tilde{x}/\theta_0=105$ compared to that of \citet{abe_direct_2019}, which may be due to different grid resolutions used in the far stream away from the wall.
However, such a difference does not affect the following comparisons for the APG zone.

\begin{figure}[!htbp]
    \centering
    \begin{subfigure}{0.32\textwidth}
        \centering
        \begin{overpic}[width=\linewidth, trim=10pt 0 10pt -10pt, clip]{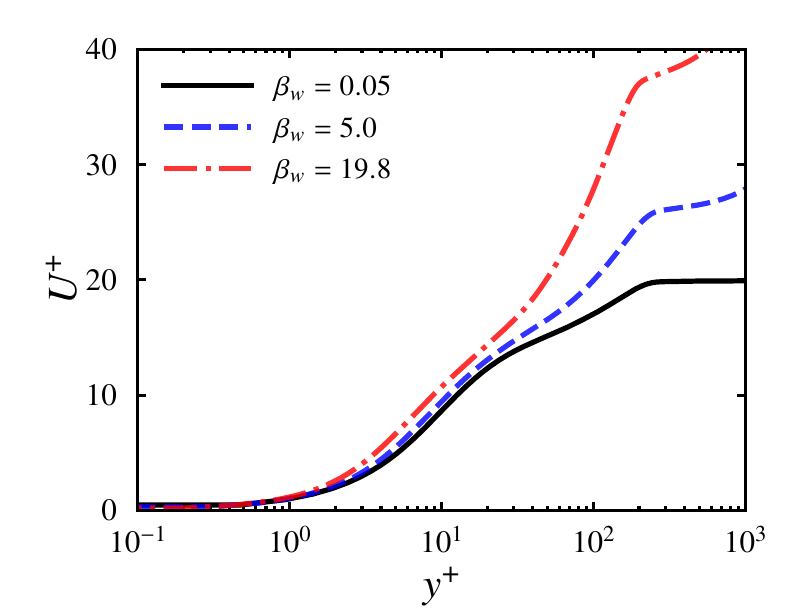}
            \put(0,75){\small(\textit{a})}
        \end{overpic}
    \end{subfigure}
    \begin{subfigure}{0.32\textwidth}
        \centering
        \begin{overpic}[width=\linewidth, trim=10pt 0 10pt -10pt, clip]{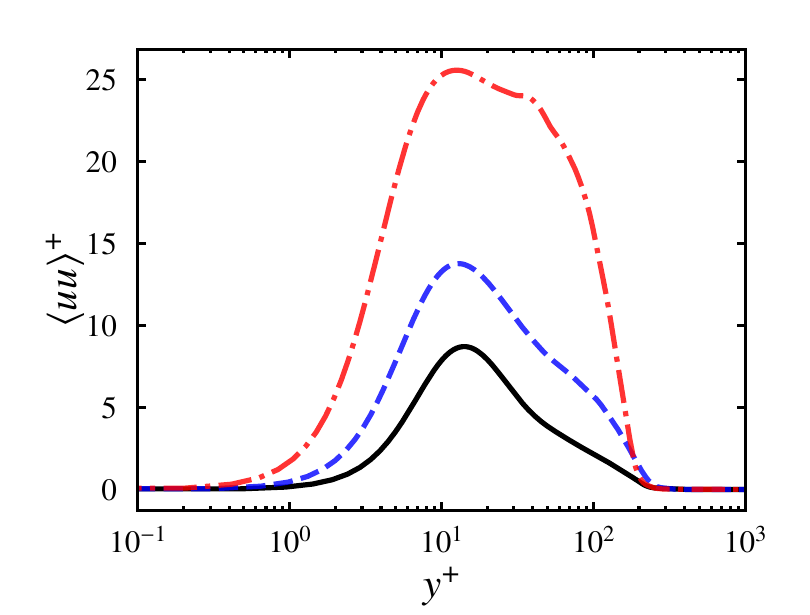}
            \put(0,75){\small(\textit{b})}
        \end{overpic}
    \end{subfigure}
    \begin{subfigure}{0.32\textwidth}
        \centering
        \begin{overpic}[width=\linewidth, trim=10pt 0 10pt -10pt, clip]{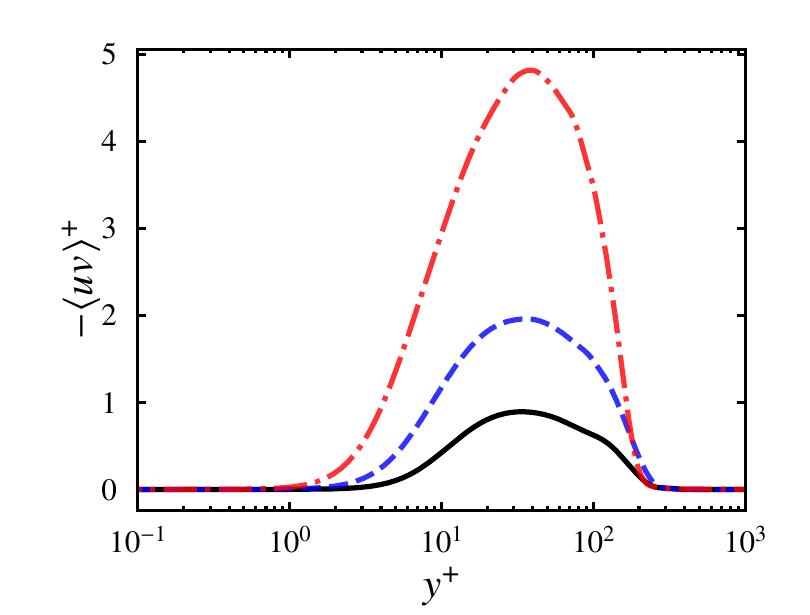}
            \put(0,75){\small(\textit{c})}
        \end{overpic}
    \end{subfigure}\\
    \begin{subfigure}{0.32\textwidth}
        \centering
        \begin{overpic}[width=\linewidth, trim=10pt 0 10pt -10pt, clip]{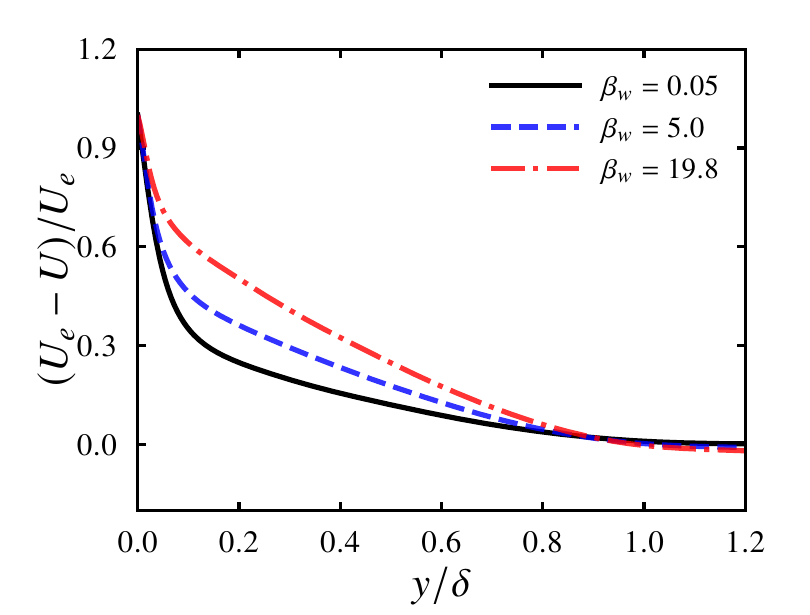}
            \put(0,75){\small(\textit{d})}
        \end{overpic}
    \end{subfigure}
    \begin{subfigure}{0.32\textwidth}
        \centering
        \begin{overpic}[width=\linewidth, trim=10pt 0 10pt -10pt, clip]{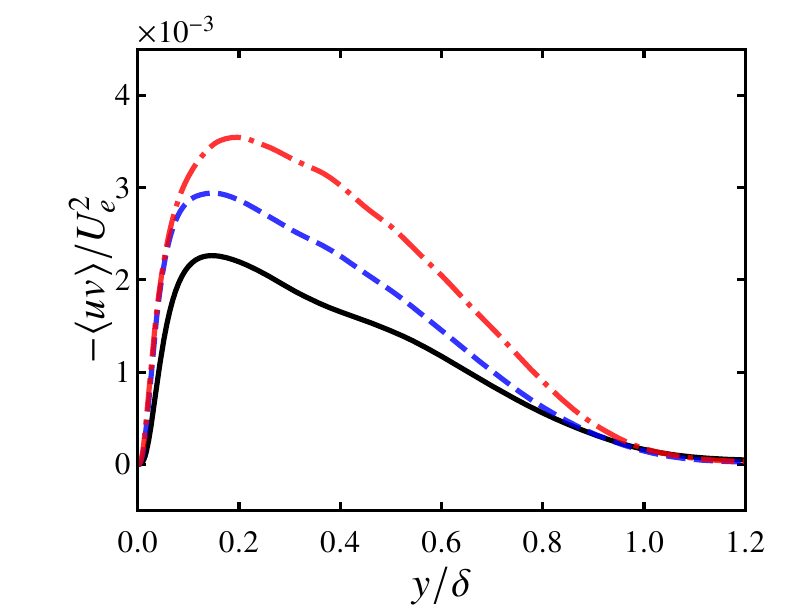}
            \put(0,75){\small(\textit{e})}
        \end{overpic}
    \end{subfigure}
    \begin{subfigure}{0.32\textwidth}
        \centering
        \begin{overpic}[width=\linewidth, trim=10pt 0 10pt -10pt, clip]{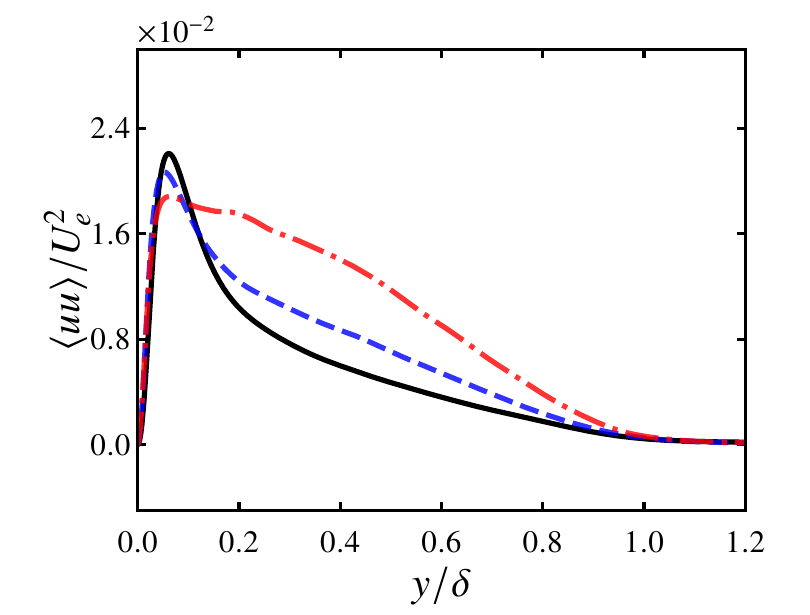}
            \put(0,75){\small(\textit{f})}
        \end{overpic}
    \end{subfigure}
    \caption{Mean velocity and Reynolds-stress profiles under (\textit{a}--\textit{c}) wall-unit and (\textit{d}--\textit{f}) freestream-velocity normalizations.}
    \label{fig:wall-unit-profiles}
    \label{fig:edge-velocity-profiles}
\end{figure}

Meanwhile, we also examine the profiles normalized by the wall-units.
In Fig.~\ref{fig:wall-unit-profiles}, the solid, dashed, and dash-dotted lines denote nearly ZPG ($\beta_w=0.05$), mild APG ($\beta_w=5.0$), and strong APG ($\beta_w=19.8$) cases, respectively; the superscript ``+'' denotes wall-unit normalization.
As shown in Fig. \ref{fig:wall-unit-profiles}(\textit{a}), the wall-unit scaling leads to large variation with different pressure gradients.
Under the mild APG of $\beta_w=5.0$, the velocity profile already exhibits a significant deviation from the standard log-law and ends up with a large wake part at the outer boundary layer.
At the strong APG, a complete deviation from the ZPG law of the wall is present right above the viscous sublayer.
Similar trends are observed in \citet{abe_direct_2019}, \citet{coleman_numerical_2018} and \citet{cite-6}.
Compared to the experimental profile of \citet{cite-13} where a logarithmic region persists under APG conditions, the velocity here deviates from the logarithmic law even at $\beta_w=5.0$, due to the relatively low Reynolds number in the simulation.
The Reynolds stress profiles $\langle uu\rangle^+$ and $-\langle uv\rangle^+$ in Figs. \ref{fig:wall-unit-profiles}(\textit{b}) and \ref{fig:wall-unit-profiles}(\textit{c}) also exhibit dramatic upward changes with APG.
Meanwhile, an increase in $\langle uu\rangle$ in the outer layer and a shift of the peak of $-\langle uv\rangle$ are observed, as noted by \citet{han_consistent_2024}.
These observations indicate that the classical viscous wall-unit scaling, due to its severe degradation, is no longer suitable for APG TBLs.

The freestream velocity $U_e$ provides another conventional outer reference, but it does not account for the strong variation of the velocity defect and turbulence intensity induced by APG.
Fig.~\ref{fig:edge-velocity-profiles}(\textit{d}) uses $U_e$ as the velocity scale to normalize the same profiles.
The normalized velocity defect increases as the APG strengthens, while $-\langle uv\rangle/U_e^2$ and $\langle uu\rangle/U_e^2$ also increase markedly in Figs.~\ref{fig:edge-velocity-profiles}(\textit{e}) and \ref{fig:edge-velocity-profiles}(\textit{f}), respectively.
The freestream-velocity normalization is therefore retained only as a baseline and is not considered a viable similarity scaling in the analysis below.

\subsection{Setting up the benchmark for APG profiles \label{sec:apg-benchmark}}
\begin{figure}[!htbp]
    \centering
    \begin{subfigure}{0.49\textwidth}
        \centering
        \begin{overpic}[width=\linewidth, trim=0 30pt -5pt 0, clip]{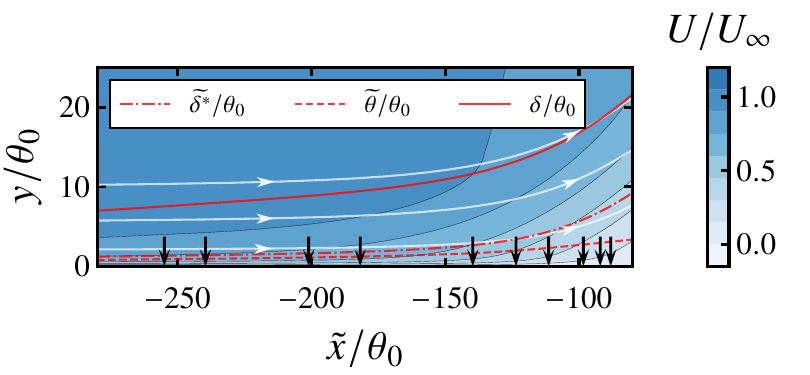}
            \put(0,31){\small(\textit{a})}
        \end{overpic}
    \end{subfigure}
    \begin{subfigure}{0.49\textwidth}
        \centering
        \begin{overpic}[width=\linewidth, trim=-5pt 30pt 0 0, clip]{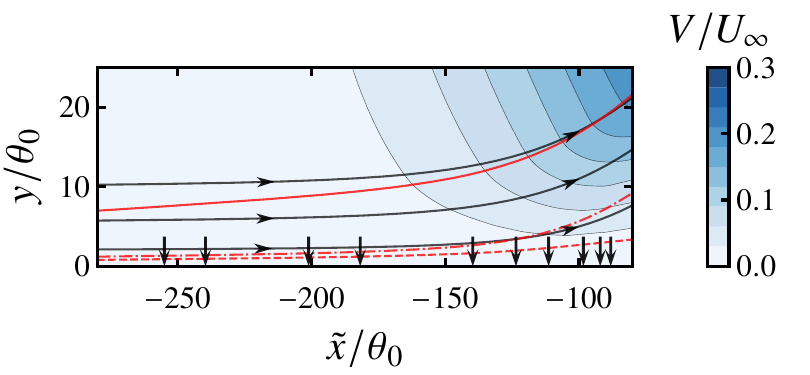}
            \put(0,31){\small(\textit{b})}
        \end{overpic}
    \end{subfigure}\\
    \begin{subfigure}{0.49\textwidth}
        \centering
        \begin{overpic}[width=\linewidth, trim=0 0 -5pt 0, clip]{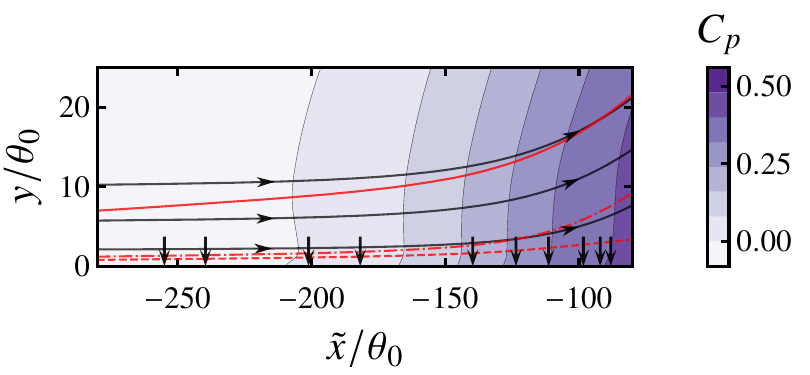}
            \put(0,38){\small(\textit{c})}
        \end{overpic}
    \end{subfigure}
    \begin{subfigure}{0.49\textwidth}
        \centering
        \begin{overpic}[width=\linewidth, trim=-5pt 0 0 0, clip]{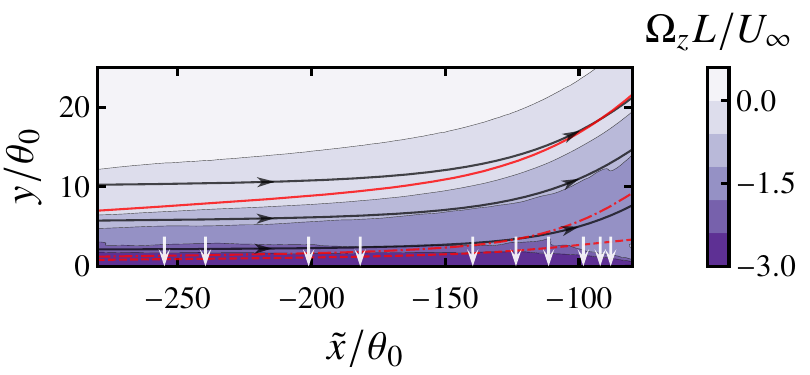}
            \put(0,38){\small(\textit{d})}
        \end{overpic}
    \end{subfigure}
    \caption{Localized contours in the APG TBL region before separation.}
    \label{fig:contour-apg}
\end{figure}

Representative profiles from the present simulation together with numerical and experimental data in the literature are used to form a comprehensive APG benchmark, as listed in table \ref{tab:simu-data} for the simulation data and table \ref{tab:exp-data} for the experimental data.
For brevity, we first adopt twenty profiles, ten from the present simulation (denoted as P01-P10 in table \ref{tab:simu-data}) and ten with matching $\beta_w$ from \citet{coleman_numerical_2018} (denoted as C01-C10 in table \ref{tab:simu-data}) to evaluate different scaling proposals.
The APG region of present DNS is shown in Fig. \ref{fig:contour-apg}, where the panels show $U$, $V$, $C_p$, and $\Omega_z$ with streamlines.
Red dash-dotted, dashed, and solid lines denote $\widetilde{\delta^\ast}$, $\widetilde{\theta}$, and $\delta$, respectively, and downward arrows over the abscissa mark the exact locations of the ten profiles from the present DNS.
The first four locations represent zero to moderate APG, and are characterized by the slowly varying freestream $U$ in Fig. \ref{fig:contour-apg}(\textit{a}) and the close-to-zero value of $V$ in Fig. \ref{fig:contour-apg}(\textit{b}).
In addition, the pressure coefficient in Fig. \ref{fig:contour-apg}(\textit{c}) also indicates that the pressure variation in these locations is very slow, while the vorticity contours in Fig. \ref{fig:contour-apg}(\textit{d}) exhibit only gentle growth.
The following six locations span moderate to very strong APG, and the flow in these locations varies more rapidly.
From the $U$ contour in Fig. \ref{fig:contour-apg}(\textit{a}), it can be observed that $U$ in the freestream decreases with APG and the near-wall low-speed region expands outward, which accounts for an increasing velocity defect in the outer region.
Similarly, Fig. \ref{fig:contour-apg}(\textit{b}) shows a noticeable increase in $V$ near the outer edge of the boundary layer, causing the streamlines to form a significant inclination.
The contours of $C_p$ in Fig. \ref{fig:contour-apg}(\textit{c}) clearly illustrate the rise of the pressure along the wall.
In addition, the vorticity-enriched layer in Fig. \ref{fig:contour-apg}(\textit{d}) thickens more rapidly at the last six locations and the contours exhibit a trend that roughly aligns with the variation of boundary-layer thickness.

Throughout the ten chosen locations, the Clauser parameter $\beta_w$ ranges from 0.05 to approximately 120, as listed in table \ref{tab:simu-data}.
These profiles cover from nearly ZPG to strong APG at the verge of separation.
For our case, the Reynolds number $Re_\theta$ increases considerably from 529 to 1847, and the shape factor $H$ increases by 70\%.
In contrast, the freestream velocity $U_e$ decreases significantly by 22\%, and the friction coefficient drops quickly to about zero near the separation.
On the other hand, the case of \citet{coleman_numerical_2018} develops from a different upstream Reynolds number and is imposed a different suction-blowing velocity.
As a consequence, the momentum Reynolds number $Re_\theta$ grows from 1834 to 8226, extending the range of our data by five times (Table~\ref{tab:simu-data}).
Note that the freestream Clauser parameter $\beta_e$ is also included in the table, which agrees closely with $\beta_w$ before the latter exceeds 80.
In addition, $p^+\equiv(\nu/u^3_\tau)(\partial P/\partial x)$, another dimensionless parameter to index the pressure gradient effect \citep{cite-13,cite-19}, is also provided in the table for reference.

Note that in \citet{abe_reynolds_2017, abe_direct_2019}, the available profiles of mean velocity and Reynolds shear stress are provided only for one streamwise location before separation (already shown in Fig.~\ref{fig:U-uv-sections}).
Meanwhile, Reynolds normal stresses have not been reported.
Therefore, in the plots below, there are no data from \citet{abe_reynolds_2017, abe_direct_2019} for a systematic comparison.
On the other hand, the direct numerical simulations of \citet{coleman_numerical_2018}, \citet{coleman_numerical_2021} and \citet{rkein_direct_2023} provide complementary sets of detailed data on different types of APG TBLs, and are thus included in the present APG benchmark.
Detailed parameters of these simulation data are presented in table \ref{tab:simu-data}.
In addition, experiments by \citet{marusic_wall_wake_1995} for non-equilibrium APG TBL and by \citet{Dengel_Fernholz_1990} for axisymmetric APG TBL are also included and provide valuable reference data for mean velocity and Reynolds stresses at high Reynolds numbers up to $Re_\theta \approx 2 \times 10^4$, as shown in table \ref{tab:exp-data}.
In the following sections, these data are used to verify the outer similarity.

\begin{table}[!htbp]
\centering
\footnotesize
\setlength{\tabcolsep}{4pt}
\caption{Data of simulations in the APG benchmark.
P01-P10 are from the present DNS described in section \ref{sec:DNS_setup}, C01-C10 are from the separation-bubble TBL DNS (case C) of \citet{coleman_numerical_2018}.
C11-C15 are from the suction-only TBL DNS (case E) of \citet{coleman_numerical_2021}, and R01-R05 are from the DNS of \citet{rkein_direct_2023}.}
\label{tab:simu-data}
\begin{tabular}{cccccccccc}
\hline
Dataset & Symbol & $\beta_w$ & $\beta_e$ & $p^+$ & $Re_\tau$ & $Re_\theta$ & $H$ & $U_e/U_\infty$ & $C_f$ \\\hline
P01 & \legenduptriangle{004BFF}{FFFFFF}{4.5pt} & 0.05 & 0.04 & 0.0012 & 230 & 529 & 1.49 & 1.01 & 0.00519\\
P02 & \legenduptriangle{1C4BE3}{FFFFFF}{4.5pt} & 0.10 & 0.10 & 0.0024 & 237 & 560 & 1.49 & 1.01 & 0.00503\\
P03 & \legenduptriangle{394BC5}{FFFFFF}{4.5pt} & 0.50 & 0.47 & 0.0108 & 249 & 654 & 1.49 & 1.00 & 0.00442\\
P04 & \legenduptriangle{544BAA}{FFFFFF}{4.5pt} & 0.99 & 0.94 & 0.0205 & 253 & 723 & 1.50 & 0.98 & 0.00398\\
P05 & \legenduptriangle{714B8E}{FFFFFF}{4.5pt} & 5.00 & 4.72 & 0.0896 & 244 & 982 & 1.61 & 0.90 & 0.00244\\
P06 & \legenduptriangle{8E4B71}{FFFFFF}{4.5pt} & 10.1 & 9.81 & 0.172 & 228 & 1176 & 1.73 & 0.86 & 0.00166\\
P07 & \legenduptriangle{A94B55}{FFFFFF}{4.5pt} & 19.8 & 20.3 & 0.340 & 199 & 1379 & 1.90 & 0.82 & 0.00100\\
P08 & \legenduptriangle{C54B39}{FFFFFF}{4.5pt} & 48.6 & 57.0 & 0.941 & 151 & 1633 & 2.16 & 0.79 & 0.00043\\
P09 & \legenduptriangle{E34B1C}{FFFFFF}{4.5pt} & 78.9 & 103 & 1.69 & 126 & 1767 & 2.33 & 0.78 & 0.00026\\
P10 & \legenduptriangle{FF4B00}{FFFFFF}{4.5pt} & 118 & 161 & 2.80 & 108 & 1847 & 2.46 & 0.78 & 0.00017\\
C01 & \legendcross{004BFF}{4.8pt} & 0.05 & 0.05 & 0.0005 & 706 & 1834 & 1.40 & 1.00 & 0.00373\\
C02 & \legendcross{1C4BE3}{4.8pt} & 0.10 & 0.09 & 0.0009 & 733 & 1947 & 1.40 & 0.99 & 0.00363\\
C03 & \legendcross{394BC5}{4.8pt} & 0.50 & 0.49 & 0.0038 & 794 & 2313 & 1.41 & 0.98 & 0.00325\\
C04 & \legendcross{544BAA}{4.8pt} & 1.00 & 0.97 & 0.0071 & 812 & 2566 & 1.43 & 0.96 & 0.00294\\
C05 & \legendcross{714B8E}{4.8pt} & 5.01 & 4.92 & 0.0287 & 809 & 3758 & 1.55 & 0.87 & 0.00178\\
C06 & \legendcross{8E4B71}{4.8pt} & 9.99 & 9.96 & 0.0523 & 768 & 4625 & 1.67 & 0.83 & 0.00122\\
C07 & \legendcross{A94B55}{4.8pt} & 20.0 & 21.0 & 0.0995 & 686 & 5700 & 1.87 & 0.79 & 0.00071\\
C08 & \legendcross{C54B39}{4.8pt} & 49.9 & 62.6 & 0.261 & 522 & 7232 & 2.29 & 0.75 & 0.00027\\
C09 & \legendcross{E34B1C}{4.8pt} & 80.2 & 114 & 0.469 & 423 & 7917 & 2.57 & 0.74 &  0.00014\\
C10 & \legendcross{FF4B00}{4.8pt} & 120 & 180 & 0.809 & 351 & 8226 & 2.72 & 0.73 &  0.00009\\
C11 & \legendcirclethick{2D1E8F}{FFFFFF}{1.8pt} & 1.03 & 1.00 & 0.0055 & 1303 & 3270 & 1.60 & 0.95 & 0.002467 \\
C12 & \legendcirclethick{5E47B6}{FFFFFF}{1.8pt} & 5.04 & 5.02 & 0.0205 & 1308 & 5195 & 1.77 & 0.86 & 0.001433 \\
C13 & \legendcirclethick{A987E7}{FFFFFF}{1.8pt} & 19.04 & 20.17 & 0.0571 & 1153 & 9757 & 1.95 & 0.78 & 0.000615 \\
C14 & \legendcirclethick{E1B0EA}{FFFFFF}{1.8pt} & 42.97 & 50.21 & 0.1171 & 947 & 14590 & 2.18 & 0.75 & 0.000266 \\
C15 & \legendcirclethick{F07A63}{FFFFFF}{1.8pt} & 65.66 & 79.45 & 0.1881 & 797 & 16997 & 2.38 & 0.73 & 0.000149 \\
R01 & \legendy{5681C7}{0.7pt} & - & 2.25 & - & 745 & 2505 & 1.48 & 0.947 & 0.00289 \\
R02 & \legendcrossthick{7E57C2}{5pt} & - & 2.64 & - & 757 & 2901 & 1.53 & 0.916 & 0.00261 \\
R03 & \legendx{C44E52}{4pt} & - & 3.01 & - & 784 & 3601 & 1.63 & 0.856 & 0.00219 \\
R04 & \legendcirclethick{00A6D6}{FFFFFF}{1.8pt} & - & 3.15 & - & 817 & 4300 & 1.76 & 0.806 & 0.00187 \\
R05 & \legendsquarethick{6A56AE}{FFFFFF}{1.8pt} & - & 3.28 & - & 865 & 4999 & 1.77 & 0.774 & 0.00165 \\\hline
\end{tabular}
\end{table}

\begin{table}[!htbp]
\centering
\footnotesize
\setlength{\tabcolsep}{4pt}
\caption{Data of experiments in the APG benchmark.
M01-M16 are the experimental data of \citet{marusic_wall_wake_1995}, and D01-D23 are from the experiments of \citet{Dengel_Fernholz_1990}.}
\label{tab:exp-data}
\begin{tabular}{cccccccccc}
\hline
Dataset & Symbol & {$\beta_w$} & $\beta_e$ & {$p^+$} & {$Re_\tau$} & $Re_\theta$ & $H$ & $U_e/U_\infty$ & $C_f$ \\\hline
M01 & \legendcircle{404040}{F5EBAE}{2pt} & - &$\approx0$& - & - & 2206 & 1.43 & - & 0.00360\\
M02 & \legendcircle{404040}{F0C284}{2pt} & - & 0.65 & - & - & 3153 & 1.44 & - & 0.00309\\
M03 & \legendcircle{404040}{EF8B67}{2pt} & - & 1.45 & - & - & 4155 & 1.49 & - & 0.00254\\
M04 & \legendcircle{404040}{E3625D}{2pt} & - & 2.90 & - & - & 5395 & 1.58 & - & 0.00201\\
M05 & \legendcircle{404040}{B54764}{2pt} & - & 4.48 & - & - & 6359 & 1.64 & - & 0.00168\\
M06 & \legendcircle{404040}{992224}{2pt} & - & 7.16 & - & - & 7257 & 1.73 & - & 0.00137\\
M11 & \legendsquare{404040}{F5EBAE}{1.8pt} & - &$\approx0$& - & - & 6430 & 1.40 & - & 0.00287\\
M12 & \legendsquare{404040}{F0C284}{1.8pt} & - & 0.71 & - & - & 8588 & 1.41 & - & 0.00251\\
M13 & \legendsquare{404040}{EF8B67}{1.8pt} & - & 1.39 & - & - & 10997 & 1.44 & - & 0.00221\\
M14 & \legendsquare{404040}{EF8B67}{1.8pt} & - & 2.74 & - & - & 14209 & 1.49 & - & 0.00185\\
M15 & \legendsquare{404040}{B54764}{1.8pt} & - & 3.96 & - & - & 16584 & 1.54 & - & 0.00162\\
M16 & \legendsquare{404040}{992224}{1.8pt} & - & 6.07 & - & - & 19133 & 1.60 & - & 0.00138\\
D01 & \legendcircle{000000}{5E47B6}{2pt} & - & 4.72 & - & - & 4000 $\sim$ 11000 & 1.46 & 0.97 & 0.003122 \\
D02 & \legendcircle{000000}{C59DF3}{2pt} & - & 11.70 & - & - & 4000 $\sim$ 11000 & 1.67 & 0.79 & 0.001129 \\
D03 & \legendcircle{000000}{F1A1B8}{2pt} & - & 17.93 & - & - & 4000 $\sim$ 11000 & 1.82 & 0.74 & 0.000701 \\
D11 & \legenddowntriangle{000000}{5E47B6}{5pt} & - & 4.38 & - & - & 4000 $\sim$ 11000 & 1.45 & 0.97 & 0.003210 \\
D12 & \legenddowntriangle{000000}{C59DF3}{5pt} & - & 10.61 & - & - & 4000 $\sim$ 11000 & 1.63 & 0.80 & 0.001281 \\
D13 & \legenddowntriangle{000000}{F1A1B8}{5pt} & - & 15.16 & - & - & 4000 $\sim$ 11000 & 1.74 & 0.75 & 0.000870 \\
D21 & \legendhexagon{000000}{5E47B6}{4.2pt} & - & 4.66 & - & - & 4000 $\sim$ 11000 & 1.46 & 0.97 & 0.003078 \\
D22 & \legendhexagon{000000}{C59DF3}{4.2pt} & - & 12.14 & - & - & 4000 $\sim$ 11000 & 1.73 & 0.78 & 0.001035 \\
D23 & \legendhexagon{000000}{F1A1B8}{4.2pt} & - & 18.76 & - & - & 4000 $\sim$ 11000 & 1.90 & 0.73 & 0.000619 \\\hline
\end{tabular}
\end{table}

\section{Component-wise similarity in the outer flow region \label{sec:self-sim}}

\subsection{Assessment of existing APG normalizations}

The wall-unit and freestream-velocity normalizations assessed in Fig.~\ref{fig:wall-unit-profiles} do not recover consistent outer-region collapse of the present APG profiles.
We therefore examine outer normalizations that have been used in previous studies of APG TBLs.
For the streamwise velocity defect, the ZS scale was originally introduced by \citet{zagarola_smits_1998} for fully developed pipe flow and was subsequently applied to APG TBLs by \citet{maciel_self-similarity_2006}:
\begin{equation}\label{eq:zs1998}
    U_{zs} = \frac{U_e}{\delta}\int_0^\delta(1-\frac{U}{U_e})dy =\frac{\delta^\ast}{\delta} U_e
\end{equation}
for which the area under the normalized defect profile is constrained to unity.
\citet{han_consistent_2024} further introduced the APG-modified scale
\begin{equation}\label{eq:zs_zpg_U}
    U_{zs\_apg}=U_{zs}\left(1+H\frac{\delta}{\rho U_e^2}\frac{dP_e}{dx}\right)=U_{zs}(1+H\beta_o)
\end{equation}
where $\beta_o=(\delta/\rho U_e^2)(dP_e/dx)$.
For the wall-normal mean velocity, $V_e-V$ is normalized by $V_e$, with $V_e=V(x,\delta)$ evaluated at the boundary-layer edge.
For assessing profile collapse, this defect form is equivalent to the $V/V_e$ normalization proposed by \citet{cite-16}.

\begin{samepage}
For the Reynolds-stress components, the hybrid scale of \citet{cite-40} was derived from the mean-momentum equation.
It combines the wall stress and the local pressure-gradient contribution:
\begin{equation}\label{eq:romero-hybrid}
    u_{\mathrm{hyb}}^2(y)=u_\tau^2+\frac{y}{\rho}\frac{dP_e}{dx}.
\end{equation}
This common, wall-distance-dependent scale is applied to all components.
\end{samepage}
Following \citet{maciel_self-similarity_2006,cite-44}, the square of the ZS velocity scale, $U_{zs}^2$, is also directly extended here to normalize each Reynolds-stress component.
The ZS\_APG formulation of \citet{han_consistent_2024} instead uses two grouped normalization amplitudes for different stress components,
\begin{subequations}\label{eq:zs-apg}
\begin{align}
    \langle uv\rangle_{zs\_apg} =\langle vv\rangle_{zs\_apg} &= U_e U_{zs}(1+H\beta_o);\\
    \langle uu\rangle_{zs\_apg} =\langle ww\rangle_{zs\_apg} &= U_eU_{zs}(1+H\beta_o) - HU_e u_\tau;
\end{align}
\end{subequations}
Table~\ref{tab:scales} summarizes these scale definitions and the component-wise reference values introduced in the next subsection.

\begin{table}[!htbp]
    \centering
    \caption{Scale variables used for the outer-profile normalizations.}
    \label{tab:scales}
    \footnotesize
    \setlength{\tabcolsep}{3.5pt}
    \renewcommand{\arraystretch}{1.35}
    \begin{tabular*}{\textwidth}{@{\extracolsep{\fill}}ccccccc@{}}
        \hline
        Method & $U_e-U$ & $V_e-V$ & $-\langle uv\rangle$ & $\langle vv\rangle$ & $\langle uu\rangle$ & $\langle ww\rangle$ \\
        \hline
        ZS & $U_{zs}=U_e\delta^\ast/\delta$ & --- & \multicolumn{4}{c}{$U_{zs}^2$} \\
        ZS\_APG & $U_{zs\_apg}=U_{zs}(1+H\beta_o)$ & --- & \multicolumn{2}{c}{$U_eU_{zs}(1+H\beta_o)$} & \multicolumn{2}{c}{$U_eU_{zs}(1+H\beta_o)-HU_eu_\tau$} \\
        Wei~\citep{cite-16} & --- & $V_e$ & \multicolumn{4}{c}{---} \\
        Romero~\citep{cite-40} & --- & --- & \multicolumn{4}{c}{$u_{\mathrm{hyb}}^2(y)=u_\tau^2+(y/\rho)(dP_e/dx)$} \\
        Present & $U_s$ & $V_s$ & $-\langle uv\rangle_s$ & $\langle vv\rangle_s$ & $\langle uu\rangle_s$ & $\langle ww\rangle_s$ \\
        \hline
    \end{tabular*}
\end{table}

\begin{figure}[!t]
    \centering
    \begin{subfigure}{0.32\textwidth}
        \centering
        \begin{overpic}[width=\linewidth, trim=10pt 45pt 25pt -10pt, clip]{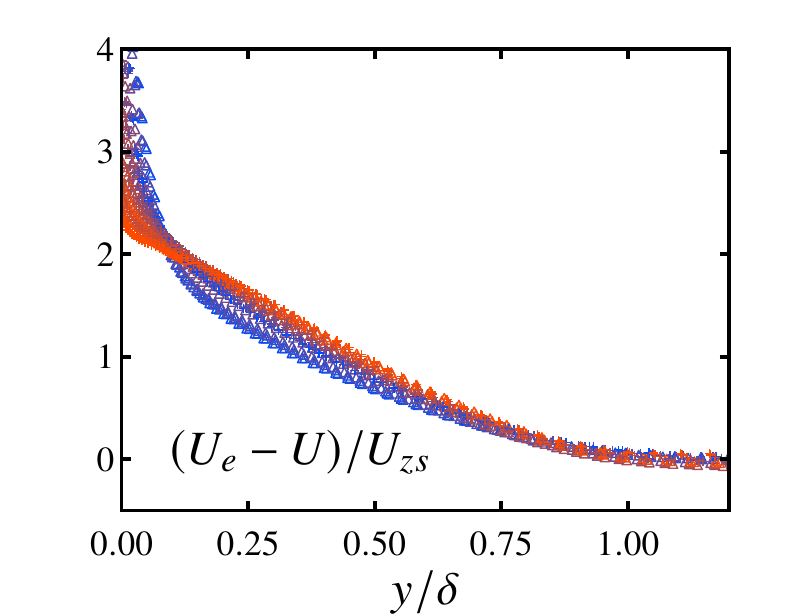}
            \put(1,70){\small(\textit{a})}
        \end{overpic}
    \end{subfigure}
    \begin{subfigure}{0.32\textwidth}
        \centering
        \begin{overpic}[width=\linewidth, trim=10pt 45pt 25pt -10pt, clip]{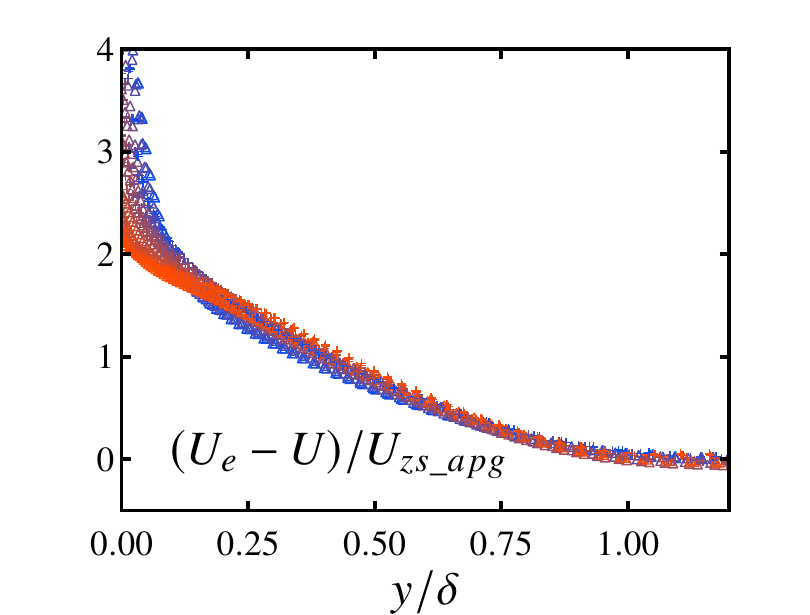}
            \put(1,70){\small(\textit{b})}
        \end{overpic}
    \end{subfigure}
    \\
    \begin{subfigure}{0.32\textwidth}
        \centering
        \begin{overpic}[width=\linewidth, trim=10pt 0pt 25pt -10pt, clip]{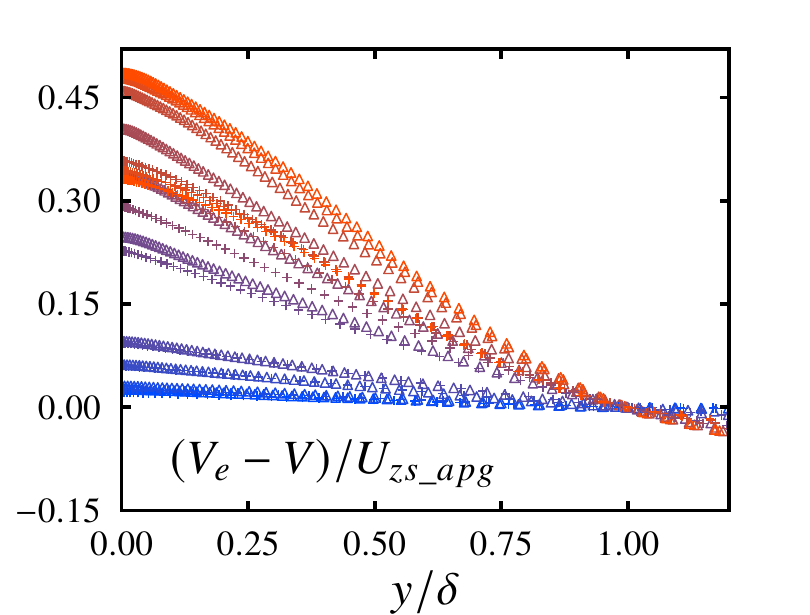}
            \put(1,84){\small(\textit{c})}
        \end{overpic}
    \end{subfigure}
    \begin{subfigure}{0.32\textwidth}
        \centering
        \begin{overpic}[width=\linewidth, trim=10pt 0pt 25pt -10pt, clip]{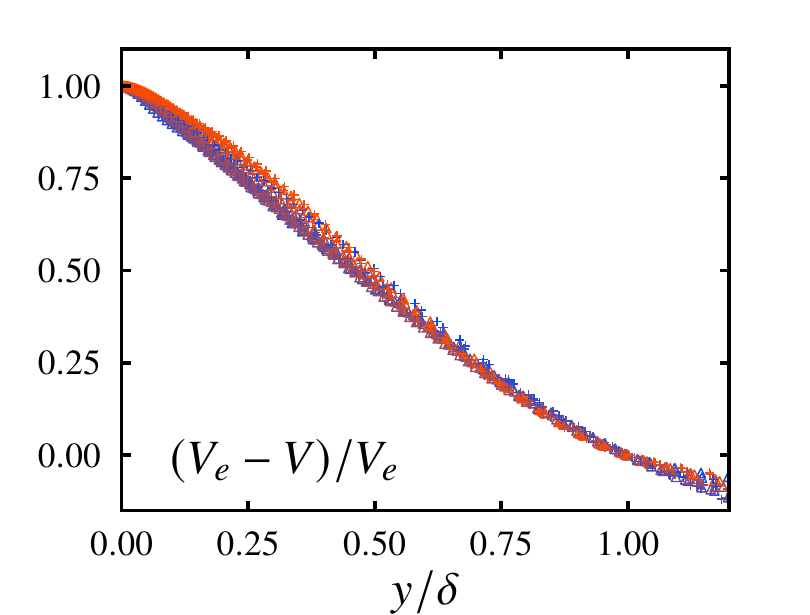}
            \put(1,84){\small(\textit{d})}
        \end{overpic}
    \end{subfigure}
    \caption{Mean velocity-defect profiles under ZS, ZS\_APG, and Wei~\citep{cite-16} normalizations.}
    \label{fig:Main-Data-1}
\end{figure}

Fig.~\ref{fig:Main-Data-1} compares the existing mean-velocity normalizations.
Triangle markers denote the present DNS and plus markers the data of \citet{coleman_numerical_2018}.
The streamwise defect normalized by $U_{zs}$ exhibits moderate scatter, with the higher-APG profiles displaced upward (Fig.~\ref{fig:Main-Data-1}(\textit{a})).
The ZS\_APG modification reduces this trend and yields a tighter group of profiles (Fig.~\ref{fig:Main-Data-1}(\textit{b})).
Applying the same $U_{zs\_apg}$ scale to $V_e-V$, however, produces pronounced APG variation (Fig.~\ref{fig:Main-Data-1}(\textit{c})).
Normalization by $V_e$ instead gives nearly common wall-normal defect profiles for the two data sets (Fig.~\ref{fig:Main-Data-1}(\textit{d})).
Thus, the mean streamwise and wall-normal components favor different normalization amplitudes under the present APG conditions.

\FloatBarrier

\begin{figure}[!t]
    \centering
    \begin{subfigure}{0.32\textwidth}
        \centering
        \begin{overpic}[width=\linewidth, trim=25pt 45pt 38pt -10pt, clip]{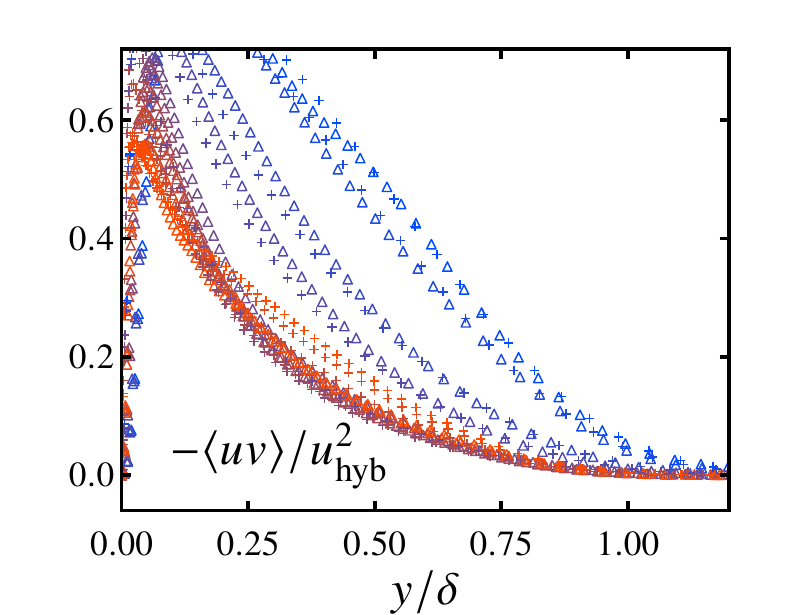}
            \put(1,74){\small(\textit{a})}
        \end{overpic}
    \end{subfigure}
    \begin{subfigure}{0.32\textwidth}
        \centering
        \begin{overpic}[width=\linewidth, trim=25pt 45pt 38pt -10pt, clip]{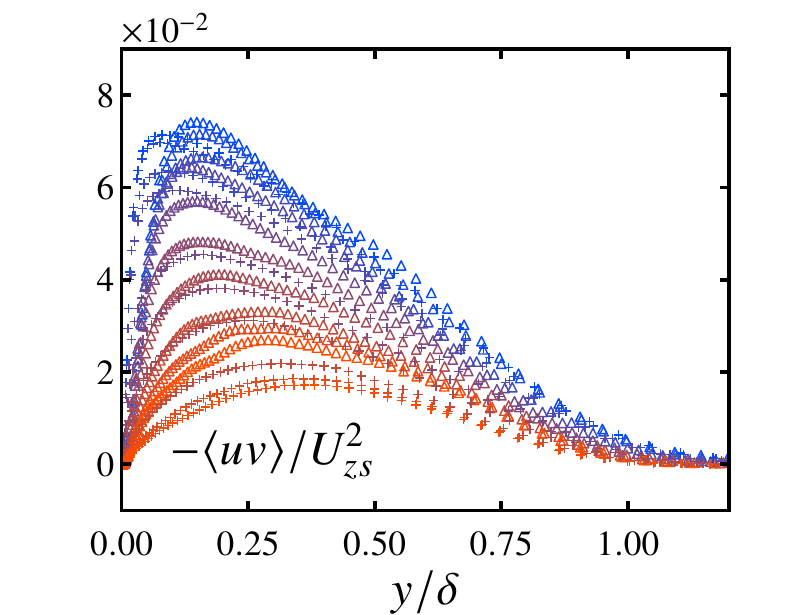}
            \put(1,74){\small(\textit{b})}
        \end{overpic}
    \end{subfigure}
    \begin{subfigure}{0.32\textwidth}
        \centering
        \begin{overpic}[width=\linewidth, trim=25pt 45pt 38pt -10pt, clip]{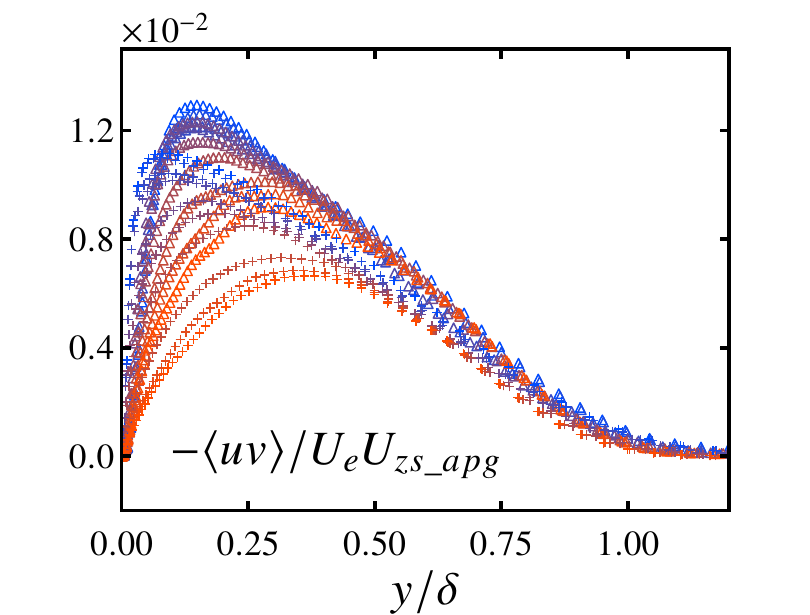}
            \put(1,74){\small(\textit{c})}
        \end{overpic}
    \end{subfigure}\\
    \begin{subfigure}{0.32\textwidth}
        \centering
        \begin{overpic}[width=\linewidth, trim=25pt 45pt 38pt -10pt, clip]{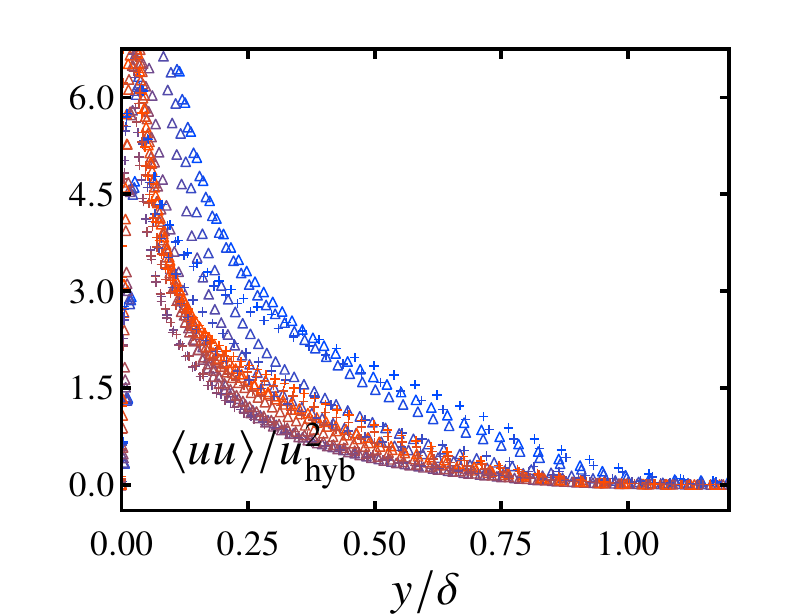}
            \put(1,74){\small(\textit{d})}
        \end{overpic}
    \end{subfigure}
    \begin{subfigure}{0.32\textwidth}
        \centering
        \begin{overpic}[width=\linewidth, trim=25pt 45pt 38pt -10pt, clip]{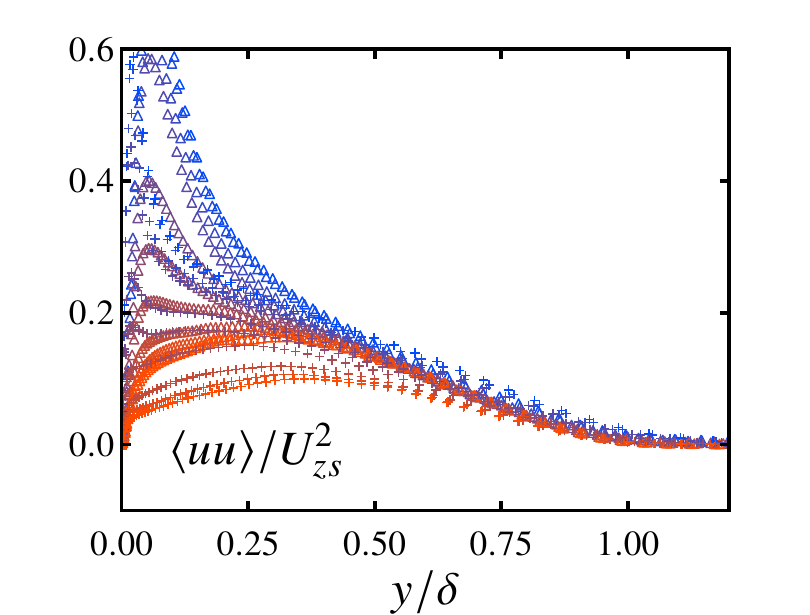}
            \put(1,74){\small(\textit{e})}
        \end{overpic}
    \end{subfigure}
    \begin{subfigure}{0.32\textwidth}
        \centering
        \begin{overpic}[width=\linewidth, trim=25pt 45pt 38pt -10pt, clip]{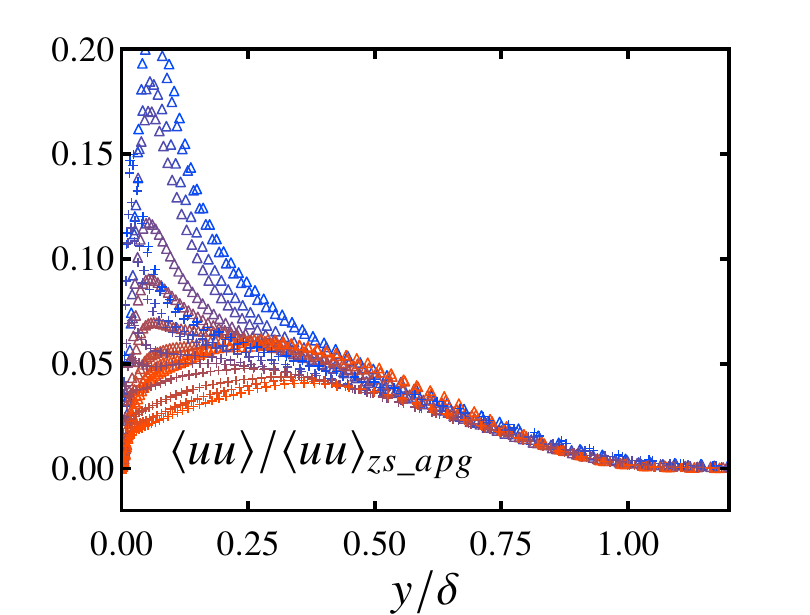}
            \put(1,74){\small(\textit{f})}
        \end{overpic}
    \end{subfigure}\\
    \begin{subfigure}{0.32\textwidth}
        \centering
        \begin{overpic}[width=\linewidth, trim=25pt 45pt 38pt -10pt, clip]{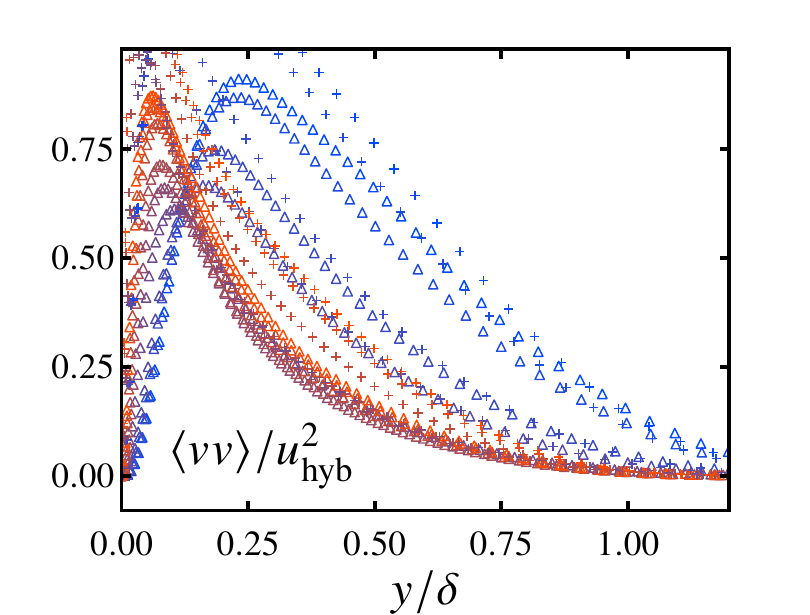}
            \put(1,74){\small(\textit{g})}
        \end{overpic}
    \end{subfigure}
    \begin{subfigure}{0.32\textwidth}
        \centering
        \begin{overpic}[width=\linewidth, trim=25pt 45pt 38pt -10pt, clip]{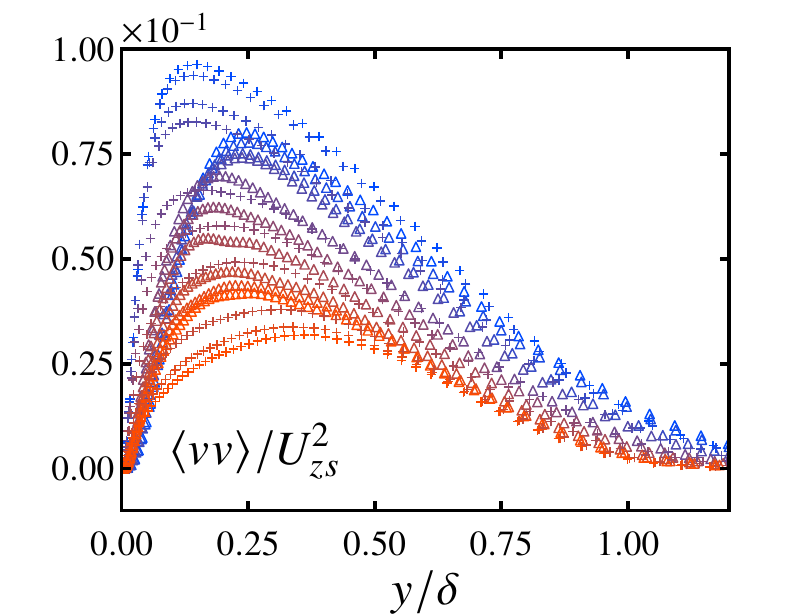}
            \put(1,74){\small(\textit{h})}
        \end{overpic}
    \end{subfigure}
    \begin{subfigure}{0.32\textwidth}
        \centering
        \begin{overpic}[width=\linewidth, trim=25pt 45pt 38pt -10pt, clip]{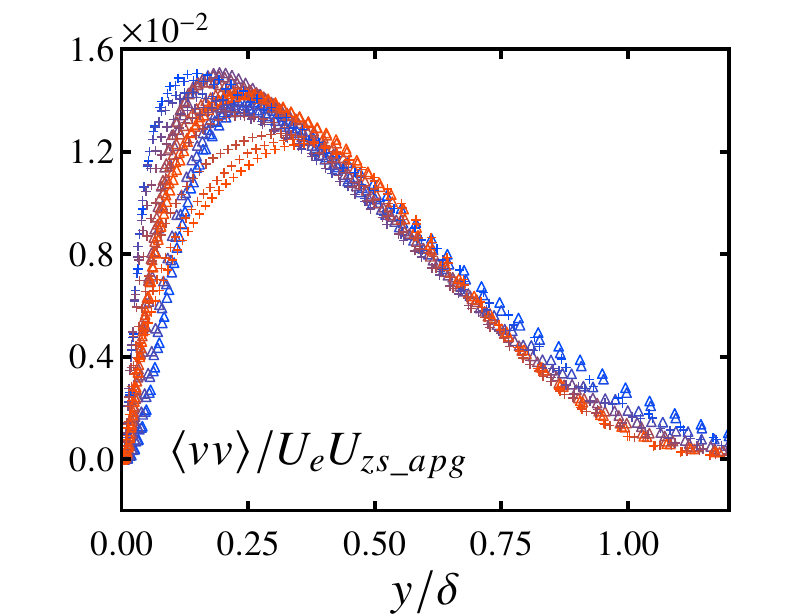}
            \put(1,74){\small(\textit{i})}
        \end{overpic}
    \end{subfigure}\\
    \begin{subfigure}{0.32\textwidth}
        \centering
        \begin{overpic}[width=\linewidth, trim=25pt 0pt 38pt -10pt, clip]{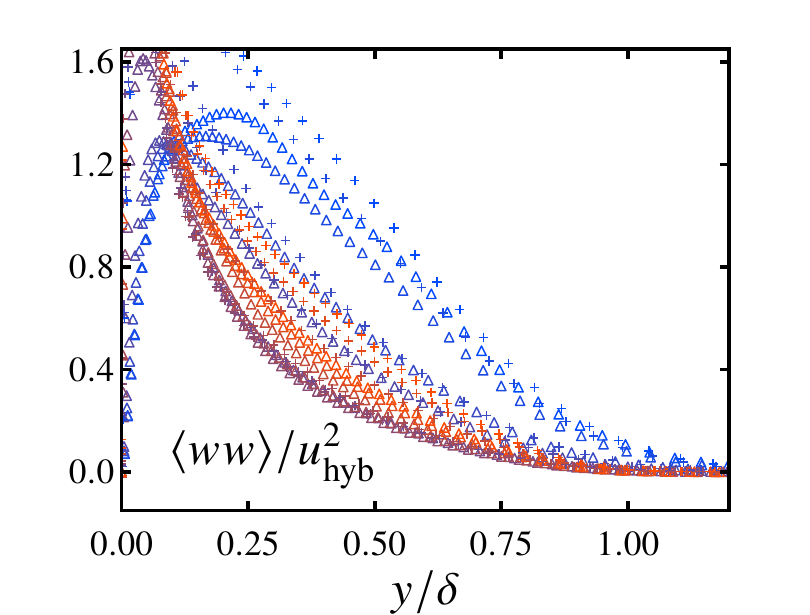}
            \put(1,88){\small(\textit{j})}
        \end{overpic}
    \end{subfigure}
    \begin{subfigure}{0.32\textwidth}
        \centering
        \begin{overpic}[width=\linewidth, trim=25pt 0pt 38pt -10pt, clip]{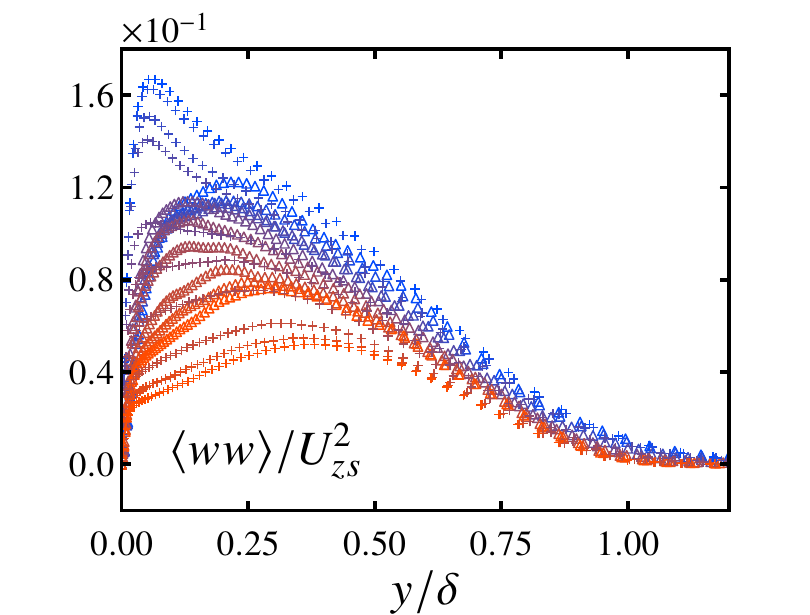}
            \put(1,88){\small(\textit{k})}
        \end{overpic}
    \end{subfigure}
    \begin{subfigure}{0.32\textwidth}
        \centering
        \begin{overpic}[width=\linewidth, trim=25pt 0pt 38pt -10pt, clip]{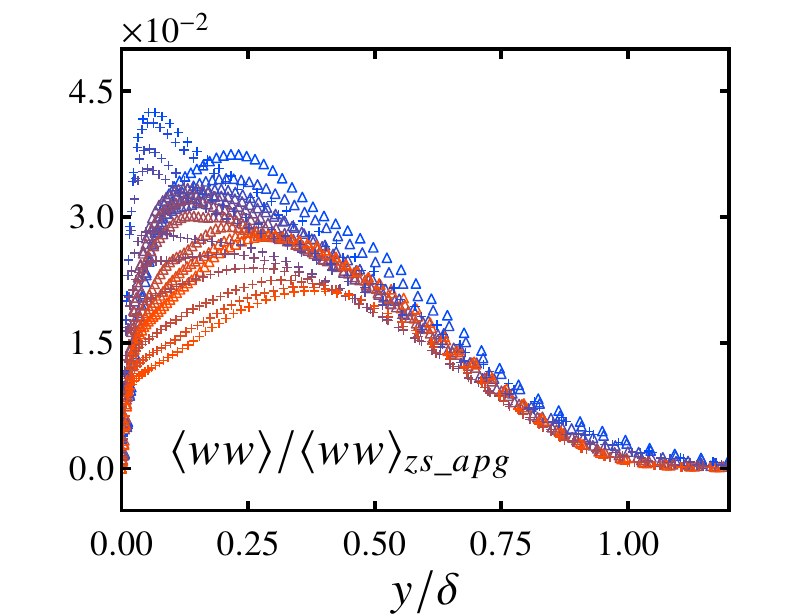}
            \put(1,88){\small(\textit{l})}
        \end{overpic}
    \end{subfigure}
    \caption{Reynolds stress profiles under Romero~\citep{cite-40}, ZS, and ZS\_APG normalizations.}
    \label{fig:Main-Data-2}
\end{figure}

Fig.~\ref{fig:Main-Data-2} assesses the Reynolds-stress normalizations, with rows corresponding to $-\langle uv\rangle$, $\langle uu\rangle$, $\langle vv\rangle$, and $\langle ww\rangle$, and columns corresponding to Romero, ZS, and ZS\_APG, respectively.
Under the Romero normalization, the profiles at moderate-to-strong APGs form relatively compact groups, whereas several conditions near the lower and upper ends of the pressure-gradient range depart from them.
Within this range, $-\langle uv\rangle$ exhibits the best outer-region collapse, with the profiles lying almost on a single curve over the assessed outer interval (Fig.~\ref{fig:Main-Data-2}(\textit{a})).
The $\langle vv\rangle$ profiles show a similar behavior with somewhat greater scatter (Fig.~\ref{fig:Main-Data-2}(\textit{g})).
By comparison, $\langle uu\rangle$ and $\langle ww\rangle$ remain less compact and retain clear APG variation (Figs.~\ref{fig:Main-Data-2}(\textit{d}) and \ref{fig:Main-Data-2}(\textit{j})).
Consistent with the trend reported by \citet{cite-40}, the profiles at appreciable APGs generally cluster below the blue near-ZPG profiles, for which $u_{\mathrm{hyb}}\approx u_\tau$.
The stress profiles normalized by the ZS scale still exhibit clear APG variation in all four components (Figs.~\ref{fig:Main-Data-2}(\textit{b}), \ref{fig:Main-Data-2}(\textit{e}), \ref{fig:Main-Data-2}(\textit{h}), and \ref{fig:Main-Data-2}(\textit{k})).
ZS\_APG gives a modest overall improvement, most visibly for $\langle vv\rangle$, but clear scatter remains in every component (Figs.~\ref{fig:Main-Data-2}(\textit{c}), \ref{fig:Main-Data-2}(\textit{f}), \ref{fig:Main-Data-2}(\textit{i}), and \ref{fig:Main-Data-2}(\textit{l})).

\FloatBarrier

Overall, none of the tested normalizations produces complete outer-profile collapse.
The different responses of the stress components show that component-wise normalization is needed to examine this similarity.

\subsection{Component-wise reference-value normalization \label{sec:present-scaling}}

We propose that, although a single normalization does not recover a common shape for all quantities, each component retains an outer self-similar form when treated separately, as expressed by Eq.~\eqref{eq:sim-define}.
This assumption implies that, across the broad range of nonequilibrium APGs considered, the outer-layer shape of each component remains similar while its amplitude can vary with the flow condition.

To explore this outer self-similar behavior, we introduce the following normalization strategy, in which each mean-velocity defect and Reynolds-stress component is normalized by its own value at a common outer reference location $y_s$:

\begin{align}\label{eq:scale}
    U_s &\equiv (U_e -U)|_{y=y_s},~~~~
    &V_s &\equiv (V_e -V)|_{y=y_s},~~~~\nonumber\\
    \langle u u\rangle_s &\equiv \langle u u\rangle|_{y=y_s}, &\langle u v\rangle_s &\equiv \langle u v\rangle|_{y=y_s}
    ,\\
    \langle vv \rangle_s &\equiv \langle vv\rangle|_{y=y_s}, &\langle ww\rangle_s &\equiv \langle ww\rangle|_{y=y_s}.\nonumber
\end{align}
Here, the subscript ``s" denotes a reference value sampled from the corresponding profile.
Equation~\eqref{eq:scale} is a profile-derived diagnostic normalization strategy for testing whether the outer profiles retain component-wise self-similar shapes over an outer wall-normal interval, rather than a new physics-based predictive scale.

\begin{figure}[!htbp]
    \centering
    \begin{subfigure}{0.32\textwidth}
        \centering
        \begin{overpic}[width=\linewidth, trim=25pt 45pt 38pt -10pt, clip]{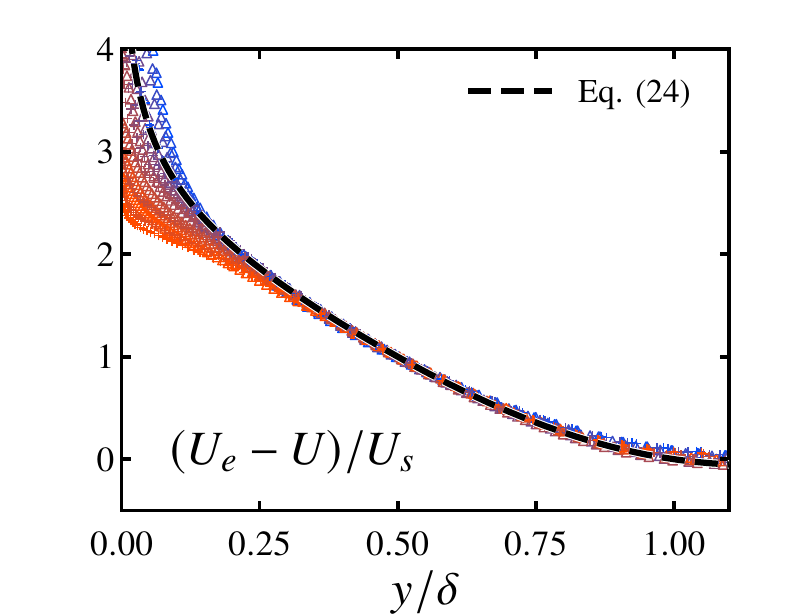}
            \put(1,74){\small(\textit{a})}
        \end{overpic}
    \end{subfigure}
    \begin{subfigure}{0.32\textwidth}
        \centering
        \begin{overpic}[width=\linewidth, trim=25pt 45pt 38pt -10pt, clip]{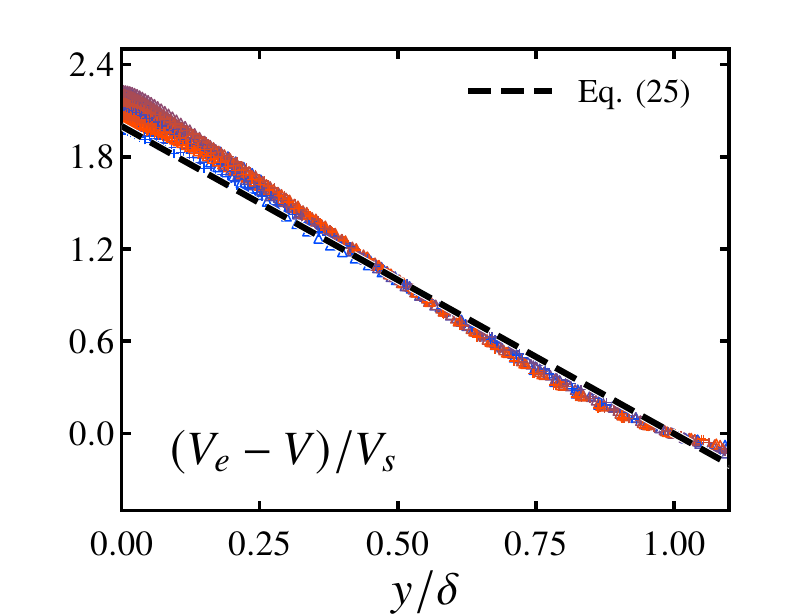}
            \put(1,74){\small(\textit{b})}
        \end{overpic}
    \end{subfigure}
    \begin{subfigure}{0.32\textwidth}
        \centering
        \begin{overpic}[width=\linewidth, trim=25pt 45pt 38pt -10pt, clip]{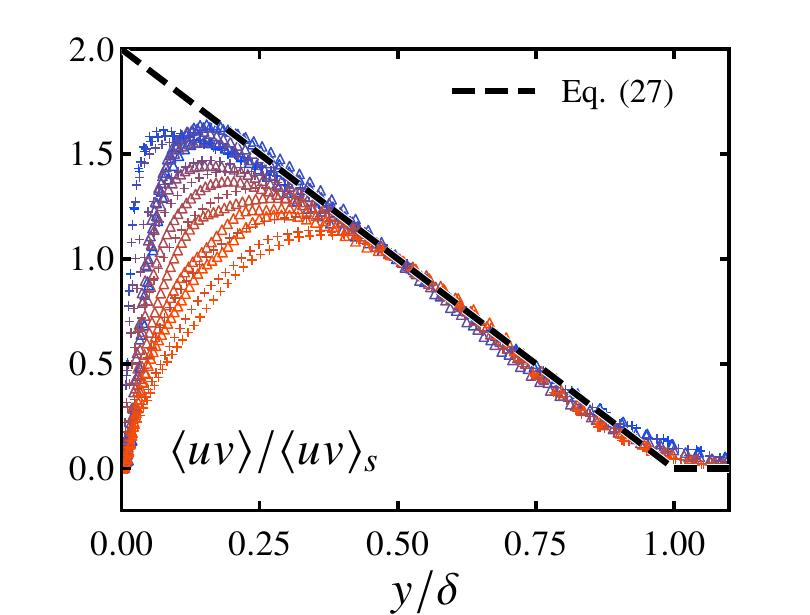}
            \put(1,74){\small(\textit{c})}
        \end{overpic}
    \end{subfigure}\\
        \begin{subfigure}{0.32\textwidth}
        \centering
        \begin{overpic}[width=\linewidth, trim=25pt 0pt 38pt -10pt, clip]{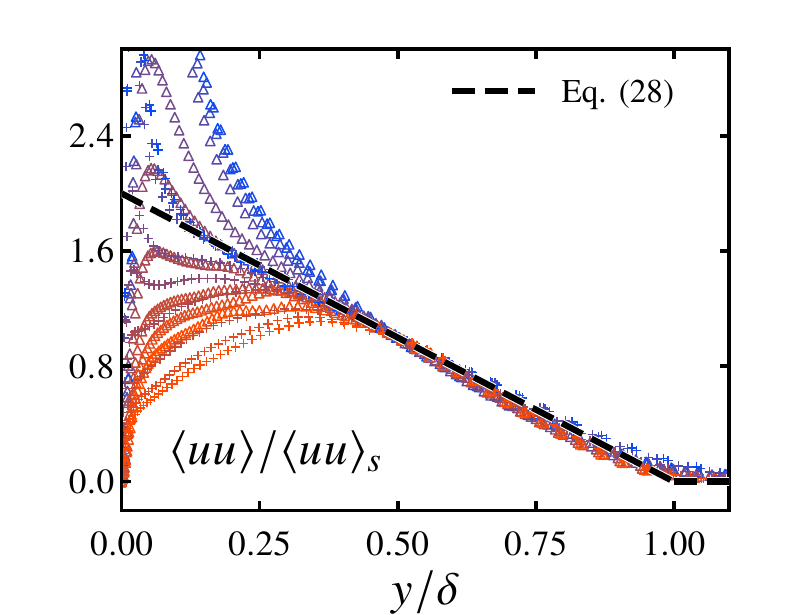}
            \put(1,88){\small(\textit{d})}
        \end{overpic}
    \end{subfigure}
    \begin{subfigure}{0.32\textwidth}
        \centering
        \begin{overpic}[width=\linewidth, trim=25pt 0pt 38pt -10pt, clip]{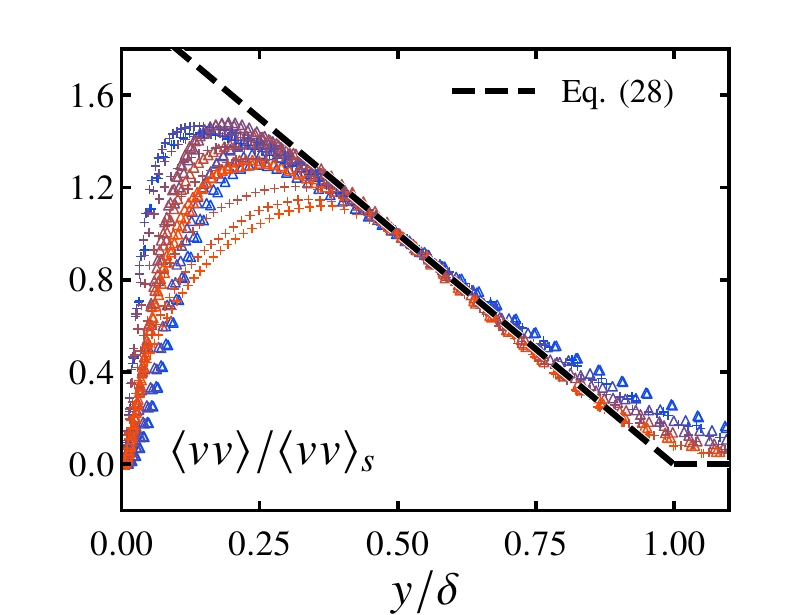}
            \put(1,88){\small(\textit{e})}
        \end{overpic}
    \end{subfigure}
    \begin{subfigure}{0.32\textwidth}
        \centering
        \begin{overpic}[width=\linewidth, trim=25pt 0pt 38pt -10pt, clip]{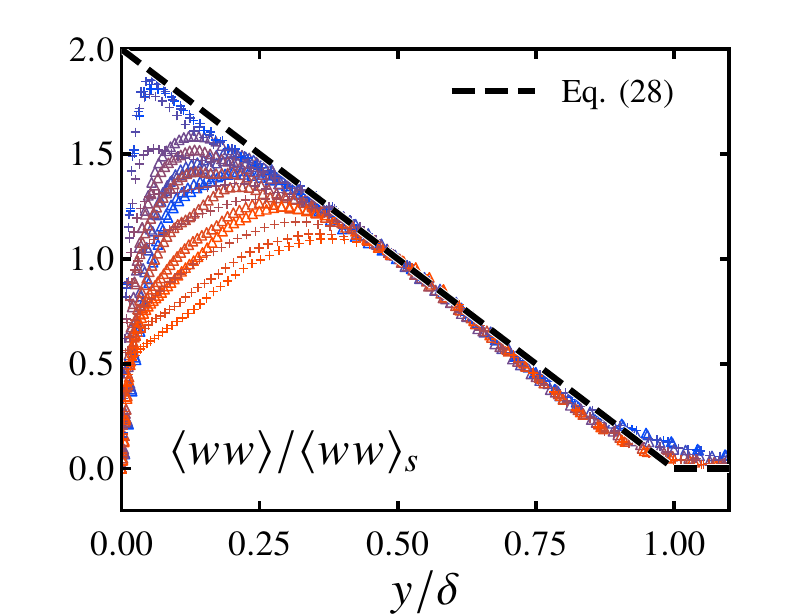}
            \put(1,88){\small(\textit{f})}
        \end{overpic}
    \end{subfigure}
    \caption{Outer similarity under component-wise reference-value normalization.}
    \label{fig:Main-Data-3}
\end{figure}

The profiles obtained with Eq.~\eqref{eq:scale}, using $y_s=0.5\delta$, are presented in Fig.~\ref{fig:Main-Data-3}.
The streamwise and wall-normal velocity defects exhibit consistent collapse for $y/\delta\gtrsim0.2$ (Figs.~\ref{fig:Main-Data-3}(\textit{a}) and \ref{fig:Main-Data-3}(\textit{b})).
Once the four stresses are normalized by their respective reference values, common outer profiles emerge for $y/\delta\gtrsim0.4$ (Figs.~\ref{fig:Main-Data-3}(\textit{c})--\ref{fig:Main-Data-3}(\textit{f})).
Their peaks shift outward from approximately $y/\delta=0.1$ at $\beta_w\approx0.05$ to $y/\delta=0.4$ at $\beta_w\approx120$, which sets the lower bound of the common-profile region.
The residual variation of $\langle vv\rangle$ near the boundary-layer edge is associated with the imposed freestream transpiration.
The analytical curves included in Fig.~\ref{fig:Main-Data-3} are derived in the next subsection.

Collapse quality is evaluated using relative $L_2$ errors.
For a quantity $\phi$, let $\varphi^{(n)}(y/\delta)$ denote its normalized profile under flow condition $n$, where $n=1,\ldots,N$, and let $\overline{\varphi}=N^{-1}\sum_{n=1}^{N}\varphi^{(n)}$ denote the ensemble-mean profile.
At each wall-normal location, the local collapse error is defined as the root-mean-square dispersion across the profiles, normalized by the root-mean-square magnitude of $\overline{\varphi}$ over the outer interval:
\begin{equation}\label{eq:local-collapse-error}
e_\phi\left(\frac{y}{\delta}\right)=
\frac{\left\{N^{-1}\sum_{n=1}^{N}\left[\varphi^{(n)}(y/\delta)-\overline{\varphi}(y/\delta)\right]^2\right\}^{1/2}}
{\left\{0.6^{-1}\int_{0.4}^{1.0}\overline{\varphi}^{\,2}(y/\delta)\,d(y/\delta)\right\}^{1/2}}.
\end{equation}
\begin{samepage}
The corresponding collapse error for the entire outer profile is defined by the relative $L_2$ norm of the deviations over the interval:
\begin{equation}\label{eq:profile-collapse-error}
E_\phi=\left\{\frac{\sum_{n=1}^{N}\int_{0.4}^{1.0}\left[\varphi^{(n)}(y/\delta)-\overline{\varphi}(y/\delta)\right]^2d(y/\delta)}
{N\int_{0.4}^{1.0}\overline{\varphi}^{\,2}(y/\delta)d(y/\delta)}\right\}^{1/2}.
\end{equation}
\end{samepage}
The interval $0.4\le y/\delta\le1.0$ is the widest common outer range over which the stress profiles can collapse after their peaks move outward with increasing APG.
Both errors vanish for exact collapse and increase with profile dispersion.
The 20 profiles from the present DNS and \citet{coleman_numerical_2018} are assigned equal weights.
Because $E_\phi$ measures dispersion over the entire outer interval, its reduction cannot result merely from the enforced agreement at the single reference location.

\begin{figure}[!htbp]
    \centering
    \includegraphics[width=\textwidth]{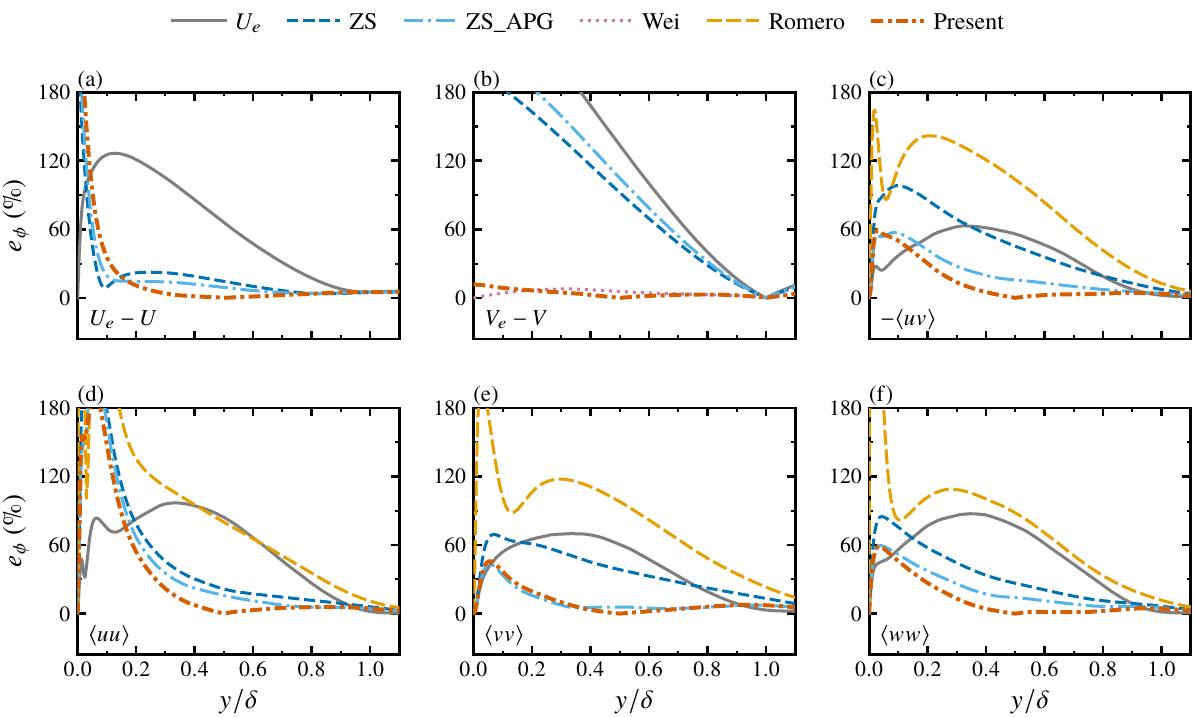}
    \caption{Local outer-profile collapse errors under the tested normalizations.}
    \label{fig:normalization-collapse-errors}
\end{figure}

Fig.~\ref{fig:normalization-collapse-errors} compares the local collapse errors under the tested normalizations.
Overall, the component-wise reference-value normalization reduces the errors of both mean-velocity defects and all four Reynolds stresses throughout the outer region.
For $U_e-U$, the $U_e$ normalization gives large errors, while the local errors under ZS and ZS\_APG are of order $10$--$20\%$ near the lower bound of the outer interval; the present normalization remains below $10\%$.
For $V_e-V$, the Wei normalization by $V_e$ already gives small errors, and the present normalization performs similarly, whereas the $U_e$, ZS, and ZS\_APG normalizations all give much larger errors.
For the four stresses, the present normalization consistently gives substantially smaller outer-region errors.
The Romero normalization yields large errors because of the profile departures at low and high APGs; the $U_e$, ZS, and ZS\_APG normalizations also retain appreciable scatter.
The notably smaller ZS\_APG error for $\langle vv\rangle$ is consistent with the preceding profile comparison.
These comparisons show that the visual collapse in Fig.~\ref{fig:Main-Data-3} represents a systematic improvement across the profile ensemble.

\FloatBarrier

\begin{figure}[!htbp]
    \centering
    \includegraphics[width=0.9\textwidth]{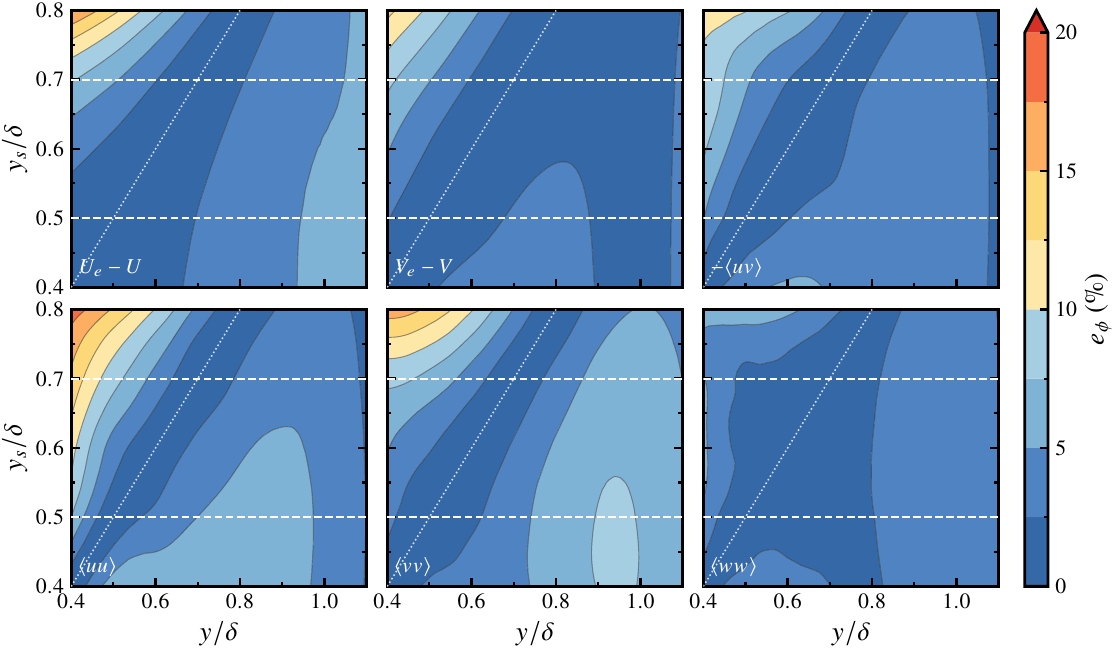}
    \caption{Collapse-error sensitivity to the outer reference location.}
    \label{fig:normalization-reference-sensitivity}
\end{figure}

The reference location $y_s=0.5\delta$ lies beyond the outer stress peaks of all the profiles examined, while remaining sufficiently far from the boundary-layer edge for the reference amplitudes to be well resolved.
We note that \citet{gungor_analysis_2014,gungor_scaling_2016} introduced the half-thickness velocity scale $U_o=2[U_e-U(y=0.5\delta)]$ for TBLs, in analogy with commonly used free-shear-layer velocity scales.
This scale is proportional to $U_s$ at $y_s=0.5\delta$ and is therefore consistent with the present reference-value normalization.
In the present study, however, $y_s=0.5\delta$ is selected as a convenient representative position and does not bind the observed similarity to a half-thickness scale.
The principal contribution is the identification of component-wise outer similarity, rather than the definition of a unique scale at a prescribed wall-normal location.
Accordingly, Fig.~\ref{fig:normalization-reference-sensitivity} tests a range of reference locations.
For each component, the contours show the local collapse error $e_\phi$ as a function of the profile coordinate $y/\delta$ on the horizontal axis and the reference location $y_s/\delta$ on the vertical axis.
The horizontal dashed lines delimit $0.5\le y_s/\delta\le0.7$, and the dotted diagonal identifies $y=y_s$.
For all six components, $E_\phi$ remains below $10\%$ throughout $0.4\le y_s/\delta\le0.8$ and below $6\%$ within the dashed interval.
The blue low-error regions also extend over substantial portions of the outer layer, rather than remaining confined to $y=y_s$.
The increase in several errors as $y_s/\delta$ approaches $0.8$ indicates the reduced suitability of reference values close to the boundary-layer edge.
Thus, the component-wise outer similarity is robust over a relatively wide range of outer reference locations and is not tied specifically to $y_s=0.5\delta$.

\FloatBarrier

In summary, the component-wise reference-value normalization reveals common outer profiles for the mean velocities and Reynolds stresses from both data sets.
Once a common shape function has been established, an outer-layer measurement at a single reference location can also supply the profile amplitude and enable rapid reconstruction of the corresponding outer profile.
This component-wise outer similarity may also inform APG TBL modeling.
Analytical expressions for these profiles are provided below as compact reference behavior for turbulence-model assessment in APG boundary layers.

\subsection{Expressions for the component-wise outer-profile functions}
For the mean streamwise velocity profile, the classical wall-wake formula proposed in the form of \citet{Coles1969PROCEEDINGSCO} reads:

\begin{equation}\label{eq:wall-wake-profile}
    \frac{U}{u_\tau} = \frac{1}{\kappa} \ln \left( \frac{y u_\tau}{\nu}\right) + B + \frac{2\Pi}{\kappa}\sin^2\left(\frac{\pi}{2}\frac{y}{\delta}\right).
\end{equation}
where $\Pi=0.6$ is the suggested ZPG wake strength in \citet{wilcox_2006_turbulence}.
As $U$ profiles scatter severely under $u_\tau$, it is rather tedious to fit (\ref{eq:wall-wake-profile}) with data by adjusting the parameters $\kappa$, $B$ and $\Pi$ for different $\beta_w$.
However, let us consider the velocity defect, which, from \eqref{eq:wall-wake-profile}, is

\begin{equation}\label{eq:Udef-wake}
    \frac{U_e - U}{u_\tau} = -\frac{1}{\kappa}\ln{\left(\frac{y}{\delta}\right)}+\frac{2\Pi}{\kappa}\left[1-\sin^2\left(\frac{\pi}{2}\frac{y}{\delta}\right)\right],
\end{equation}
hence, after substituting in $y/\delta=0.5$,
\begin{equation}\label{eq:Udef-wake2}
    \frac{U_s}{u_\tau} = -\frac{1}{\kappa}\ln{\left(0.5\right)}+\frac{2\Pi}{\kappa}\left[1-\sin^2\left(\frac{\pi}{2} \cdot 0.5\right)\right].
\end{equation}
By the ratio of \eqref{eq:Udef-wake} and \eqref{eq:Udef-wake2}, the functional expression of $f_1 $ is readily obtained:

\begin{equation} \label{eq:Udef-sim}
    \frac{U_e-U}{U_s}=f_1 \left(\frac{y}{\delta}\right) = \frac{-\ln{\left(\frac{y}{\delta}\right)}+2\Pi\left[1-\sin^2(\frac{\pi}{2}\frac{y}{\delta})\right]}{\ln2+\Pi},
\end{equation}
which is marked by the dashed line in Fig. \ref{fig:Main-Data-3}(\textit{a}) and adheres closely to the data.
Compared to \eqref{eq:wall-wake-profile}, there is only one parameter in \eqref{eq:Udef-sim}, \textit{i.e.} the aforementioned $\Pi=0.6$, which is universal for ZPG and APG cases, whilst the coefficient $2$ in the denominator comes from the defined reference location of $y_s=0.5\delta$ such that $f_1(0.5)=1$.

Note that once the streamwise defect in Eq.~\eqref{eq:Udef-sim} is specified, integration of the incompressible mean continuity equation yields $V_e-V(y)=-\int_y^{\delta}\frac{\partial}{\partial x}[U_e-U_s f_1(y'/\delta)]\,dy'$, from which $V_s$ follows by setting $y=y_s$.
Evaluation of this integral requires accurately resolved streamwise derivatives of $U_e$, $U_s$, and $\delta$, which cannot be obtained from many of the benchmark data sets.
Calculations using the present DNS and the DNS data of \citet{coleman_numerical_2018} show that the resulting values of $V_s$ generally differ from the directly evaluated $V_e-V(y_s)$ by only a few percent.
At a few stations under near-ZPG conditions, however, the numerical and approximation errors associated with evaluating $V_s$ become non-negligible relative to the magnitude of the wall-normal velocity component.
We therefore retain the direct definition of $V_s$ at $y_s$ so that the same component-wise normalization procedure can be applied consistently to all quantities and data sets.

For the mean wall normal velocity, we follow the analysis of \citet{Chen_Hussain_She_2019} that the leading contribution to the dissipation term in the transport equation for $V^2/2$, $\varepsilon_V=\nu(\partial V/\partial y)^2$, scales as $\nu V^2_e/\delta^2$.
Since our present simulations adopt suction on the freestream, we consider the leading order expansion of this term as $\varepsilon_V\sim \text{const.}$, hence that $\partial V/\partial y\sim \text{const.}$ ($\sim$ indicates to scale with).
Given the freestream condition $V=V_e$ at $y/\delta=1$, one has $V/V_e\approx y/\delta$, and therefore

\begin{equation}\label{eq:Vdf-sim}
     \frac{V_e-V}{V_s}=f_2 \left(\frac{y}{\delta}\right) = 2\left(1-\frac{y}{\delta}\right).
\end{equation}
Again, the coefficient $2$ is determined by the definition of $y_s$ such that $f_2(0.5)=1$.
Fig. \ref{fig:Main-Data-3}(\textit{b}) illustrates that the linear function \eqref{eq:Vdf-sim} satisfactorily represents the outer-region data and remains accurate well toward the wall.

Next, the similarity function of $-\langle uv\rangle$ is developed based on \eqref{eq:Udef-sim} and \eqref{eq:Vdf-sim}.
Recall the approximate mean momentum equation in the outer flow (neglecting the density variation),

\begin{equation}\label{eq:zpg-balance}
    - U \frac{\partial V}{\partial y} + V \frac{\partial U}{\partial y} + \frac{1}{\rho}\frac{dP_e}{dx} \approx - \frac{\partial \langle uv \rangle}{\partial y},
\end{equation}
where in the first term, the original streamwise derivative $\partial U/\partial x $ has been replaced by $-\partial V/\partial y $.
The pressure-gradient term is independent of $y$.
Substituting $V\propto y$ from \eqref{eq:Vdf-sim} and $\partial U/\partial y \propto y^{-1} $ from the logarithmic term in \eqref{eq:Udef-sim}, one has  $V\partial U/\partial y \propto y^0$.
For $U \partial V/\partial y$, let us invoke a power-law approximation of \eqref{eq:Udef-sim}, \textit{i.e.}, $U\propto(y/\delta)^{\alpha}$, where, according to \citet{barenblatt_scaling_1993}, the exponent $\alpha-1\rightarrow-1$ or $\alpha\rightarrow0$.
Then $U \partial V/\partial y\propto y^\alpha\rightarrow y^0$.
Therefore, the convection terms on the left hand side of \eqref{eq:zpg-balance} are approximated by $y^0$ and, together with the pressure-gradient term, give ${\partial \langle uv \rangle}/{\partial y} \sim \text{const.}$.
Hence, we have a linear profile $\langle uv\rangle = c_1\left({y}/{\delta}\right)+c_2$ with $c_1$ and $c_2$ being two coefficients.
With  $\langle uv\rangle=0$ at ${y}/{\delta}=1$, we have $c_2=-c_1$, and hence

\begin{equation}\label{eq:uv-sim}
         \frac{\langle uv \rangle}{\langle uv \rangle_s}=g_0 \left(\frac{y}{\delta}\right) = 2\left(1-\frac{y}{\delta}\right).
\end{equation}
Data verification of \eqref{eq:uv-sim} is given in Fig. \ref{fig:Main-Data-3}(\textit{c}), showing good agreement for $y/\delta \gtrsim 0.4$.

Finally, according to \citet{townsend_shear_flow}, the ratios of $\langle uv\rangle$ to each normal stress component and the turbulent kinetic energy can be considered constants.
We take this view and hence:

\begin{equation}\label{eq:sim-uiuj}
   g_0= g_1  = g_2  = g_3 = 2\left(1-\frac{y}{\delta}\right),
\end{equation}
where $g_0$, $g_1$, $g_2$ and $g_3$ are similarity functions for $\langle uv\rangle$, $\langle uu\rangle$, $\langle vv\rangle$ and $\langle ww\rangle$, respectively.
The linear behavior is well supported by the outer-region collapse of the Reynolds normal stresses, as presented in Fig. \ref{fig:Main-Data-3}(\textit{d})-(\textit{f}).
Alternatively, one may assume a first-order Taylor expansion for $g_j$ and obtain \eqref{eq:sim-uiuj} as well.
However, the present procedure provides insight into the interconnections between the outer behaviors of mean velocities and Reynolds stresses.
Despite the pressure gradient significantly altering the overall development of the boundary layer, the wake law remains robust when represented in terms of defect form.
From the perspective of momentum balance, \eqref{eq:zpg-balance} relates the outer mean-flow convection to the pressure gradient and Reynolds shear stress gradient.
As demonstrated by \citet{panton_review_2005,panton_composite_2007}, the form described by \eqref{eq:Udef-wake} can be systematically expressed as a composite asymptotic expansion.
More recent studies on APG TBLs \citep{cite-18, cite-19} further discuss how the relationship between velocity defect and Reynolds shear stress influences scaling in asymptotic accuracy, and it is worth integrating the above functions $f_i$ and $g_j$ into asymptotic expansions in the future.

\section{Further verifications of component-wise outer similarity\label{sec:further-verif}}

It is natural to ask whether the component-wise outer similarity identified above holds for more general flow conditions.
To test this, we perform further verification using the experimental and numerical data from the APG benchmark described in subsection \ref{sec:apg-benchmark}.

\subsection{Verifications with experimental data}

The experimental data include the non-equilibrium APG experiments of \citet{marusic_wall_wake_1995} and the axisymmetric APG TBL experiment of \citet{Dengel_Fernholz_1990}.
Since the experiments do not give vorticity information, defining $\widetilde{U}$ is infeasible for these experimental data.
Therefore, $U_e$ is kept as the local freestream mean velocity $U$ and the boundary layer thickness $\delta$ is simply determined by the wall distance of $0.99U_e$  for \citet{marusic_wall_wake_1995}, and taken as provided for \citet{Dengel_Fernholz_1990}.

\begin{figure}[!htbp]
    \centering
    \begin{subfigure}{0.32\textwidth}
        \centering
        \begin{overpic}[width=\linewidth, trim=25pt 45pt 38pt -10pt, clip]{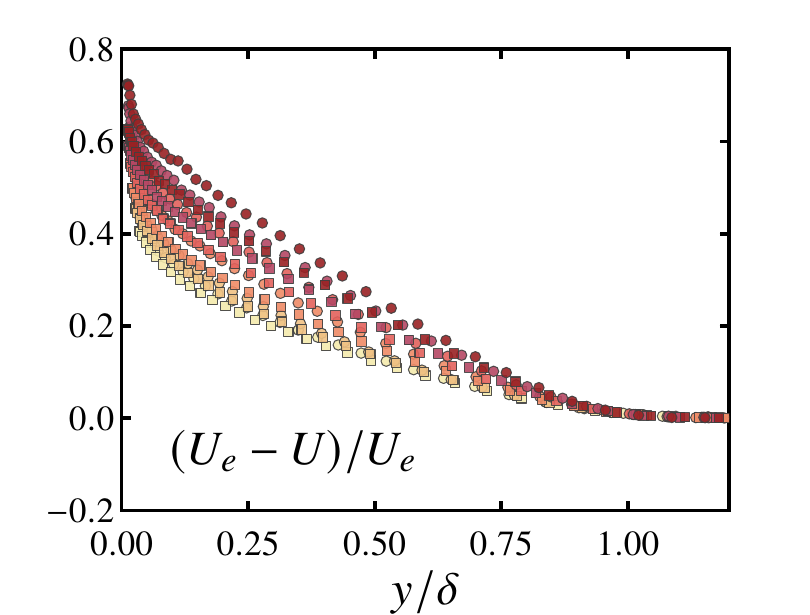}
            \put(1,74){\small(\textit{a})}
        \end{overpic}
    \end{subfigure}
        \begin{subfigure}{0.32\textwidth}
        \centering
        \begin{overpic}[width=\linewidth, trim=25pt 45pt 38pt -10pt, clip]{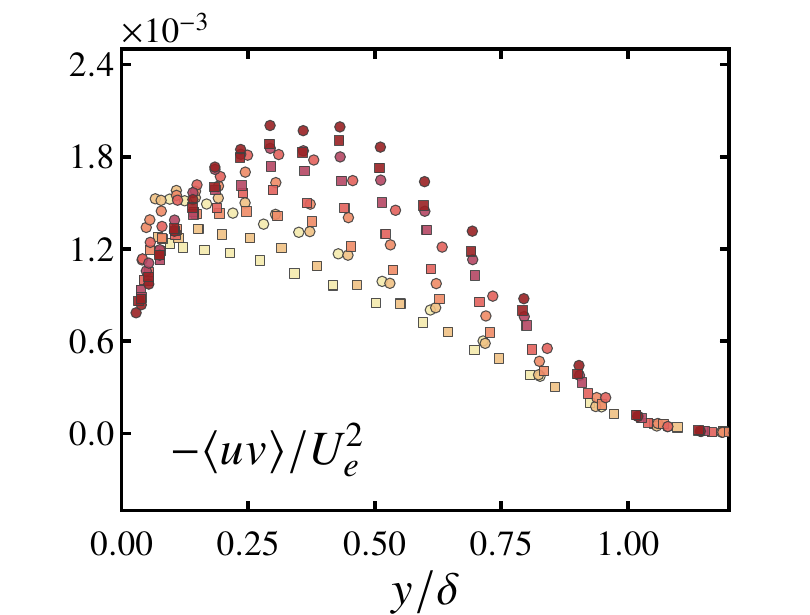}
            \put(1,74){\small(\textit{b})}
        \end{overpic}
    \end{subfigure}
    \begin{subfigure}{0.32\textwidth}
        \centering
        \begin{overpic}[width=\linewidth, trim=25pt 45pt 38pt -10pt, clip]{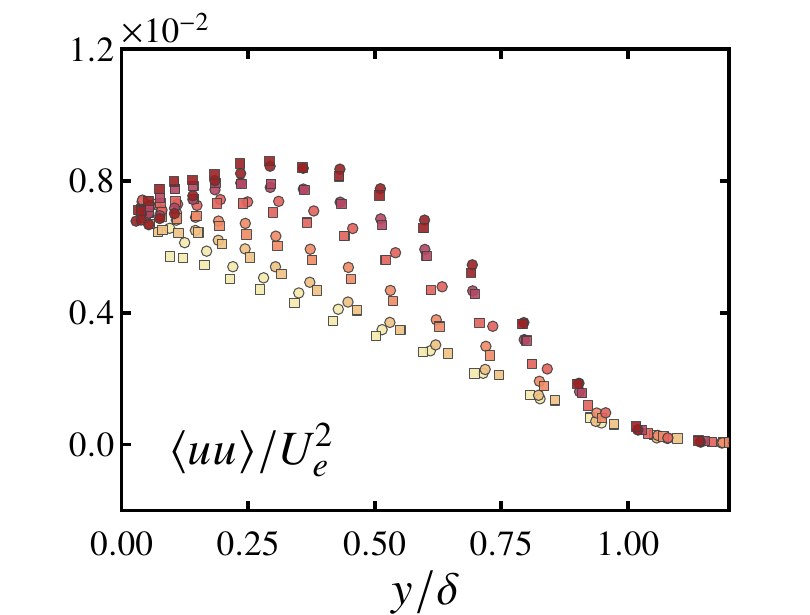}
            \put(1,74){\small(\textit{c})}
        \end{overpic}
    \end{subfigure}\\
        \begin{subfigure}{0.32\textwidth}
        \centering
        \begin{overpic}[width=\linewidth, trim=25pt 45pt 38pt -10pt, clip]{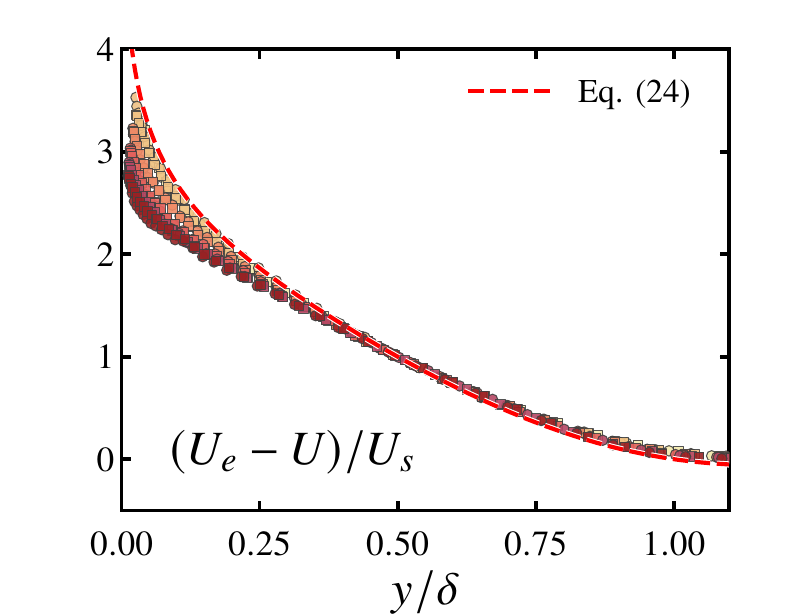}
            \put(1,74){\small(\textit{d})}
        \end{overpic}
    \end{subfigure}
        \begin{subfigure}{0.32\textwidth}
        \centering
        \begin{overpic}[width=\linewidth, trim=25pt 45pt 38pt -10pt, clip]{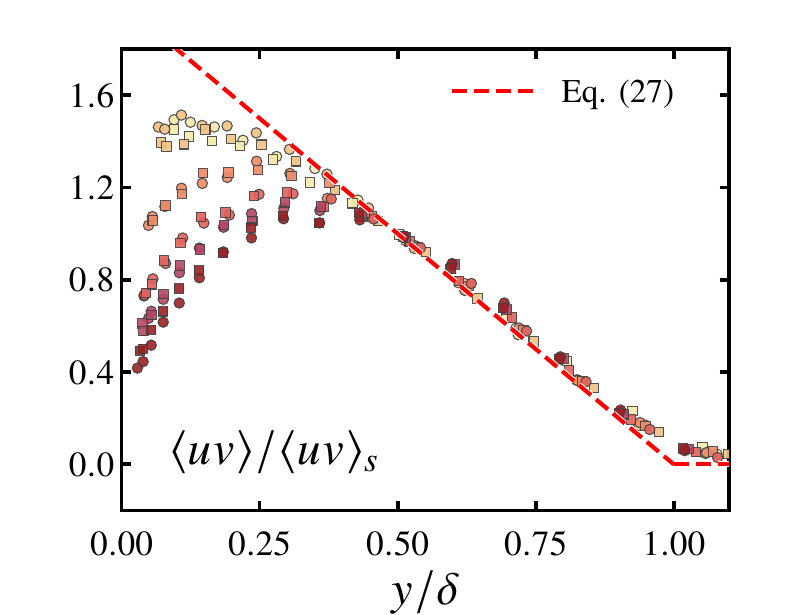}
            \put(1,74){\small(\textit{e})}
        \end{overpic}
    \end{subfigure}
    \begin{subfigure}{0.32\textwidth}
        \centering
        \begin{overpic}[width=\linewidth, trim=25pt 45pt 38pt -10pt, clip]{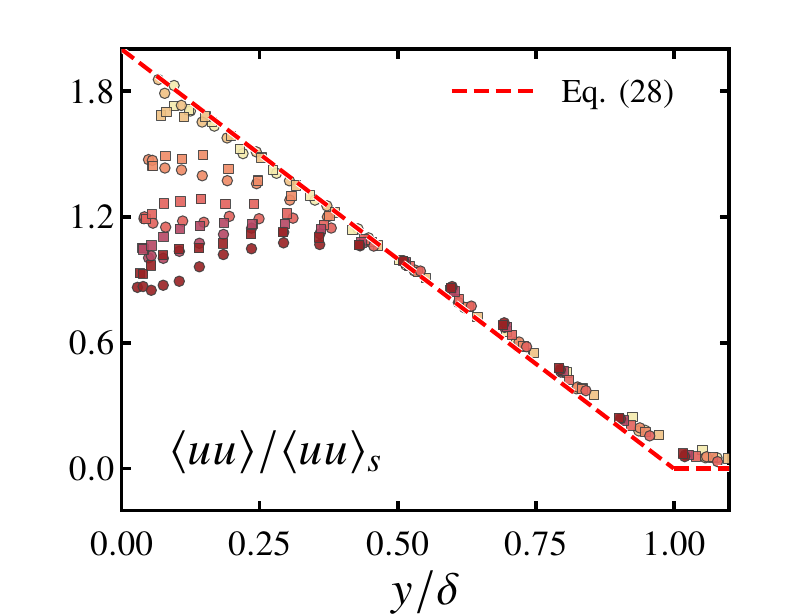}
            \put(1,74){\small(\textit{f})}
        \end{overpic}
    \end{subfigure}\\
    \begin{subfigure}{0.32\textwidth}
        \centering
        \begin{overpic}[width=\linewidth, trim=25pt 45pt 38pt -10pt, clip]{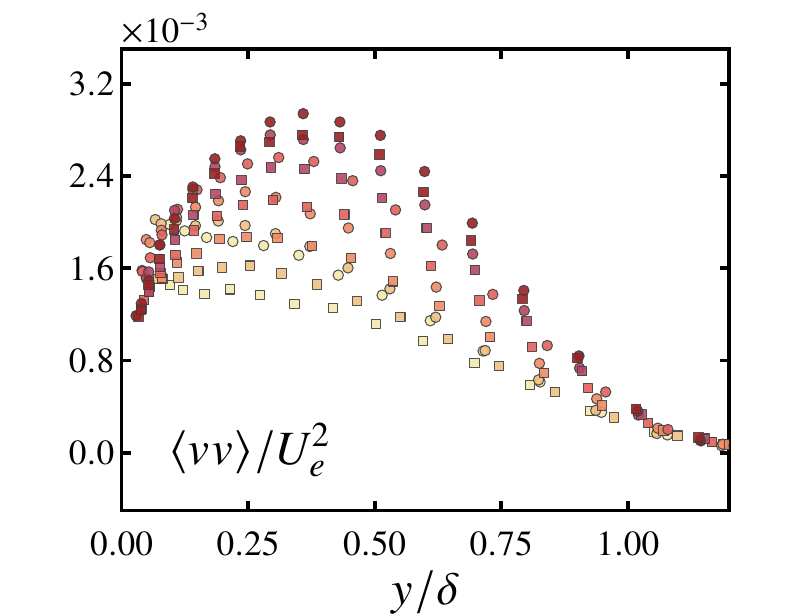}
            \put(1,74){\small(\textit{g})}
        \end{overpic}
    \end{subfigure}
        \begin{subfigure}{0.32\textwidth}
        \centering
        \begin{overpic}[width=\linewidth, trim=25pt 45pt 38pt -10pt, clip]{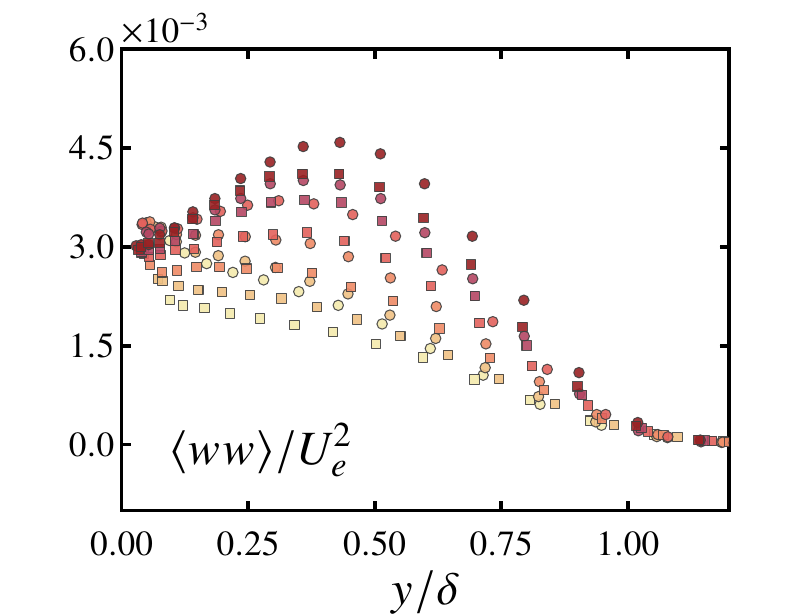}
            \put(1,74){\small(\textit{h})}
        \end{overpic}
    \end{subfigure}
        \begin{subfigure}{0.32\textwidth}
        \centering
        \begin{overpic}[width=\linewidth, trim=25pt 45pt 38pt -10pt, clip]{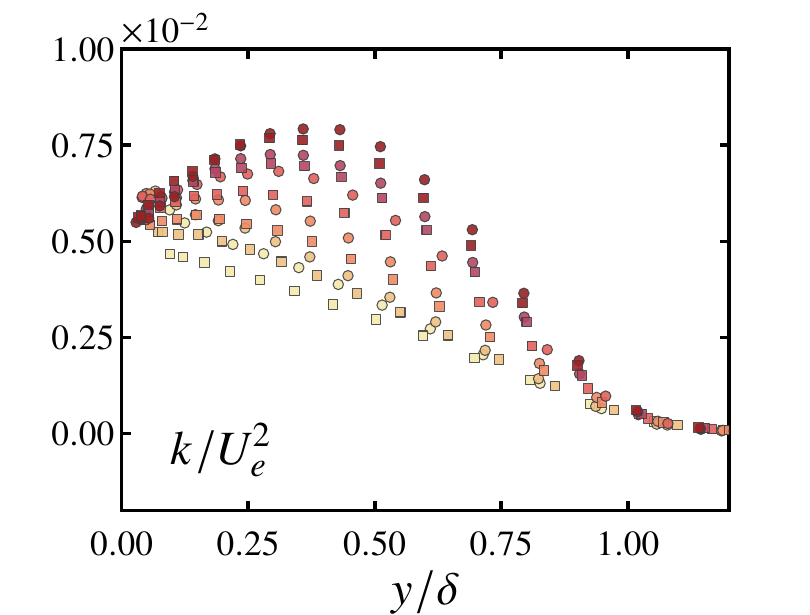}
            \put(1,74){\small(\textit{i})}
        \end{overpic}
    \end{subfigure}\\
        \begin{subfigure}{0.32\textwidth}
        \centering
        \begin{overpic}[width=\linewidth, trim=25pt 0pt 38pt -10pt, clip]{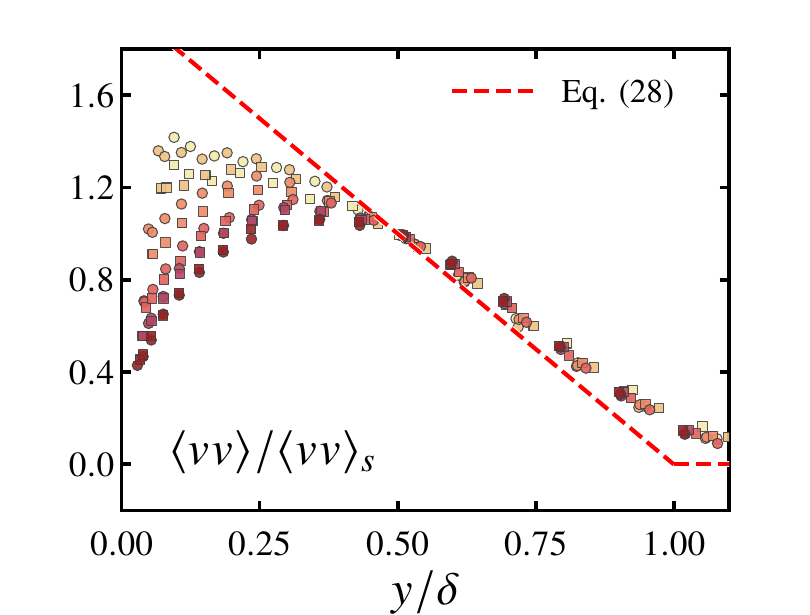}
            \put(1,88){\small(\textit{j})}
        \end{overpic}
    \end{subfigure}
        \begin{subfigure}{0.32\textwidth}
        \centering
        \begin{overpic}[width=\linewidth, trim=25pt 0pt 38pt -10pt, clip]{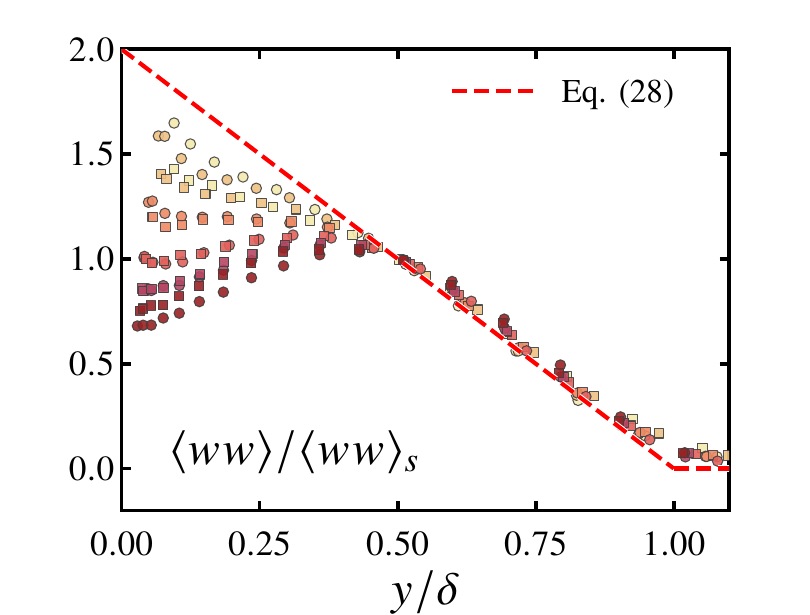}
            \put(1,88){\small(\textit{k})}
        \end{overpic}
    \end{subfigure}
        \begin{subfigure}{0.32\textwidth}
        \centering
        \begin{overpic}[width=\linewidth, trim=25pt 0pt 38pt -10pt, clip]{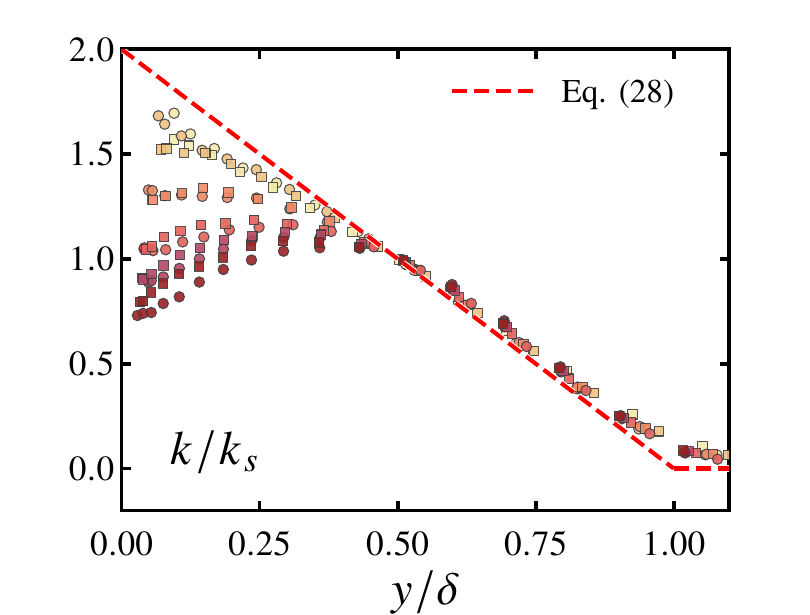}
            \put(1,88){\small(\textit{l})}
        \end{overpic}
    \end{subfigure}
    \caption{Comparisons with the experimental data of \citet{marusic_wall_wake_1995}.}
    \label{fig:Marusic95-1}
\end{figure}

\begin{figure}[!htbp]
    \centering
    \begin{subfigure}{0.32\textwidth}
        \centering
        \begin{overpic}[width=\linewidth, trim=25pt 45pt 38pt -10pt, clip]{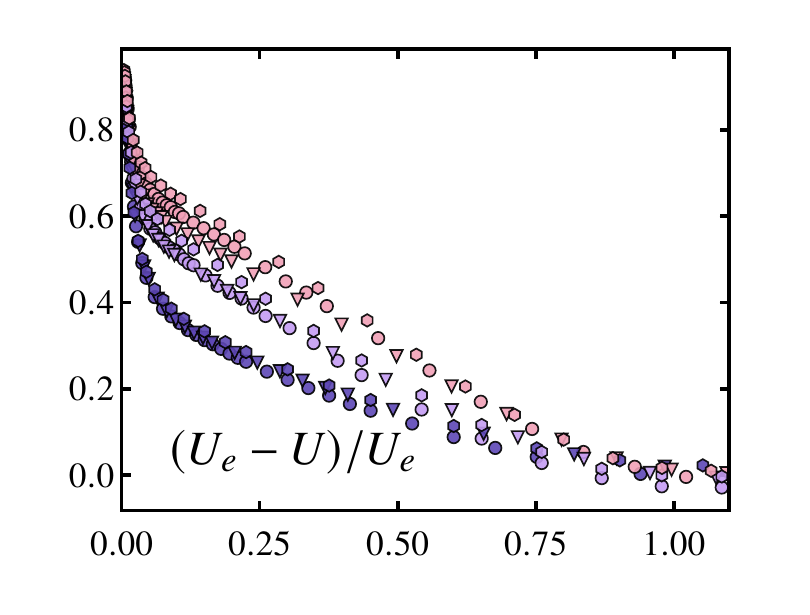}
            \put(1,74){\small(\textit{a})}
        \end{overpic}
    \end{subfigure}
        \begin{subfigure}{0.32\textwidth}
        \centering
        \begin{overpic}[width=\linewidth, trim=25pt 45pt 38pt -10pt, clip]{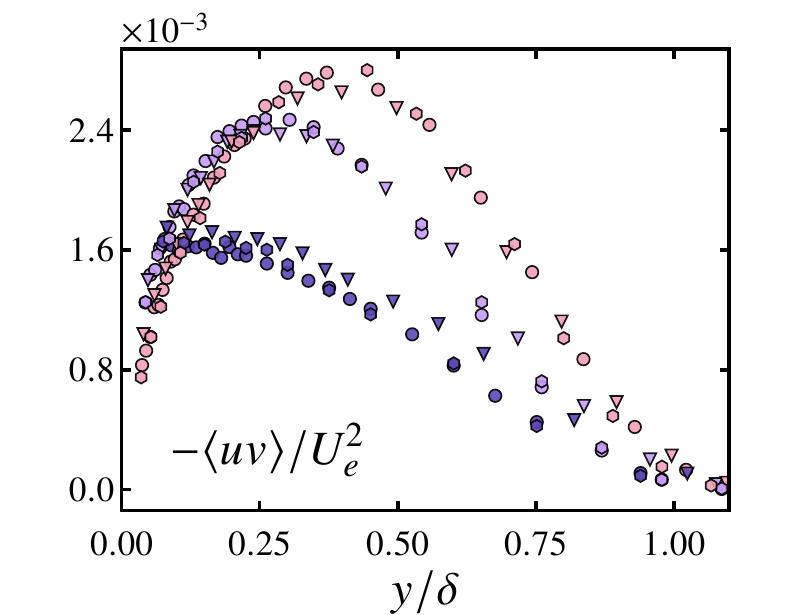}
            \put(1,74){\small(\textit{b})}
        \end{overpic}
    \end{subfigure}
    \begin{subfigure}{0.32\textwidth}
        \centering
        \begin{overpic}[width=\linewidth, trim=25pt 45pt 38pt -10pt, clip]{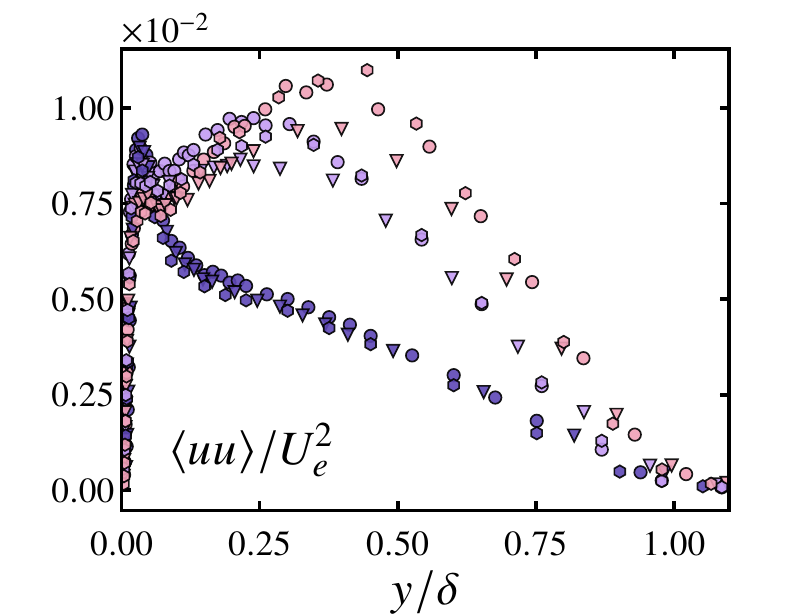}
            \put(1,74){\small(\textit{c})}
        \end{overpic}
    \end{subfigure}\\
        \begin{subfigure}{0.32\textwidth}
        \centering
        \begin{overpic}[width=\linewidth, trim=25pt 45pt 38pt -10pt, clip]{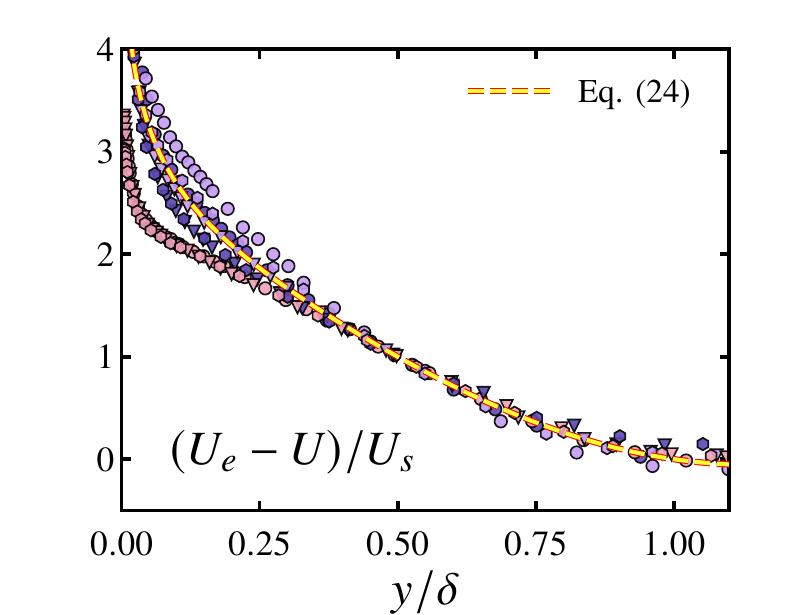}
            \put(1,74){\small(\textit{d})}
        \end{overpic}
    \end{subfigure}
        \begin{subfigure}{0.32\textwidth}
        \centering
        \begin{overpic}[width=\linewidth, trim=25pt 45pt 38pt -10pt, clip]{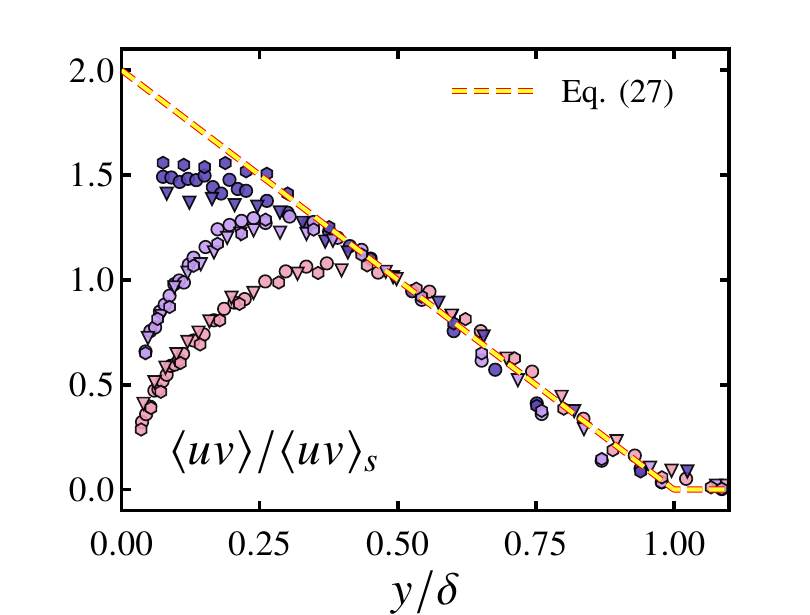}
            \put(1,74){\small(\textit{e})}
        \end{overpic}
    \end{subfigure}
    \begin{subfigure}{0.32\textwidth}
        \centering
        \begin{overpic}[width=\linewidth, trim=25pt 45pt 38pt -10pt, clip]{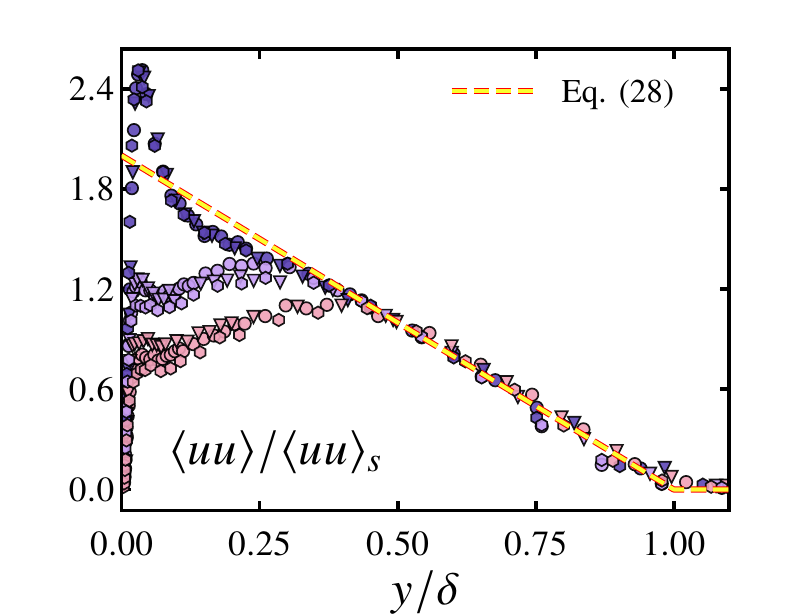}
            \put(1,74){\small(\textit{f})}
        \end{overpic}
    \end{subfigure}\\
    \begin{subfigure}{0.32\textwidth}
        \centering
        \begin{overpic}[width=\linewidth, trim=25pt 45pt 38pt -10pt, clip]{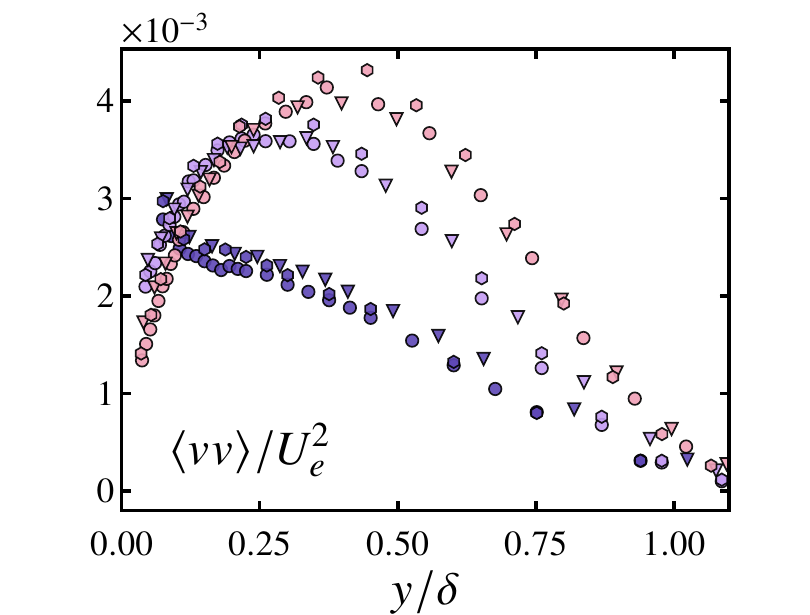}
            \put(1,74){\small(\textit{g})}
        \end{overpic}
    \end{subfigure}
    \begin{subfigure}{0.32\textwidth}
        \centering
        \begin{overpic}[width=\linewidth, trim=25pt 45pt 38pt -10pt, clip]{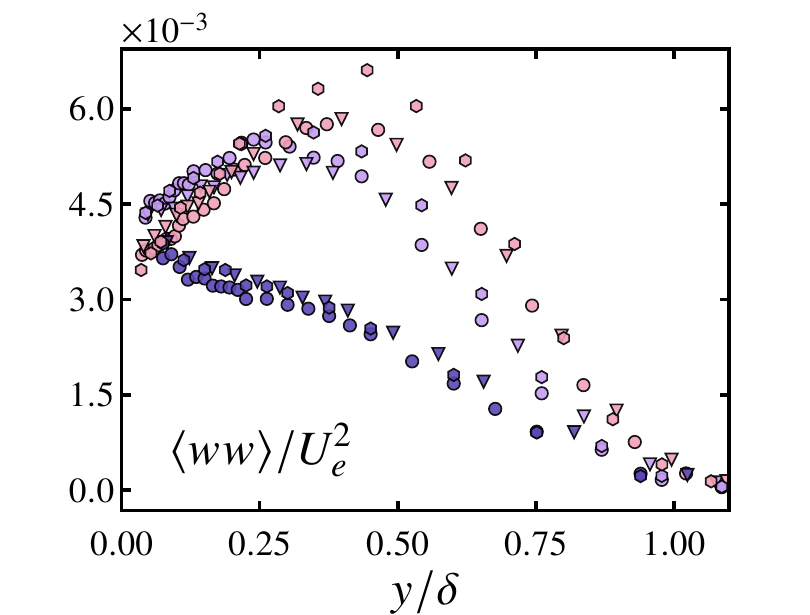}
            \put(1,74){\small(\textit{h})}
        \end{overpic}
    \end{subfigure}
    \begin{subfigure}{0.32\textwidth}
        \centering
        \begin{overpic}[width=\linewidth, trim=25pt 45pt 38pt -10pt, clip]{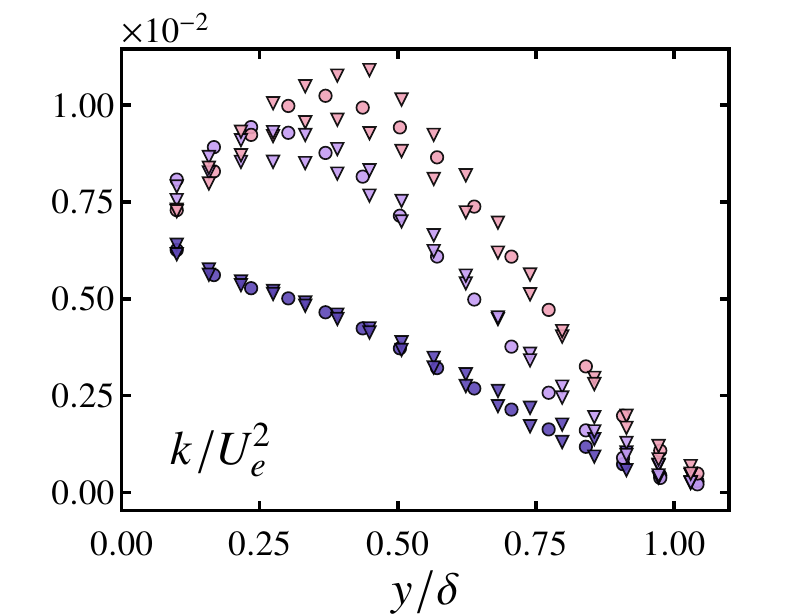}
            \put(1,74){\small(\textit{i})}
        \end{overpic}
    \end{subfigure}\\
        \begin{subfigure}{0.32\textwidth}
        \centering
        \begin{overpic}[width=\linewidth, trim=25pt 0pt 38pt -10pt, clip]{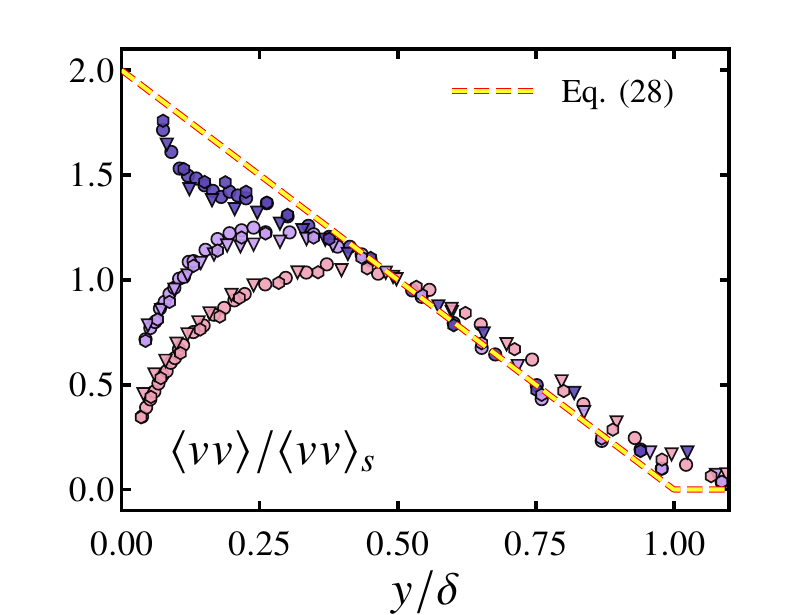}
            \put(1,88){\small(\textit{j})}
        \end{overpic}
    \end{subfigure}
    \begin{subfigure}{0.32\textwidth}
        \centering
        \begin{overpic}[width=\linewidth, trim=25pt 0pt 38pt -10pt, clip]{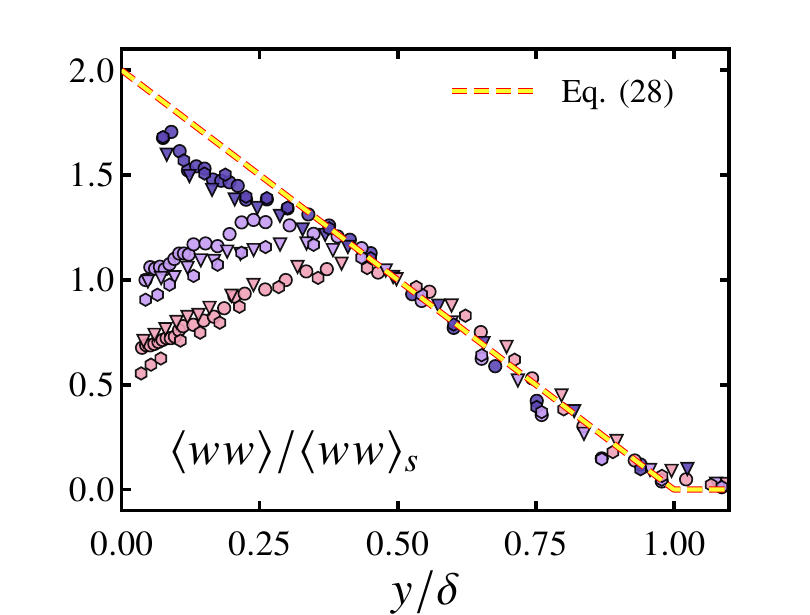}
            \put(1,88){\small(\textit{k})}
        \end{overpic}
    \end{subfigure}
    \begin{subfigure}{0.32\textwidth}
        \centering
        \begin{overpic}[width=\linewidth, trim=25pt 0pt 38pt -10pt, clip]{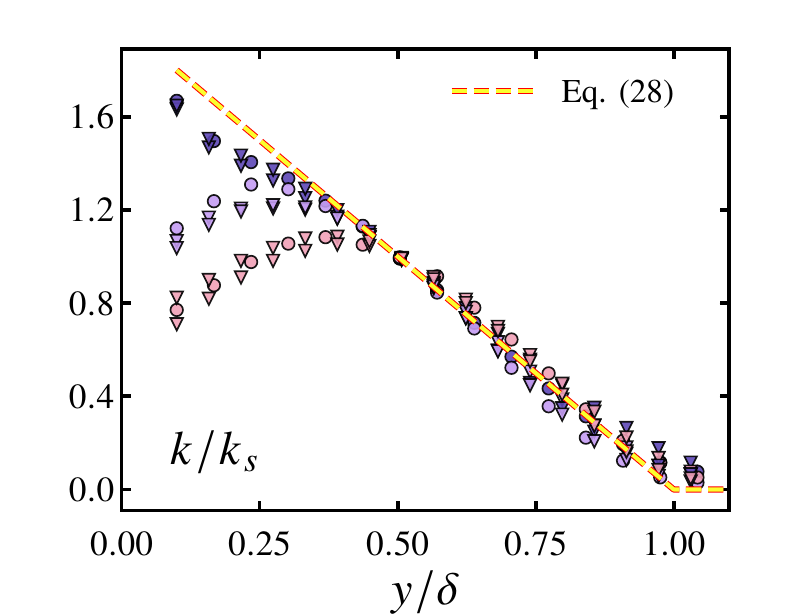}
            \put(1,88){\small(\textit{l})}
        \end{overpic}
    \end{subfigure}
    \caption{Comparisons with the experimental data of \citet{Dengel_Fernholz_1990}.}
    \label{fig:Dengel90-1}
\end{figure}

In the measurements of \citet{marusic_wall_wake_1995}, there are two cases with similar APG conditions but different Reynolds numbers, denoted by M01-M06 and M11-M16 in table \ref{tab:exp-data}, respectively.
Both cases develop from $\beta_{e} \approx 0$ to moderate $\beta_{e}$ values at the end of the measured region ($\beta_e=7.16$ for M06 and $\beta_e=6.07$ for M16), with $Re_\theta$ increasing from 2206 to 7257 for the first case and 6430 to 19133 for the second.
Note that the original pressure gradient parameter $\beta$ is regarded as $\beta_e$ in the present definition because it is obtained based on the freestream velocity gradient.
The data of \citet{Dengel_Fernholz_1990} were measured in an axisymmetric TBL at the verge of separation, with $Re_\theta$ ranging from 4000 to 11000 and $\beta_{e}$ from 4 to 18, as summarized in Table \ref{tab:exp-data}.
In Figs.~\ref{fig:Marusic95-1} and \ref{fig:Dengel90-1}, dashed lines represent \eqref{eq:Udef-sim}, \eqref{eq:uv-sim}, and \eqref{eq:sim-uiuj} for the velocity defect, Reynolds shear stress, and Reynolds normal stresses, respectively.

Comparisons with the measurements of \citet{marusic_wall_wake_1995} are shown in Fig.~\ref{fig:Marusic95-1}.
The mean velocity defect $U_e-U$, the Reynolds stresses $-\langle uv\rangle$ and $\langle uu \rangle$ are presented in Figs.~\ref{fig:Marusic95-1}(\textit{a-c}) scaled by $U_e$, and in Figs.~\ref{fig:Marusic95-1}(\textit{d-f}) under the component-wise normalization of \eqref{eq:scale}.
Under the component-wise normalization, the APG-dependent scatter is substantially reduced in the outer region $y/\delta>0.4$, and the profiles align closely with the expressions \eqref{eq:Udef-sim} for $U$, \eqref{eq:uv-sim} for $-\langle uv \rangle$ and \eqref{eq:sim-uiuj} for $\langle uu\rangle$.
Similarly, Figs.~\ref{fig:Marusic95-1}(\textit{g-i}) display the Reynolds stresses $\langle vv\rangle$ and $\langle ww \rangle$ and the kinetic energy $k=(\langle uu\rangle+\langle vv\rangle+\langle ww\rangle)/2$ in $U_e$ scale, and (\textit{j-l}) under component-wise normalization {(with $k_s$ defined at the same reference location as in \eqref{eq:scale} and sharing the Reynolds-stress profile expression in \eqref{eq:sim-uiuj})}.
The outer-region APG scatter is likewise reduced by the normalization in \eqref{eq:scale}.

The same conclusion holds for the dataset of \citet{Dengel_Fernholz_1990}, as shown in Fig.~\ref{fig:Dengel90-1}.
Using the same layout, the velocity defect and Reynolds stresses exhibit consistent outer-region collapse for $y/\delta>0.4$ in Figs.~\ref{fig:Dengel90-1}(\textit{a-f}) under the component-wise normalization of \eqref{eq:scale}, rather than under $U_e$ scaling.
In this region, the profiles follow the expressions \eqref{eq:Udef-sim}, \eqref{eq:uv-sim} and \eqref{eq:sim-uiuj} closely.
The remaining components $\langle vv\rangle$, $\langle ww\rangle$ and $k$ are presented in Figs.~\ref{fig:Dengel90-1}(\textit{g-l}), where the APG scatter is likewise reduced.
For $\langle vv\rangle$, the prediction of \eqref{eq:sim-uiuj} slightly underestimates the data of \citet{marusic_wall_wake_1995} but agrees closely with the data of \citet{Dengel_Fernholz_1990}.
Based on our numerical setup, this discrepancy may be attributed to the influence of the outer boundary condition on wall-normal fluctuations.


\subsection{Verifications with numerical data}
\begin{figure}[!htbp]
    \centering
    \begin{subfigure}{0.32\textwidth}
        \centering
        \begin{overpic}[width=\linewidth, trim=25pt 45pt 38pt -10pt, clip]{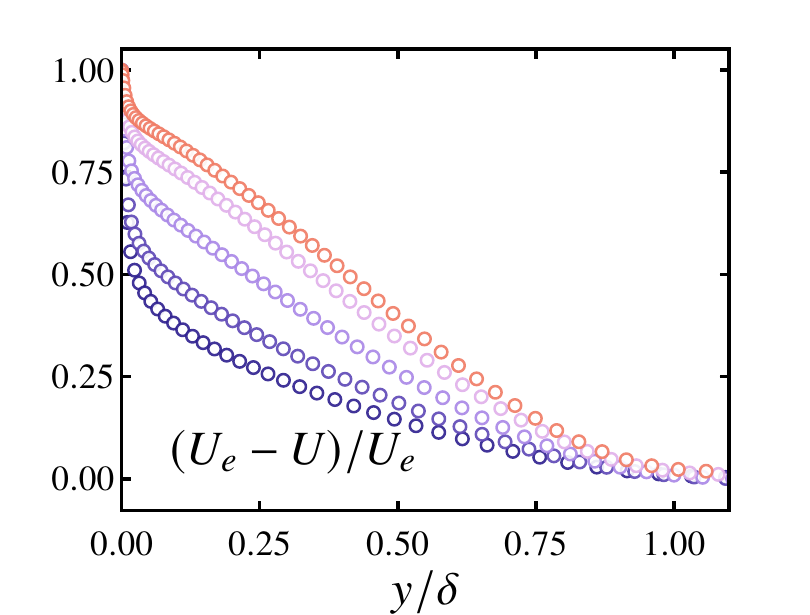}
            \put(1,74){\small(\textit{a})}
        \end{overpic}
    \end{subfigure}
    \begin{subfigure}{0.32\textwidth}
        \centering
        \begin{overpic}[width=\linewidth, trim=25pt 45pt 38pt -10pt, clip]{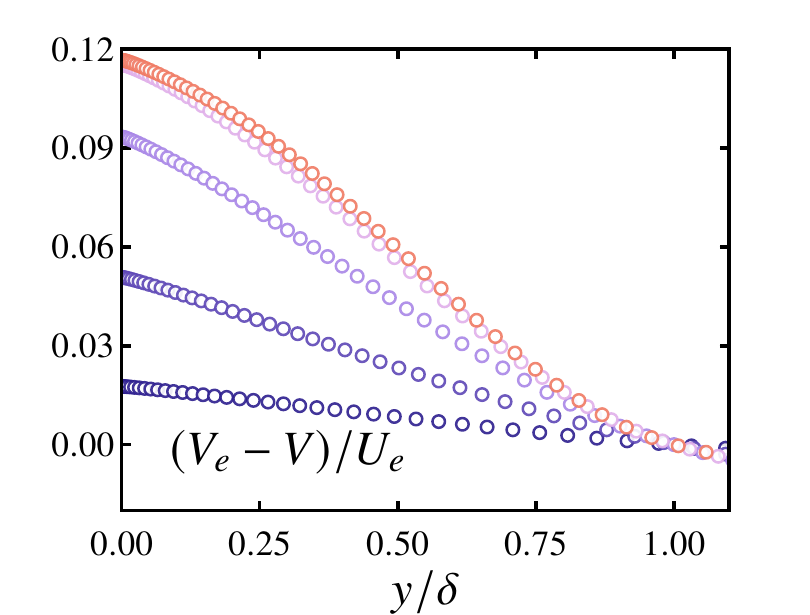}
            \put(1,74){\small(\textit{b})}
        \end{overpic}
    \end{subfigure}
    \begin{subfigure}{0.32\textwidth}
        \centering
        \begin{overpic}[width=\linewidth, trim=25pt 45pt 38pt -10pt, clip]{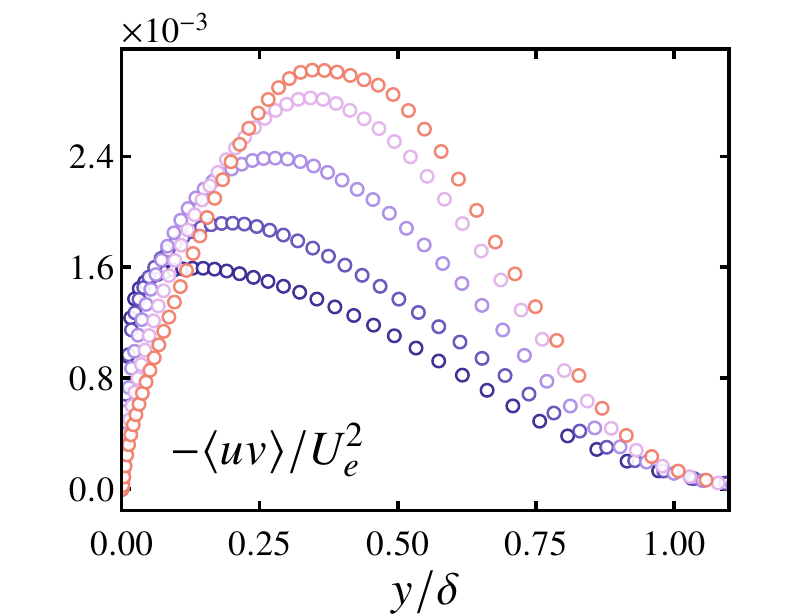}
            \put(1,74){\small(\textit{c})}
        \end{overpic}
    \end{subfigure}\\
        \begin{subfigure}{0.32\textwidth}
        \centering
        \begin{overpic}[width=\linewidth, trim=25pt 45pt 38pt -10pt, clip]{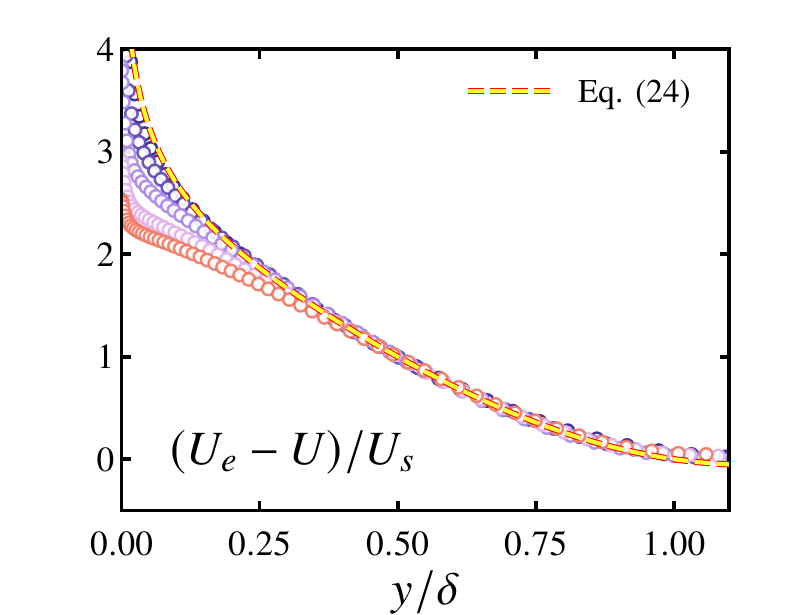}
            \put(1,74){\small(\textit{d})}
        \end{overpic}
    \end{subfigure}
    \begin{subfigure}{0.32\textwidth}
        \centering
        \begin{overpic}[width=\linewidth, trim=25pt 45pt 38pt -10pt, clip]{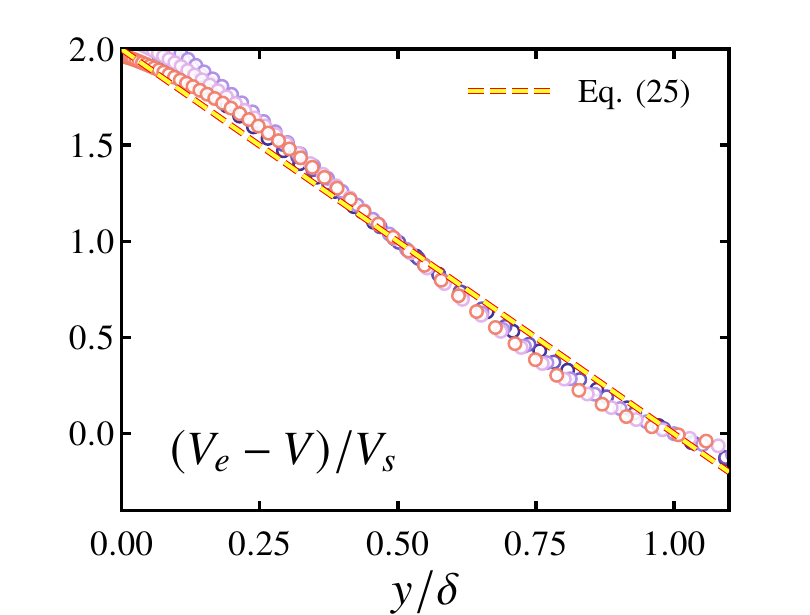}
            \put(1,74){\small(\textit{e})}
        \end{overpic}
    \end{subfigure}
    \begin{subfigure}{0.32\textwidth}
        \centering
        \begin{overpic}[width=\linewidth, trim=25pt 45pt 38pt -10pt, clip]{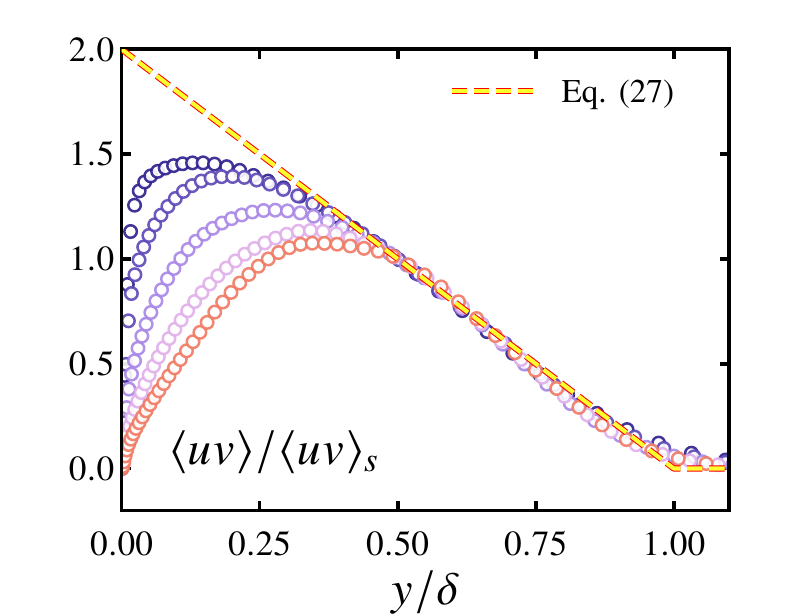}
            \put(1,74){\small(\textit{f})}
        \end{overpic}
    \end{subfigure}\\
    \begin{subfigure}{0.32\textwidth}
        \centering
        \begin{overpic}[width=\linewidth, trim=25pt 45pt 38pt -10pt, clip]{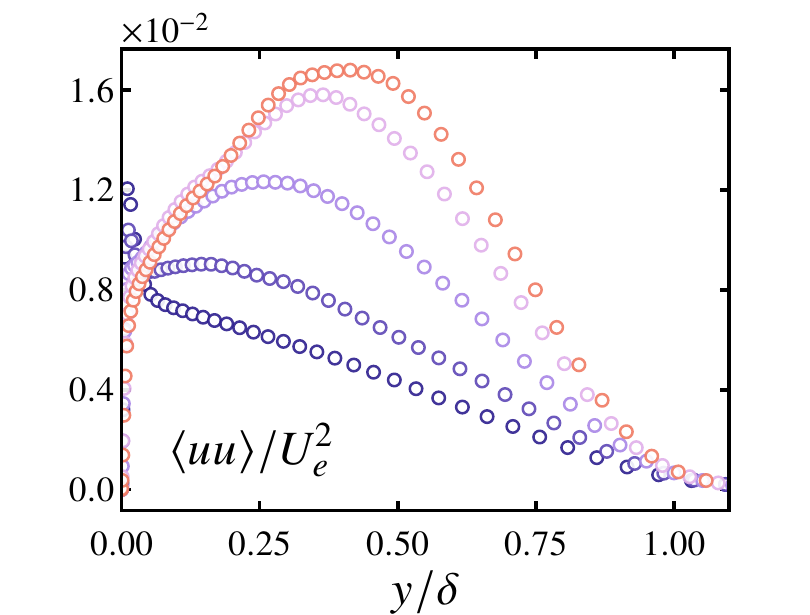}
            \put(1,74){\small(\textit{g})}
        \end{overpic}
    \end{subfigure}
    \begin{subfigure}{0.32\textwidth}
        \centering
        \begin{overpic}[width=\linewidth, trim=25pt 45pt 38pt -10pt, clip]{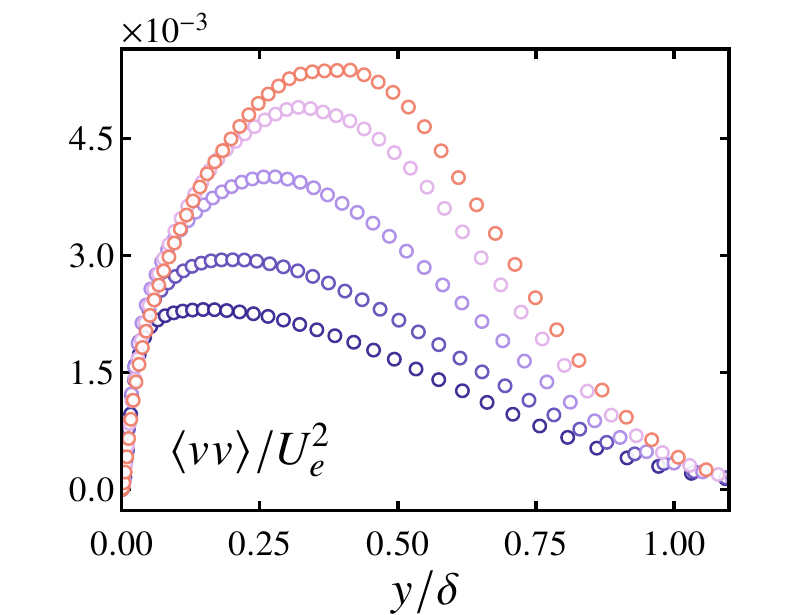}
            \put(1,74){\small(\textit{h})}
        \end{overpic}
    \end{subfigure}
    \begin{subfigure}{0.32\textwidth}
        \centering
        \begin{overpic}[width=\linewidth, trim=25pt 45pt 38pt -10pt, clip]{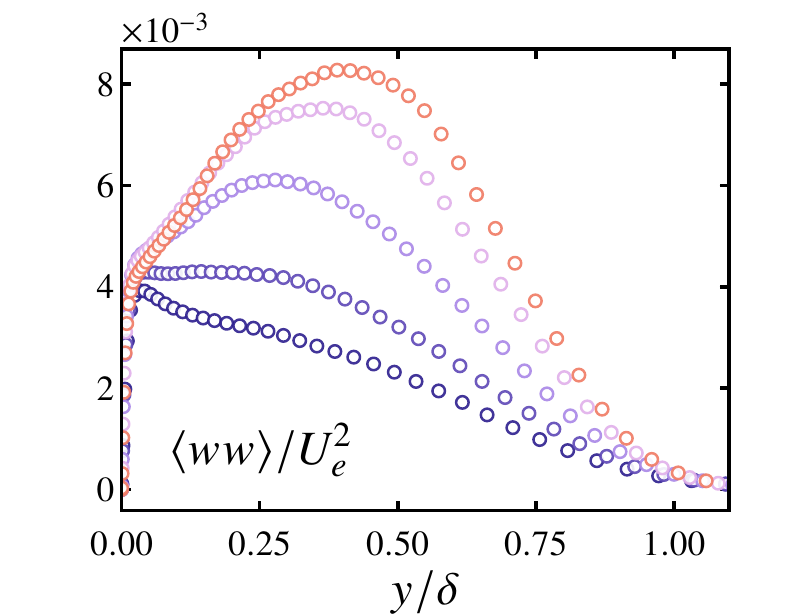}
            \put(1,74){\small(\textit{i})}
        \end{overpic}
    \end{subfigure}\\
        \begin{subfigure}{0.32\textwidth}
        \centering
        \begin{overpic}[width=\linewidth, trim=25pt 0pt 38pt -10pt, clip]{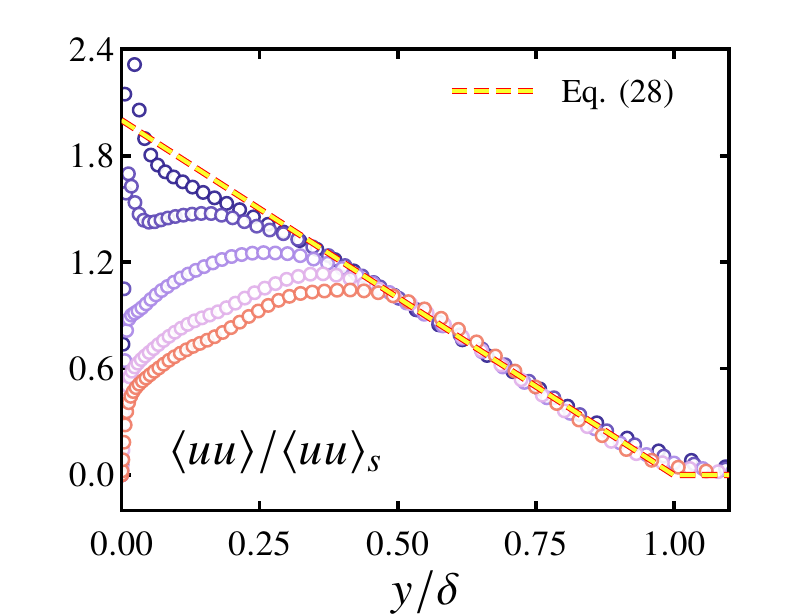}
            \put(1,88){\small(\textit{j})}
        \end{overpic}
    \end{subfigure}
    \begin{subfigure}{0.32\textwidth}
        \centering
        \begin{overpic}[width=\linewidth, trim=25pt 0pt 38pt -10pt, clip]{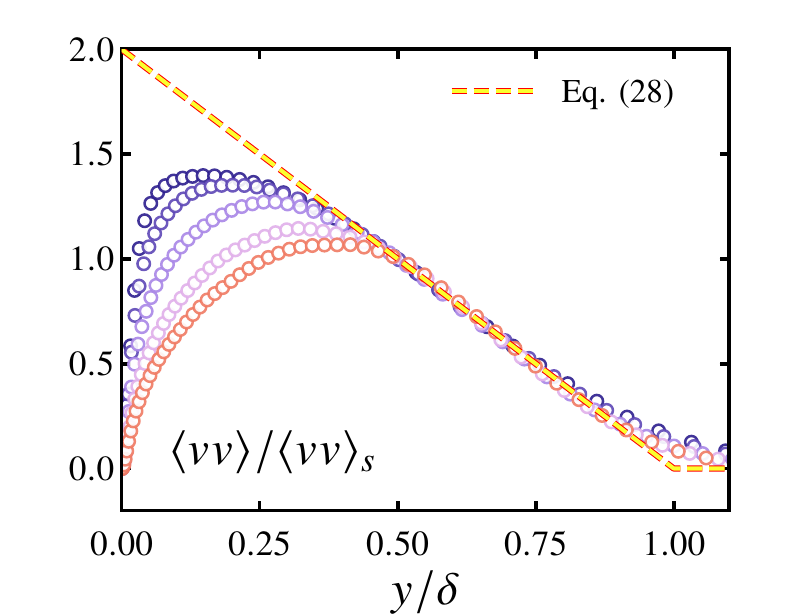}
            \put(1,88){\small(\textit{k})}
        \end{overpic}
    \end{subfigure}
    \begin{subfigure}{0.32\textwidth}
        \centering
        \begin{overpic}[width=\linewidth, trim=25pt 0pt 38pt -10pt, clip]{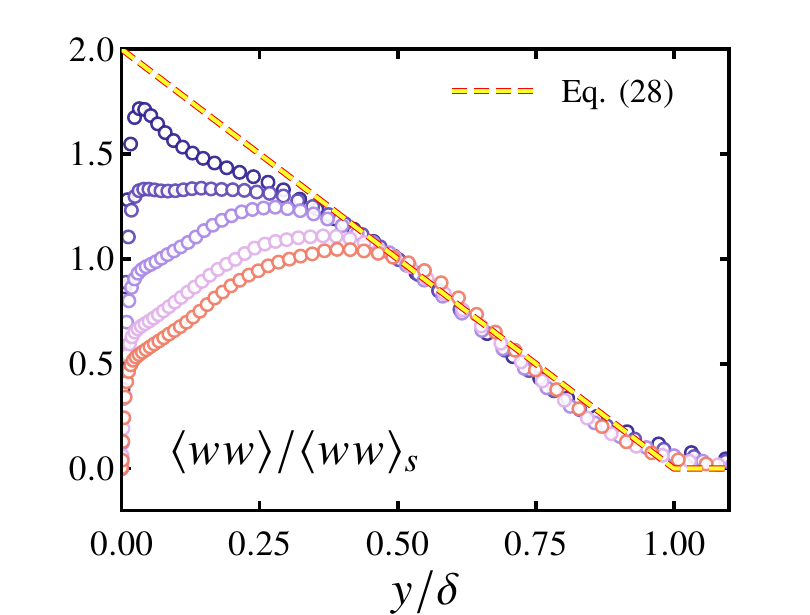}
            \put(1,88){\small(\textit{l})}
        \end{overpic}
    \end{subfigure}
    \caption{Comparisons with the numerical data of \citet{coleman_numerical_2021}.}
    \label{fig:Coleman21-1}
\end{figure}

\begin{figure}[!htbp]
    \centering
    \begin{subfigure}{0.32\textwidth}
        \centering
        \begin{overpic}[width=\linewidth, trim=25pt 45pt 38pt -10pt, clip]{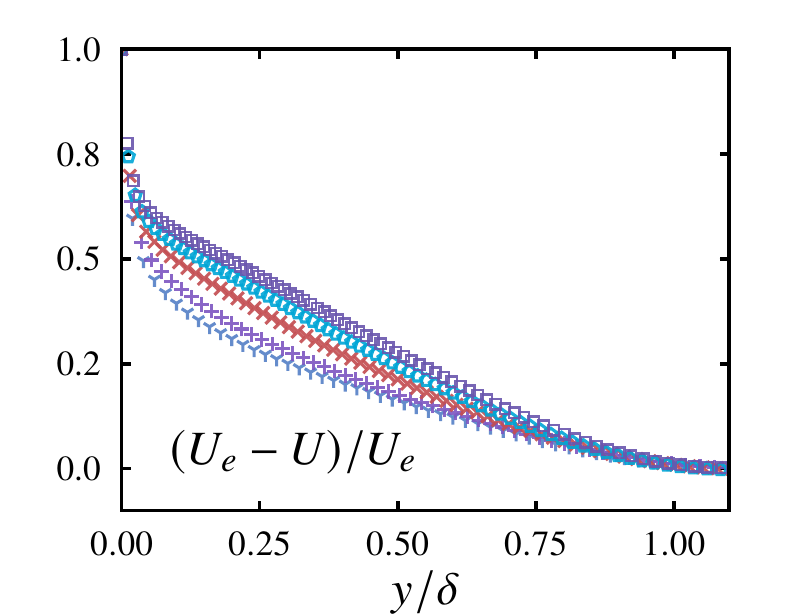}
            \put(1,74){\small(\textit{a})}
        \end{overpic}
    \end{subfigure}
    \begin{subfigure}{0.32\textwidth}
        \centering
        \begin{overpic}[width=\linewidth, trim=25pt 45pt 38pt -10pt, clip]{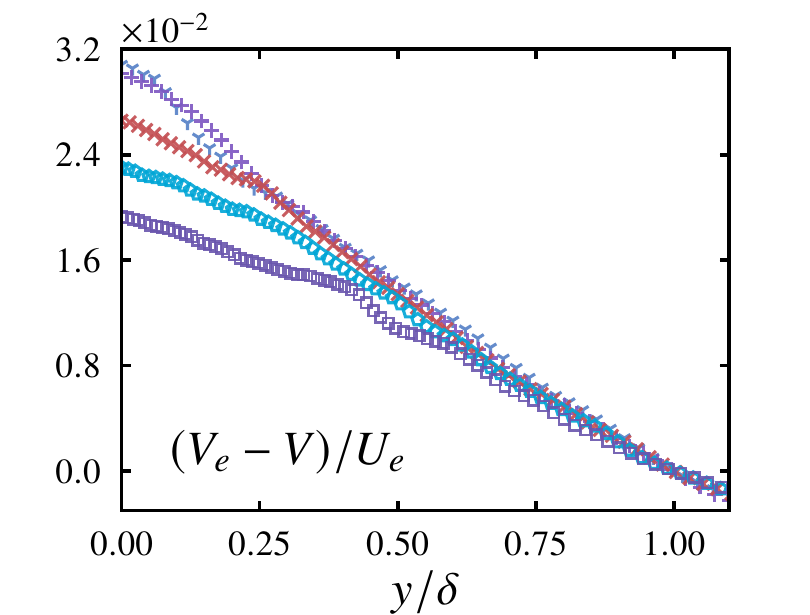}
            \put(1,74){\small(\textit{b})}
        \end{overpic}
    \end{subfigure}
    \begin{subfigure}{0.32\textwidth}
        \centering
        \begin{overpic}[width=\linewidth, trim=25pt 45pt 38pt -10pt, clip]{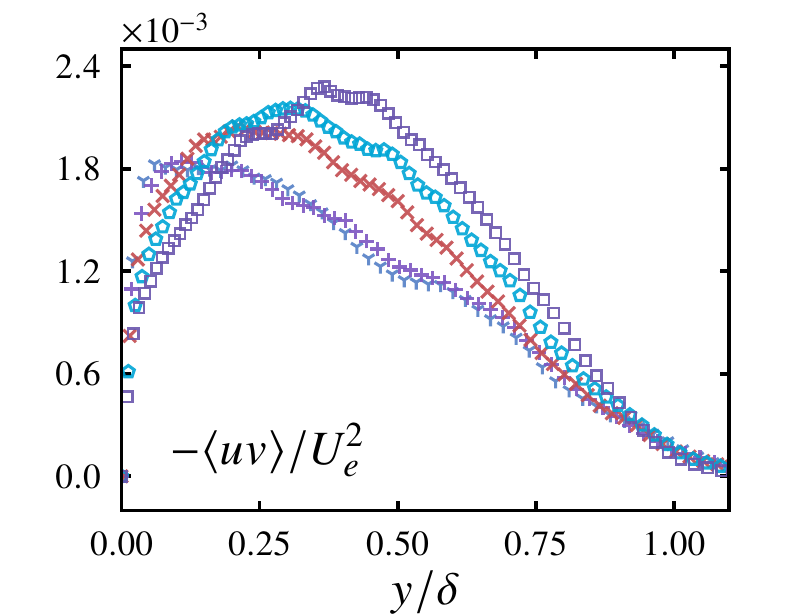}
            \put(1,74){\small(\textit{c})}
        \end{overpic}
    \end{subfigure}\\
        \begin{subfigure}{0.32\textwidth}
        \centering
        \begin{overpic}[width=\linewidth, trim=25pt 45pt 38pt -10pt, clip]{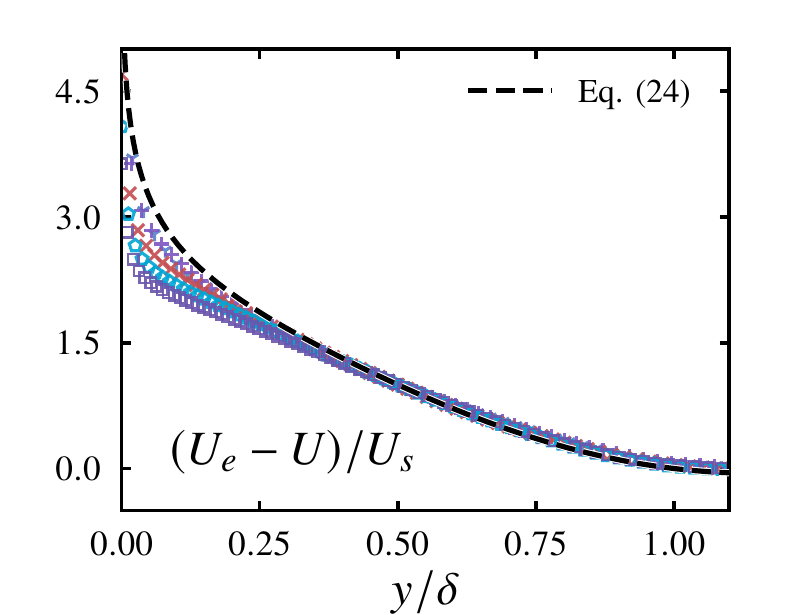}
            \put(1,74){\small(\textit{d})}
        \end{overpic}
    \end{subfigure}
    \begin{subfigure}{0.32\textwidth}
        \centering
        \begin{overpic}[width=\linewidth, trim=25pt 45pt 38pt -10pt, clip]{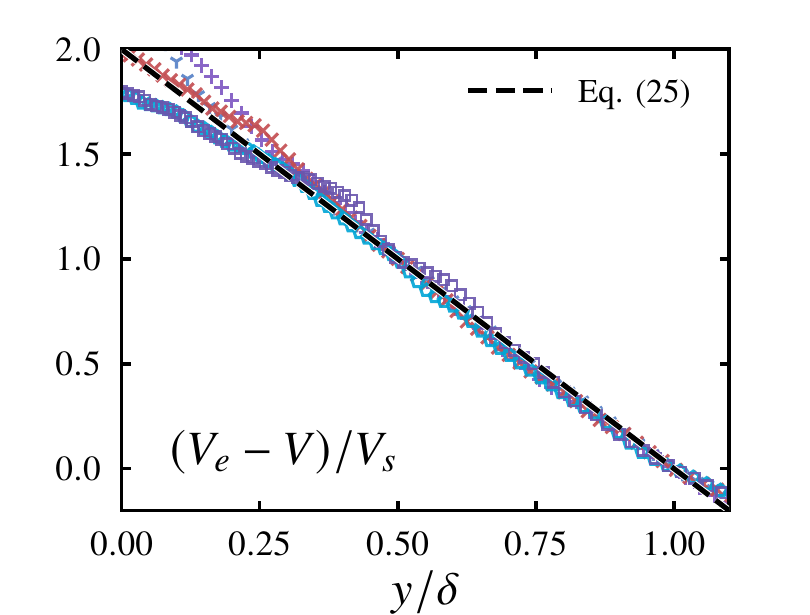}
            \put(1,74){\small(\textit{e})}
        \end{overpic}
    \end{subfigure}
    \begin{subfigure}{0.32\textwidth}
        \centering
        \begin{overpic}[width=\linewidth, trim=25pt 45pt 38pt -10pt, clip]{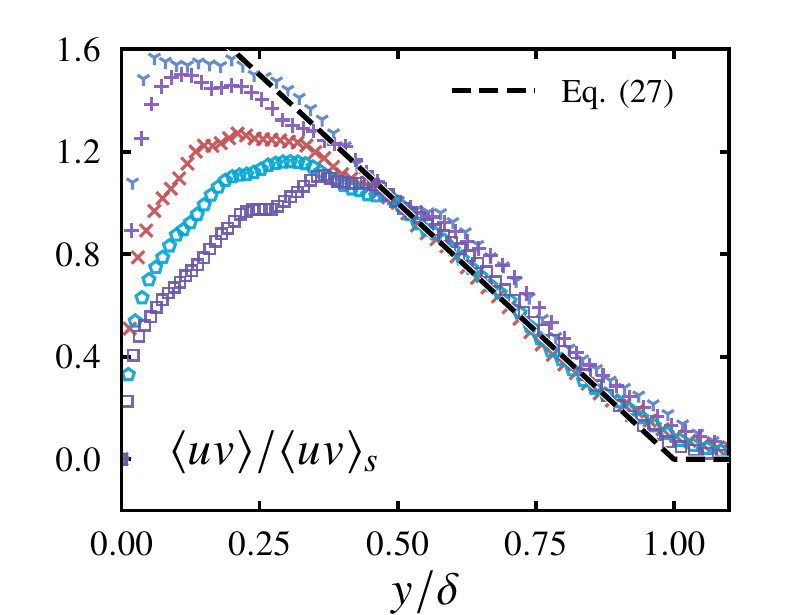}
            \put(1,74){\small(\textit{f})}
        \end{overpic}
    \end{subfigure}\\
      \begin{subfigure}{0.32\textwidth}
        \centering
        \begin{overpic}[width=\linewidth, trim=25pt 45pt 38pt -10pt, clip]{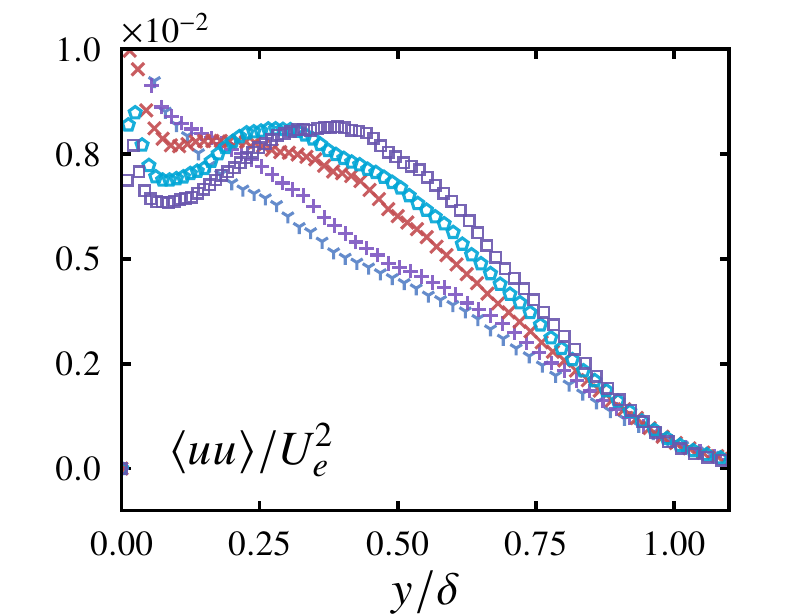}
            \put(1,74){\small(\textit{g})}
        \end{overpic}
    \end{subfigure}
    \begin{subfigure}{0.32\textwidth}
        \centering
        \begin{overpic}[width=\linewidth, trim=25pt 45pt 38pt -10pt, clip]{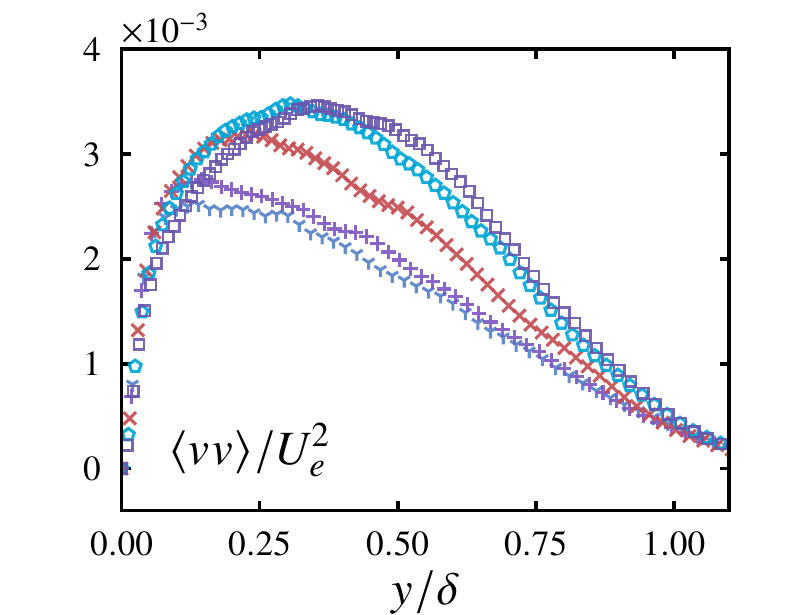}
            \put(1,74){\small(\textit{h})}
        \end{overpic}
    \end{subfigure}
    \begin{subfigure}{0.32\textwidth}
        \centering
        \begin{overpic}[width=\linewidth, trim=25pt 45pt 38pt -10pt, clip]{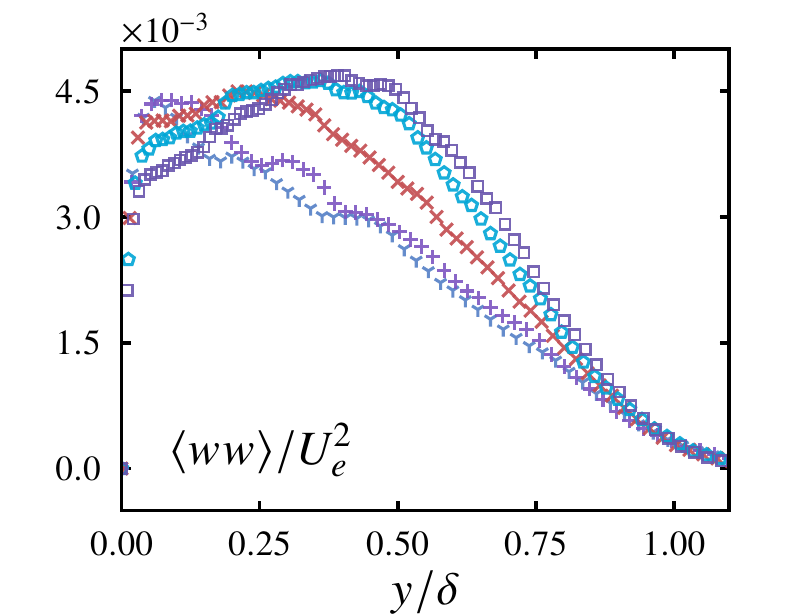}
            \put(1,74){\small(\textit{i})}
        \end{overpic}
    \end{subfigure}\\
        \begin{subfigure}{0.32\textwidth}
        \centering
        \begin{overpic}[width=\linewidth, trim=25pt 0pt 38pt -10pt, clip]{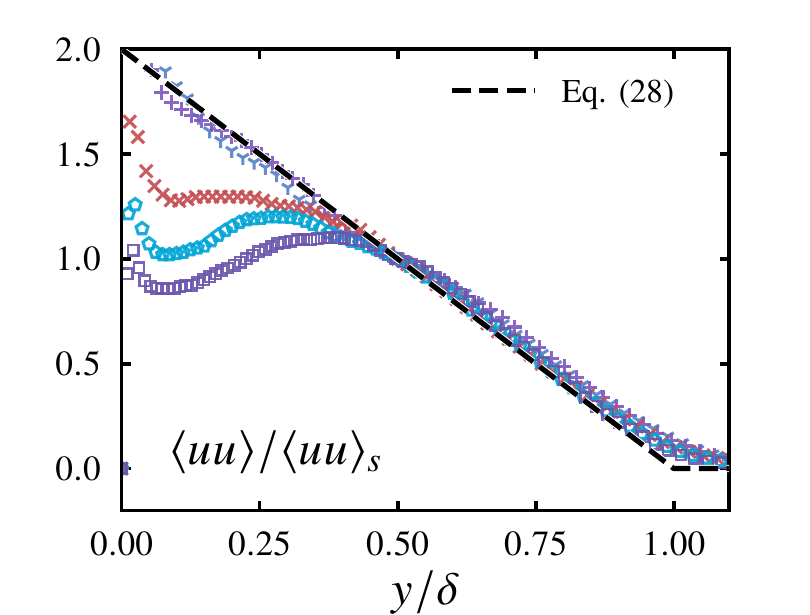}
            \put(1,88){\small(\textit{j})}
        \end{overpic}
    \end{subfigure}
    \begin{subfigure}{0.32\textwidth}
        \centering
        \begin{overpic}[width=\linewidth, trim=25pt 0pt 38pt -10pt, clip]{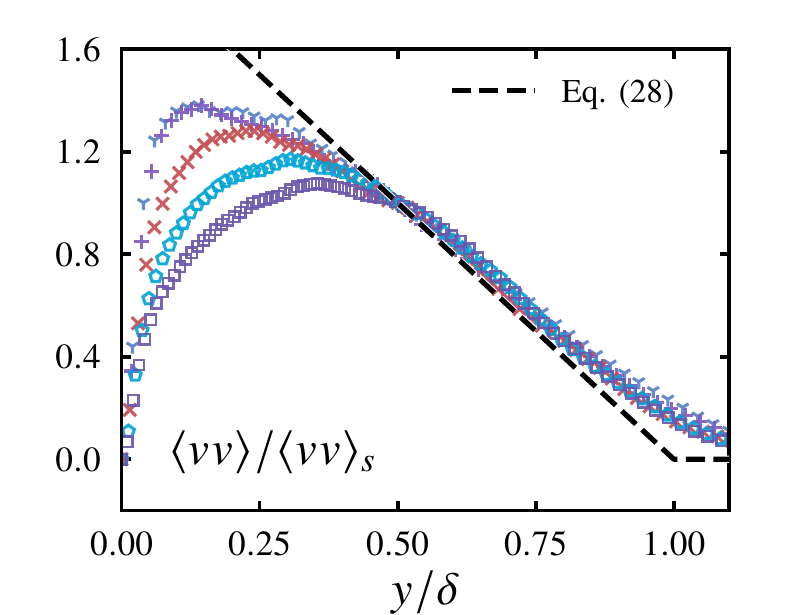}
            \put(1,88){\small(\textit{k})}
        \end{overpic}
    \end{subfigure}
    \begin{subfigure}{0.32\textwidth}
        \centering
        \begin{overpic}[width=\linewidth, trim=25pt 0pt 38pt -10pt, clip]{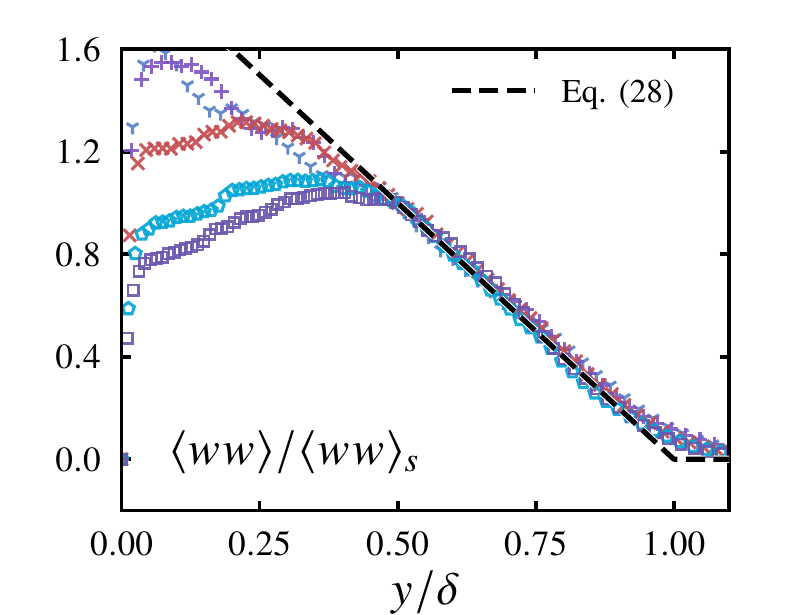}
            \put(1,88){\small(\textit{l})}
        \end{overpic}
    \end{subfigure}
    \caption{Comparisons with the numerical data of \citet{rkein_direct_2023}.}
    \label{fig:Rkein23-1}
\end{figure}

This subsection further assesses the component-wise outer similarity and the associated reference-value normalization using two additional numerical data sets from \citet{coleman_numerical_2021} and \citet{rkein_direct_2023}.
These data sets provide independent tests over two different APG configurations: a near-separation DNS with strong streamwise variation of the pressure gradient, and a near-equilibrium DNS with moderate APG strength across a relatively broad Reynolds-number range.
In Figs.~\ref{fig:Coleman21-1} and \ref{fig:Rkein23-1}, panels (\textit{a})--(\textit{c}) show $U_e-U$, $V_e-V$, and $-\langle uv \rangle$ scaled by freestream quantities, with panels (\textit{d})--(\textit{f}) showing the corresponding component-wise-normalized results; panels (\textit{g})--(\textit{l}) use the same arrangement for $\langle uu\rangle$, $\langle vv \rangle$, and $\langle ww \rangle$.
Dashed lines represent \eqref{eq:Udef-sim}, \eqref{eq:Vdf-sim}, \eqref{eq:uv-sim}, and \eqref{eq:sim-uiuj}.

The DNS of \citet{coleman_numerical_2021} tests the robustness of the component-wise normalization under strong APG conditions.
It considers a turbulent boundary layer subject to a designed suction-only condition, under which the pressure gradient remains adverse along the streamwise direction.
The boundary layer is maintained close to incipient separation, while no mean separation bubble is formed.
Several streamwise stations from Case E of \citet{coleman_numerical_2021} are included and denoted as C11--C15 in table \ref{tab:simu-data}.
These stations cover a wide range of APG strengths, with the pressure-gradient parameter increasing from a mild value of $\beta_w=1.03$ to large values of $\beta_w$ up to 80, while the momentum-thickness Reynolds number increases from approximately 8000 to 17000.
As a complementary near-equilibrium case, the DNS data of \citet{rkein_direct_2023} are further included to provide additional validation of the component-wise outer similarity in APG boundary layers at moderate pressure-gradient strength.
In this data set, the pressure-gradient parameter is approximately maintained in the range $\beta_e\approx2$--$3$, while the Reynolds number spans a relatively broad range.
Five profiles with momentum-thickness Reynolds numbers between 2500 and 5000 are selected and denoted as R01--R05 in table \ref{tab:simu-data}.

Fig. \ref{fig:Coleman21-1} presents the comparison based on the DNS data of \citet{coleman_numerical_2021}.
Panels (\textit{a})--(\textit{c}) show the mean velocity defects $U_e-U$ and $V_e-V$ and the Reynolds shear stress $-\langle uv\rangle$ normalized by the freestream scales, while panels (\textit{d})--(\textit{f}) show the corresponding results under component-wise normalization.
Panels (\textit{g})--(\textit{l}) further compare the Reynolds normal stresses $\langle uu\rangle$, $\langle vv\rangle$, and $\langle ww\rangle$ using the same arrangement, with the freestream-scaled results followed by the component-wise-normalized results.
Under freestream normalization, all profiles retain clear APG-dependent trends similar to those observed in section \ref{sec:self-sim}.
In contrast, component-wise normalization produces a consistent collapse of the outer-region profiles: the collapse is observed approximately for $y/\delta \gtrsim 0.4$ in $U_e-U$ and the Reynolds stresses and extends farther toward the wall for $V_e-V$.
This consistent outer-region collapse across the full range of APG strengths considered here supports the component-wise outer similarity.
For the Reynolds stresses, the lower wall-normal limit of the collapsed region remains associated with the outer peak location, which shifts away from the wall with increasing APG strength, as directly seen in Fig. \ref{fig:Coleman21-1}, and also exhibits a Reynolds-number-dependent outward shift when Fig. \ref{fig:Main-Data-3} is compared with Fig. \ref{fig:Coleman21-1}.

Fig. \ref{fig:Rkein23-1} presents the comparison based on the DNS data of \citet{rkein_direct_2023}, following the same panel arrangement as Fig. \ref{fig:Coleman21-1}.
Panels (\textit{a})--(\textit{f}) compare $U_e-U$, $V_e-V$, and $-\langle uv\rangle$ under freestream and component-wise normalizations, while panels (\textit{g})--(\textit{l}) show the corresponding results for $\langle uu\rangle$, $\langle vv\rangle$, and $\langle ww\rangle$.
With the component-wise normalization, the selected profiles exhibit clear outer-region collapse and agree well with the profile functions proposed in subsection \ref{sec:present-scaling}.
This provides further support for the component-wise outer similarity and its associated reference-value normalization in near-equilibrium APG TBLs at moderate pressure-gradient strength over a relatively broad Reynolds-number range.
Despite the small variation in the pressure-gradient parameter, the Reynolds-stress peaks shift clearly outward with increasing Reynolds number, suggesting a possible influence of accumulated upstream history.
If this shift moves the peak beyond $y/\delta=0.5$ at higher Reynolds numbers, the reference location used in the present normalization should be shifted farther from the wall accordingly; nevertheless, the outer similarity remains robust beyond the stress-peak region.
A minor deviation appears in the $\langle vv\rangle$ profiles near the outer edge, where the scaled profiles are slightly elevated relative to the proposed expression.
Together with Fig. \ref{fig:Coleman21-1}, these results support the component-wise outer similarity and the applicability of its associated reference-value normalization to both strongly non-equilibrium and near-equilibrium APG TBLs, provided that the relevant streamwise dependencies are properly incorporated.
The present evidence supports this component-wise outer similarity for APG TBLs with statistically spanwise-uniform mean flow, over conditions ranging from mild APG to near separation in the benchmark data.
Extensions to flows with strongly three-dimensional mean fields, highly compressible regimes, shock/boundary-layer interaction, or substantial surface curvature require additional validation.

\FloatBarrier
\section{Conclusions \label{conclusion} }
To study the outer similarity of TBLs with strong adverse-pressure gradient (APG) effects, we performed a DNS with a separation bubble under quasi-incompressible conditions.
The pressure gradient is imposed through an upper-boundary transpiration velocity, and the simulation results agree well with comparable published data through detailed comparisons of the boundary-layer thickness, momentum thickness, displacement thickness, shape factor, and related integral quantities.

Furthermore, component-wise outer similarity is observed for the mean-velocity defects and Reynolds stresses in the present simulation and is further supported by comparisons with published experimental and numerical data.
These data cover pressure-gradient parameters spanning four orders of magnitude ($|\beta_w|$ from $10^{-2}$ to $10^{2}$) and Reynolds numbers up to $Re_\theta=2\times10^4$.
In their respective outer regions, the quantities in \eqref{eq:sim-define} can be expressed as

\begin{equation}\label{eq:sim-general}
    \phi(x,y) = \phi_s(x)f_\phi(y/\delta),
\end{equation}
in which $\phi$ represents any of the following: $U_e-U$, $V_e-V$, $\langle uu\rangle$, $\langle vv\rangle$, $\langle ww\rangle$ and $\langle uv\rangle$.
The reference value $\phi_s$ is taken at a common wall-normal height $y_s$, \textit{e.g.}, $y_s=0.5\delta$; the general function $f_\phi$ depends on $\phi$ and $y/\delta$ but is independent of $x$.
Alternative choices such as $y_s=0.6\delta$ or $0.7\delta$ are also applicable and retain the component-wise outer similarity in the form \eqref{eq:sim-general}.
It should be noted that the present normalization strategy is not a direct replacement for the scaling arguments assessed in this study; however, it reveals clear outer self-similarity of the individual components.

Finally, explicit expressions for the normalized outer profiles are provided for all six quantities.
Under the present component-wise normalization, the streamwise velocity defect follows the classical wake formula across the APG range considered, and linear outer formulations are given for the wall-normal velocity defect and Reynolds stresses.
A slight discrepancy between the linear profile and $\langle vv \rangle$ appears in part of the benchmark data, but good agreement is found with our simulation and other published data sets.
Overall, these results support component-wise outer similarity under strong APG conditions, whereas no tested common normalization recovers all six outer-profile collapses.
The resulting profile functions may inform turbulence-model assessment in APG boundary layers, and extensions to more complex configurations require further validation.

\section*{Funding Sources}
This study is supported by grants from the National Natural Science Foundation of China (grant no. 92252201).

\section*{Acknowledgments}
 Special thanks go to B. Cantwell for helpful discussions in the initial stage of this work.
 The authors also express their gratitude to X.L. Li for accessing numerical simulation code, and to J.C He, M.M. Han, and T.T. Du and C. Yan for discussions.

\FloatBarrier
\bibliography{MyLibrary}


\end{document}